\documentclass[preprint]{elsarticle}
\usepackage[margin=2.5cm]{geometry} 

\usepackage{hyperref}

\usepackage{amsmath} 
\usepackage{amssymb} 
\usepackage{subcaption}
\usepackage{hyperref}
\hypersetup{
     colorlinks   = true,
     citecolor    = blue,
     linkcolor    = blue}
\usepackage{ulem}

\renewcommand{\v}[1]{\ensuremath{\mathbf{#1}}} 
 
\newcommand{\uv}[1]{\ensuremath{\mathbf{\hat{#1}}}} 
\newcommand{\abs}[1]{\left| #1 \right|} 
\newcommand{\avg}[1]{\left< #1 \right>} 
\newcommand{\pd}[2]{\frac{\partial #1}{\partial #2}} 
 
\newcommand{\grad}[1]{\nabla #1} 
\renewcommand{\div}[1]{\nabla \cdot #1} 
\let\baraccent=\= 
\renewcommand{\=}[1]{\stackrel{#1}{=}} 

\newcommand{\vpar}{v_\parallel}
\newcommand{\upar}{u_\parallel}

\newcommand{\Lx}{L_x}
\newcommand{\Ly}{L_y}
\newcommand{\Lz}{L_z}
\newcommand{\Nx}{N_x}
\newcommand{\Ny}{N_y}

\newcommand{\Nvpar}{N_{\vpar}}
\newcommand{\Nmu}{N_{\mu}}
\newcommand{\Dx}{\Delta x}
\newcommand{\Dy}{\Delta y}
\newcommand{\Dz}{\Delta z}
\newcommand{\numB}{N_b}

\newcommand{\dmu}{\text{d}\mu}
\newcommand{\dvpar}{\text{d}\vpar}

\newcommand{\dx}{\text{d}x}
\newcommand{\dy}{\text{d}y}
\newcommand{\dz}{\text{d}z}
\newcommand{\dxi}{\text{d}\xi}
\newcommand{\deta}{\text{d}\eta}
\newcommand{\bhat}{\uv{b}}

\newcommand{\kpar}{k_\parallel}
\newcommand{\kperp}{k_\perp}

\newenvironment{eqnal}{\equation\aligned}{\endaligned\endequation}

\newcommand{\ysh}{\mathcal{S}}
\newcommand{\yshh}{\mathcal{S}_h}

\newcommand{\yshhik}{\mathcal{S}_{hi,k}}

\newcommand{\ftar}{f_{\mathrm{tar}}}
\newcommand{\fdo}{f_{\mathrm{do}}}
\newcommand{\xilo}{\xi_{\mathrm{lo}}}
\newcommand{\xiup}{\xi_{\mathrm{up}}}

\newcommand{\xiloh}{\xi_{\mathrm{lo},h}}
\newcommand{\xiuph}{\xi_{\mathrm{up},h}}
\newcommand{\xiuphk}{\xi_{\mathrm{up},h,k}}
\newcommand{\etalo}{\eta_{\mathrm{lo}}}
\newcommand{\etaup}{\eta_{\mathrm{up}}}

\newcommand{\etaloh}{\eta_{\mathrm{lo},h}}

\newcommand{\itar}{i_{\mathrm{tar}}}
\newcommand{\jtar}{j_{\mathrm{tar}}}
\newcommand{\ido}{i_{\mathrm{do}}}
\newcommand{\jdo}{j_{\mathrm{do}}}
\newcommand{\Ndo}{N_{\mathrm{do}}}

\newcommand{\gdb}{{\large\texttt{GDB}}}
\newcommand{\gkeyll}{{\large\texttt{Gkeyll}}}

\newcommand{\ignore}[1]{}  

\journal{Journal of \LaTeX\ Templates}

\begin{document}

\begin{frontmatter}

\title{Mapped discontinuous Galerkin interpolations and sheared boundary conditions}

\author{Manaure Francisquez$^a$\corref{mycorrespondingauthor}}
\cortext[mycorrespondingauthor]{Corresponding author}
\ead{mfrancis@pppl.gov}

\author{Noah R. Mandell$^b$}

\author{Ammar Hakim$^a$}

\author{Gregory W. Hammett$^a$}

\address[pppl]{Princeton Plasma Physics Laboratory, Princeton, NJ 08543}
\address[mit]{MIT Plasma Science and Fusion Center, Cambridge, MA, 02139}

\begin{abstract}
Translations or, more generally, coordinate transformations of scalar fields arise in several applications, such as weather, accretion disk and magnetized plasma turbulence modeling. In local studies of accretion disks and magnetized plasmas these coordinate transformations consist of an analytical mapping and enter via sheared-shift boundary conditions. This work introduces a discontinous Gakerkin algorithm to compute these coordinate transformations or boundary conditions based on projections and quadrature-free integrals. The procedure is high-order accurate, preserves certain moments exactly and works in multiple dimensions. Tests of the proposed approach with increasing complexity are presented, beginning with translations of one and two dimensional fields, followed by 3D and 5D simulations with sheared (twist-shift) boundary conditions. The results show that the algorithm is $(p+1)$-order accurate in the DG representation and $(p+2)$-order accurate in the cell averages, with $p$ being the order of the polynomial basis functions. Quantification of the algorithm's diffusion and, for shearing boundary conditions, discussion of aliasing errors are provided.
\end{abstract}

\begin{keyword}
sheared box \sep coordinate transformation \sep accretion disk \sep gyrokinetics \sep flux-tube \sep twist-and-shift \sep plasma \sep fusion \sep high-order \sep interpolation \sep overset mesh \sep chimera \sep ITG \sep cyclone
\end{keyword}

\end{frontmatter}


\section{Introduction} \label{sec:intro}

The solution of partial differential equations often involves complex geometries or a large number of degrees of freedom. The former is sometimes tackled by block-structured or overset (Chimera) grids; these employ disconnected meshes that may be structured, unstructured, mapped or Cartesian, covering separate parts of the computational domain and intersecting in overlap regions~\cite{Steger1987,Sherer2005}. These intersections typically require interpolating the dynamic fields (e.g. particle density, fluid velocity, pressure) from one mesh to the other. The literature on these grids and interpolation methods is vast, historically targeting finite difference (FD), finite volume (FV) and finite element (FEM) discretizations~\cite{Barth1993,Nejat2008}. Some interpolation schemes for multiblock and overset grids using discontinuous Galerkin (DG) discretizations exist as well~\cite{Hall2013}. In both FD/FV/FEM and DG schemes such interpolations are often based on the construction of interpolants given neighboring point-wise or cell-average values. There also exists weak formulations of the operation that lead to integrals which are then computed via numerical quadrature~\cite{Marshall2014}. Constructing interpolants with neighboring cells can lead to large stencils, and quadrature-based interpolation can have a (computational) complexity that is greater than would be desired, especially for high-order accurate schemes.

In magnetized plasma physics mapped multiblock approaches appeared a number of years ago amongst FD~\cite{Stegmeir2018} and FV~\cite{Dorf2018} codes, although the former uses a slightly different formulation and terminology (i.e. the flux-coordinate independent approach (FCI)~\cite{Stegmeir2018,Hariri2013}). Prior to the development of such codes, and still today, the prevalent approach in magnetized plasma turbulence modeling involved not multiblock or overset grids, bur rather reducing the number of degrees within a minimal computational volume with a single mesh. This reduction was accomplished by employing computational coordinates aligned with the magnetic field and by tailoring the computational domain to the anisotropy in these environments, using only a small domain in the plane perpendicular to the magnetic field and an elongated but coarsely meshed domain along the magnetic field~\cite{Hammett1993,Beer1995}. These ``flux-tubes'' have been used to simulate magnetized plasma turbulence with reduced two-fluid~\cite{Rogers1998} and gyrofluid~\cite{Beer1996} models, as well as Fokker-Planck equations averaged over the gyromotion around magnetic field lines called gyrokinetic equations~\cite{Dorland2000}. Gyrokinetic models provide great computational savings over 6D Fokker-Planck (Boltzmann) models since they reduce the problem to 5D phase-space and remove faster time-scales than what is needed to describe low-frequency processes like turbulence.  Presently, nearly every (continuum) gyrokinetic code can use a field-aligned flux-tube domain~\cite{Candy2016,Jenko2000,Dorland2000,Watanabe2005}.

Field-aligned flux-tubes use $(x,y)$ coordinates to identify the location on the plane locally perpendicular to the magnetic field, and $z$ to label the location along the field line. This domain is finite in $z$ and, due to the ergodic nature of the magnetic field, its ends may be at physically distinct locations. Additionally, magnetic shear causes the cross section of the flux-tube to change, say, from a rectangle to a sheared parallelogram as one moves along the field line in $z$. Therefore these domains are typically combined with twist-and-shift $z$-boundary conditions (BCs)~\cite{Hammett1993,Beer1995}, that exploit toroidal symmetry in fusion devices and assume that turbulence is statistically indistinguishable at locations with the same poloidal angle ($\theta$) and different toroidal angle ($\phi$). The recipe for flux-tubes with twist-shift BCs is to have a domain with $z$-ends at the same $\theta$ and enforce $z$-periodicity holding $x$ and $\phi$ constant, leading to $z$-periodicity with a $y$-shift (sketched in figure~\ref{fig:periodicityExample}). This $y$-shift is in general sheared because of the sheared magnetic field, so turbulent structures shift and twist as they pass from one $z$-end to the other, hence the name twist-and-shift. Most gyrokinetic solvers also use a Fourier representation in the perpendicular plane, for which twist-shift BCs are cast as a re-scaling of Fourier coefficients. Real-space codes however must interpolate dynamic fields at one $z$-end of the box onto a shifted mesh that is then identified with the mesh at the other $z$-end. This procedure has been implemented for FD and FV codes, but to our knowledge it does not exist in FEM or DG solvers.

Boundary conditions with a (sheared) shift are not unique to fusion plasma modeling. Another example is simulations of the magnetorotational instability and other processes in accretion disks~\cite{Hawley1995} that use a local sheared-box. These local sheared-boxes are motivated by the colossal size and broad wave-number spectrum of accretion disks, motivating a minimum simulation volume in which $(x,y,z)$ coordinates correspond to radial, azimuthal and vertical directions, respectively. Equations such as those of a magnetohydrodynamic (MHD) model are cast in the frame of reference of the rotating disk, which is to lowest order sheared in the $x$ direction. These simulations employ radial periodicity but over time the lower $x$-end of the box drifts in the azimuthat direction ($y$) relative to the upper $x$-boundary. Radial periodicity in accretion disk sheared-boxes thus entail a shift in $y$ that is proportional to time and the strength of the mean flow shear. This is a similar situation to that arising in the twist-shift BCs of magnetized plasma turbulence modeling, albeit now involving $y$-shifts in $x$-BCs rather than in $z$-BCs. It is also akin to newly proposed non-twisting domains for magnetized plasma modeling which contain $x$-BCs with a $y$-shift~\cite{Ball2021}. In all cases one must resort to interpolation of fields onto curved meshes, much in the same way that multiblock and overset grids require interpolations between two curvilinear meshes.

Some of the inter-grid interpolation schemes cited above are quite general and intended for unstructured meshes without mapped blocks, i.e. grid blocks without a mapping between computationally Cartesian and physical curvilinear coordinates. In the case of mapped blocks however, the relationship between coordinates in adjacent grids may be analytic and static, as is the case for twist-shift and sheared-box BCs. We here specialize in such applications, for which a numerical-quadrature-free DG algorithm can be devised. We present such algorithm in the context of sheared BCs and later discuss its relevance to and challenges arising from more complicated applications. This paper thus begins with a brief summary of field-aligned coordinates and twist-shift BCs (section~\ref{sec:coords}), and is followed by a description of the algorithm (section~\ref{sec:algorithm}). We test this approach with static 1D and 2D interpolations as well as time-dependent simulations in 3D and 5D, results of which are discussed in section~\ref{sec:results}. Additional remarks regarding the relationship between these BC interpolations and those arising in other applications are offered in section~\ref{sec:beyond}, prior to closing with a summary in section~\ref{sec:conclusion}.

\section{Field-aligned coordinates and boundary conditions} \label{sec:coords}

The strong background magnetic field in magnetized fusion devices (e.g. tokamaks and stellarators) endows plasma turbulence with a highly anisotropic character. Fluctuations have parallel wavelengths that are much longer than perpendicular wavelengths, i.e. $\kpar\ll\kperp$ where $\kpar$ ($\kperp$) denotes the wavenumber parallel (perpendicular) to the background magnetic field $\v{B}$. The minimum computational volume thus consists of a thin (flux) tube wrapping and following a bundle of magnetic field lines, with small perpendicular and large parallel extents, each being several correlation lengths wide.

These flux-tubes were designed for background magnetic fields that are axisymmetric in the toroidal angle $\phi$, and can be represented as 
\begin{equation}
    \v{B} = R B_\phi \nabla \phi + \nabla \psi \times \nabla \phi,
\end{equation}
where $B_\phi$ is the toroidal component of the magnetic field and $\psi$ is the poloidal flux. One can also define a coordinate system $(\psi,\chi,\phi)$, where $\chi$ is a poloidal-like angle, in which the magnetic-field appears as straight lines, defined such that 
\begin{equation}
   \frac{\v{B}\cdot\nabla\phi}{\v{B}\cdot\nabla\chi} = q(\psi) .
\end{equation}
This gives field lines that are straight lines with slope $q(\psi)$ in the $(\chi,\phi)$ plane at constant $\psi$, parametrized by $q\chi-\phi=\text{const}$. Here, $q(\psi)$ is the safety factor, that represents the number of toroidal revolutions required to complete a single poloidal revolution when following a field line on flux surface $\psi$. A field-aligned coordinate system can then be defined as~\cite{Lapillonne2009,Mandell2021}
\begin{equation}
    x = \psi-x_0, \qquad y = C_y(q\chi - \phi)-y_0, \qquad z = \chi, \label{eq:coords}
\end{equation}
with $C_y$ a normalization constant,
so that the background magnetic field can be expressed in Clebsch form as
\begin{equation}
    \v{B} = {C_y}^{-1}\nabla x\times \nabla y.
\end{equation}
Here, $x$ is a radial-like coordinate, $y$ is a field-line-labeling coordinate, and $z$ is the parallel coordinate measuring the location along the field line. This coordinate system allows us to account for and study the highly anisotropic tokamak turbulence with a fine grid perpendicular to the background field (i.e. in $x$-$y$) and a coarse grid parallel to it (i.e. in $z$).

\begin{figure}
  \begin{subfigure}[b]{0.49\textwidth}
    \centering
    \includegraphics[width=\textwidth]{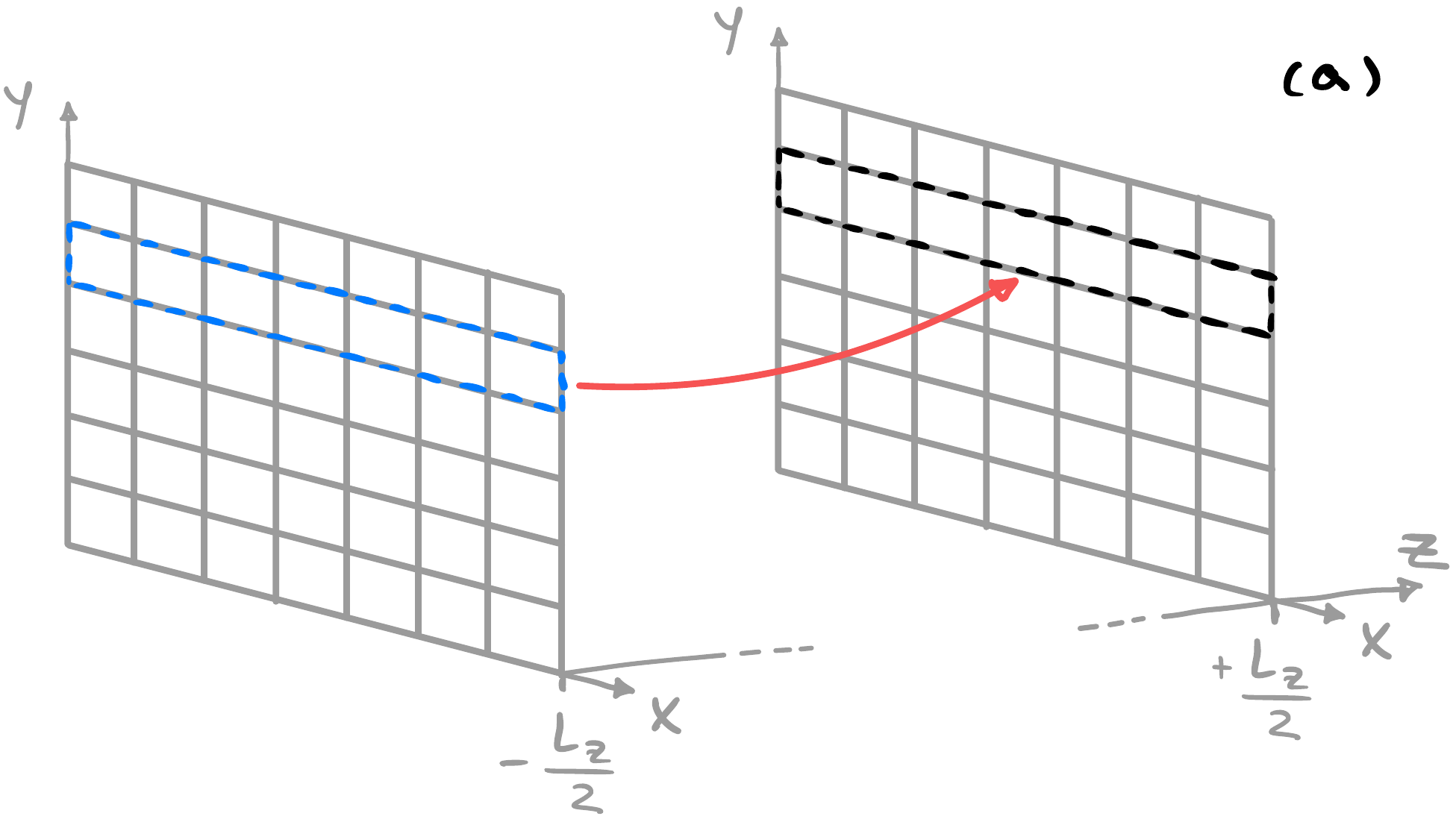}
  \end{subfigure}
    \begin{subfigure}[b]{0.49\textwidth}
    \centering
    \includegraphics[width=\textwidth]{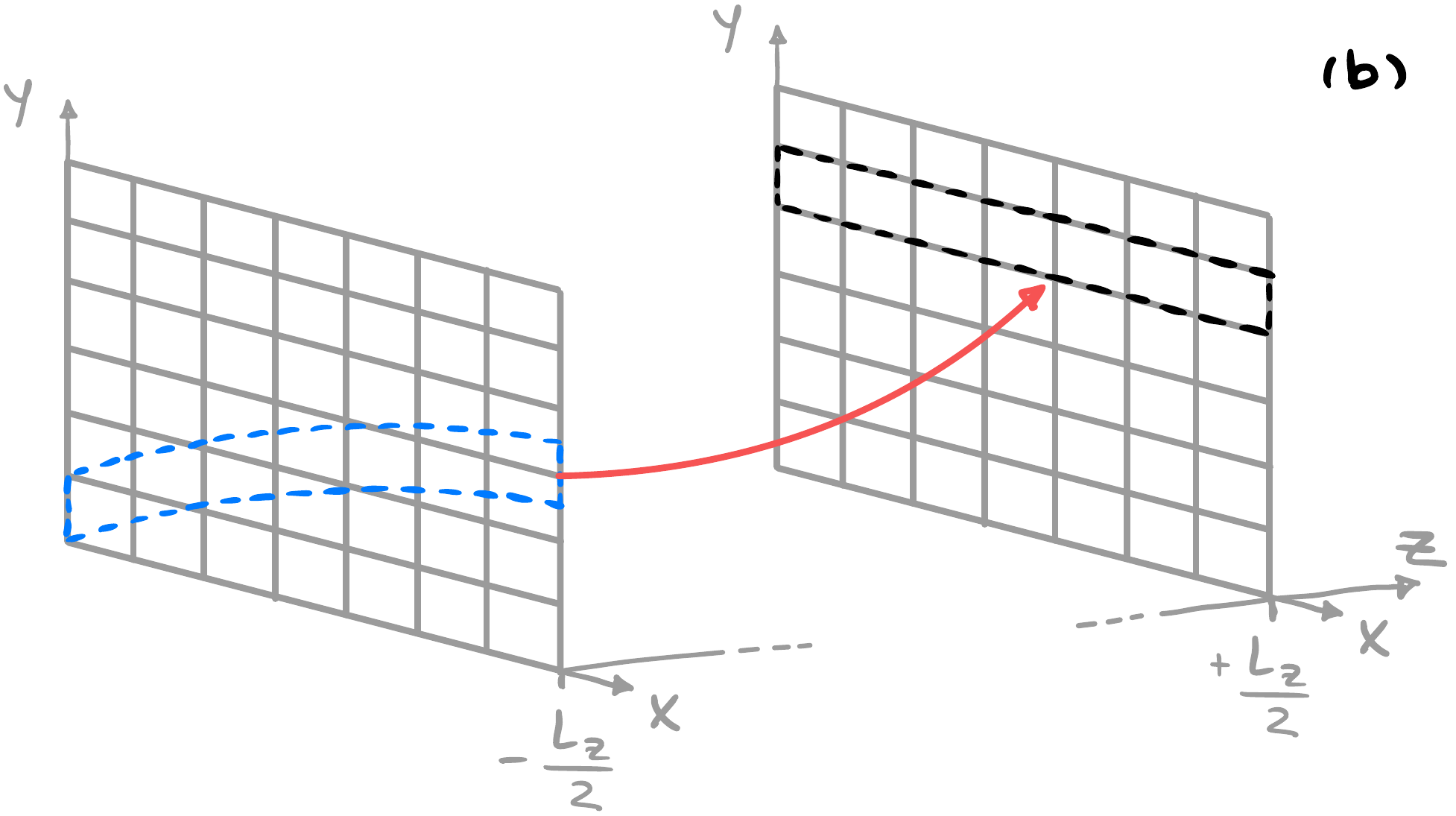}
  \end{subfigure}
   \caption{(a) Regular $z$-periodicity: the field at $z=\Lz/2$ is identified with the field at $z=-\Lz/2$ holding $x$ and $y$ constant.(b) $z$-periodicity with a sheared $y$-shift: the field at $z=\Lz/2$ equals the field at $z=-\Lz/2$ holding $x$ constant but shifted in $y$.}
   \label{fig:periodicityExample}
\end{figure}

The angles $\phi$ and $\chi$ are naturally periodic, so that for any physical quantity $f(\psi,\chi,\phi)$ we must have
\begin{align}
    f(\psi, \chi, \phi+2\pi) &= f(\psi, \chi, \phi), \label{eq:phiPer} \\
    f(\psi, \chi+2\pi, \phi) &= f(\psi, \chi, \phi). \label{eq:chiPer}
\end{align}
Given the definitions in equation~\ref{eq:coords} one can show that $\phi$-periodicity does imply regular $y$-periodicity: $f(x,y+\Ly,z)=f(x,y,z)$ for $\Ly=2\pi C_y$. Yet, given that the computational domain has a finite extent in $z$ with boundaries that correspond to physically distinct locations (except for a few rational flux-surfaces on which the flux-tube closes back on itself), one would in general be wrong to apply regular periodicity in $z$, i.e. $f(x,y,z=-\Lz/2)=f(x,y,z=+\Lz/2)$ holding $(x,y)$ constant in a computational domain with $z\in\left[-\Lz/2,\Lz/2\right]$ (see figure~\ref{fig:periodicityExample}(a)). Instead we must apply $z$-periodicity holding the toroidal angle $\phi$ constant:
\begin{equation}
    f(x,y(z+\Lz)|_{\phi=\mathrm{const.}},z+\Lz) = f(x,y,z)
\end{equation}
which owing to equation~\ref{eq:coords} states
\begin{equation}
    f(x,y,z+\Lz) = f(x,y-\Lz C_yq,z).
\end{equation}
Therefore in the parallel ($z$) direction, the boundary condition results in a shift of the $y$ coordinate by $\mathcal{S}(x)=2\pi C_y q(x)$. The shift also gives some twist due to the dependence of the shift on $x$. For this reason, this boundary condition is sometimes called the twist-and-shift boundary condition. We can use the same logic with $z-\Lz$ in order to obtain the lower-$z$ BC, and combine them both into a single equation as
\begin{align} 
    f(x,y,z\pm\Lz) &= f(x,(y\mp\Lz C_y q)\,\mathrm{mod}\,L_y,z), \label{eq:twistShiftBC}
\end{align}
where the modulo operation accounts for cases where $y-\Lz C_y q$ is outside the $y$ domain by applying periodicity in $y$.
\ignore{
\begin{align}
    f(x, y+2\pi C_y, z ) &= f(x,y,z), \\
    f(x, y, z + 2\pi) &= f(x, y - 2\pi C_y q, z).
\end{align}

In many cases, the correlation length of the turbulence in the toroidal direction is smaller than the full toroidal domain, so one can instead take a toroidal wedge of width $\Delta \phi$ and assume statistical periodicity,
\begin{equation}
    f(\psi, \chi, \phi+ \Delta\phi) = f(\psi, \chi, \phi).
\end{equation}
The $y$ domain then has length $L_y = C_y \Delta \phi$, and we modify the periodicity conditions in the field-aligned coordinates as
\begin{align} 
    f(x, y+L_y, z ) &= f(x,y,z), \\
    f(x, y, z + \Lz) &= f(x, (y - \Lz C_y q)\,\mathrm{mod}\, L_y, z), \label{eq:twistShiftBC}
\end{align}
where $\Lz$ is most commonly $2\pi$ and the modulo operation accounts for cases where $y-\Lz C_y q$ is outside the $y$ domain by applying periodicity in $y$.
}
Another way to interpret equation~\ref{eq:twistShiftBC} that illustrates its connection to interpolations between mapped-grids is that given the dynamic field $f(x,y,z=-\Lz/2)$ originating from the lower $z$-boundary, an interpolation onto a curved grid (relative to the $x$-$y$ grid) with coordinates $x'=x$, $y'=y-\ysh(x)$ and $z'=z$ must take place.

\section{Discontinuous Galerkin twist-shift BC algorithm} \label{sec:algorithm}

Boundary conditions with twist and shear such as equation~\ref{eq:twistShiftBC}, or its sheared-box equivalent, have been implemented in finite difference, finite volume and spectral codes. However we are here concerned with an algorithm for carrying out these interpolations in DG solvers. To our knowledge there is no prior work on developing a recipe for enforcing these BCs using DG discretizations, and there are aspects of the interpolation algorithm below that are novel and relevant to other forms of inter-grid transfers arising in  simulations with DG.

In order to formulate and describe the algorithm for applying twist-shift BCs in $z$ to 3D and 5D DG fields, we will consider an analogous, slightly simpler operation. Consider two 2D fields covering the $y$-periodic domain $\Omega=\left[-\Lx/2,\Lx/2\right]\times\left[-\Ly/2,\Ly/2\right]$. Given the {\it donor} field $\fdo(x,y)$ we wish to compute the {\it target} field $\ftar(x,y)$ via a sheared translation of the donor field according to
\begin{equation} \label{eq:perShifted}
\ftar(x,y) = \fdo(x,y-\ysh(x)),
\end{equation}
where the $y$-shift $\ysh(x)$ is as of now general and unspecified, although we make some restrictions below. Place on this domain $\Omega$ the mesh (tesselation) $\mathcal{T}$ with cells $K_{i,j}\in\mathcal{T}$ labeled by $i=1,\dots,\Nx$ and $j=1,\dots,\Ny$; their indices along $x$ and $y$. We adopt a modal DG discretization, for which we introduce the $\mathcal{V}^p_{i,j}$ polynomial space of order $p$ in the $(i,j)$-th cell with cardinality $|\mathcal{V}^p_{i,j}|=\numB$. These $\psi_{i,j,k}\in\mathcal{V}^p_{i,j}$ polynomials have compact support $\mathrm{supp}(\psi_{i,j,k}(x,y)) = K_{i,j}$ restricted to $K_{i,j}\equiv\left[x_{i-1/2},x_{i+1/2}\right]\times\left[y_{j-1/2},y_{j+1/2}\right]$, and are orthogonal and normalized such that for  $\psi_{m,n,\ell}\in\mathcal{V}^p_{m,n}$ one has $\int_{K_{i,j}}\psi_{i,j,k}\psi_{m,n,\ell}\dx\,\dy=\delta_{im}\delta_{jn}\delta_{k\ell}\Dx\Dy/4$, where $\Dx=x_{i+1/2}-x_{i-1/2}$ and $\Dy=y_{j+1/2}-y_{j-1/2}$. 

A 2D field $f(x,y)$ is therefore represented by an polynomial expansion in each of these disconnected basis sets:
\begin{equation} \label{eq:dgExp}
    f(x,y) = \sum_{i=1}^{\Nx}\sum_{j=1}^{\Ny}\sum_{k=1}^{\numB}f_{i,j,k}\psi_{i,j,k}(x,y).
\end{equation}


The first step in formulating the DG algorithm to compute $\ftar(x,y)$ consists of performing a weak (Galerkin) projection of equation~\ref{eq:perShifted} onto the $k$-th DG basis function in the $(i,j)$-th cell, $\psi_{i,j,k}(x,y)$. This projection over the whole domain is
\begin{equation} \label{eq:perShiftedProjG}
\int_{-\Lx/2}^{\Lx/2}\int_{-\Ly/2}^{\Ly/2}\dx\,\dy~\psi_{i,j,k}(x,y)\ftar(x,y) = \int_{-\Lx/2}^{\Lx/2}\int_{-\Ly/2}^{\Ly/2}\dx\,\dy~\psi_{i,j,k}(x,y)\fdo(x,y-\ysh(x)).
\end{equation}
Inserting the DG expansion of the target field in the left side of this equation yields
\begin{equation}
\int_{-\Lx/2}^{\Lx/2}\int_{-\Ly/2}^{\Ly/2}\dx\,\dy~\psi_{j,k}(x,y)\sum_{i',j',k'}f_{\mathrm{tar},i',j',k'}\psi_{i',j',k'}(x,y) = \int_{-\Lx/2}^{\Lx/2}\int_{-\Ly/2}^{\Ly/2}\dx\,\dy~\psi_{i,j,k}(x,y)\fdo(x,y-\ysh(x)).
\end{equation}
Due to the compact support of $\psi_{i',j',k'}$, the left side integral is nonzero only when $i=i'$ and $j=j'$, and hence the $y$-integral is restricted to the $(i,j)$-th cell:
\begin{eqnal} \label{eq:restrictlhs}
\int_{x_{i-1/2}}^{x_{i+1/2}}\int_{y_{j-1/2}}^{y_{j+1/2}}\dx\,\dy~\psi_{i,j,k}(x,y)\sum_{k'}f_{\mathrm{tar},i,j,k'}\psi_{i,j,k'}(x,y) &= \int_{-\Lx/2}^{\Lx/2}\int_{-\Ly/2}^{\Ly/2}\dx\,\dy~\psi_{i,j,k}(x,y)\fdo(x,y-\ysh(x)).
\end{eqnal}
For algorithmic convenience we introduce the logical coordinates $\xi,\eta\in[-1,1]$ defined via
\begin{equation} \label{eq:logicalCoords}
    x = x_i + \frac{\Dx}{2}\xi,
    \qquad\qquad
    y = y_j + \frac{\Dy}{2}\eta,
\end{equation}
where $x_i$ and $y_j$ are the cell center coordinates and we assume a uniform grid with constant cell lengths $\Dx$ and $\Dy$. In terms of logical coordinates we can write the left side of~\ref{eq:restrictlhs} as
\begin{eqnal}
\frac{\Dx}{2}\frac{\Dy}{2}\int_{-1}^{1}\int_{-1}^{1}\dxi\,\deta~\psi_{i,j,k}(\xi,\eta)\sum_{k'}f_{\mathrm{tar},i,j,k'}\psi_{i,j,k'}(\xi,\eta) = \frac{\Dx\Dy}{4} f_{\mathrm{tar},i,j,k} = \mathrm{RHS~of~equation}~\ref{eq:restrictlhs},
\end{eqnal}
where we used the orthogonality of ${\psi_{i,j,k}}$.

On the right side of these equations we employ the inverse mapping given by $y'=y-\ysh(x)$ such that our integral becomes, after substituting the expansion of the donor field,
\begin{eqnal} \label{eq:interpLyOld}
f_{\mathrm{tar},i,j,k} &= \frac{4}{\Dx\Dy}\int_{-\Lx/2}^{\Lx/2}\int_{(-\Ly/2-\ysh)\mathrm{mod}\Ly}^{(\Ly/2-\ysh)\mathrm{mod}\Ly}\dx\,\mathrm{d}y'~\psi_{i,j,k}(x,y'+\ysh)\sum_{i',j',k'}f_{\mathrm{do},i',j',k'}\psi_{i',j',k'}(x,y').
\end{eqnal}
We cannot use the support and orthonormality of the basis set to simplify the $y$-integral on the right side because the shift $\ysh(x)$ changes the support of the basis functions. In spite of that, $\ysh(x)$ only depends on $x$, is the same at all $y$, and does not change the support along $x$, $\mathrm{supp}_x\left(\psi_{i,j,k}\right)=\left[x_{i-1/2},x_{i+1/2}\right]$. So we can invoke the disconnectedness of $\psi_{i,j,k}$ and $\psi_{i',j',k'}$ along $x$ in order to limit the $x$-integral to the $i$-th cell. The algorithm described below does not change from one $x$-cell to the next, so for notational ease we will drop the $i$ subscripts on basis and DG expansion coefficients and assume that we are computing DG coefficients in the $i$-th cell. Equation~\ref{eq:interpLyOld} then becomes
\begin{eqnal} \label{eq:interpLy}
f_{\mathrm{tar},j,k} &= \frac{4}{\Dx\Dy}\int_{x_{i-1/2}}^{x_{i+/2}}\int_{(-\Ly/2-\ysh)\mathrm{mod}\Ly}^{(\Ly/2-\ysh)\mathrm{mod}\Ly}\dx\,\mathrm{d}y'~\psi_{j,k}(x,y'+\ysh)\sum_{j',k'}f_{\mathrm{do},j',k'}\psi_{j',k'}(x,y').
\end{eqnal}

Equation~\ref{eq:interpLy} hints at a way forward in order to compute the $f_{\mathrm{tar},j,k}$ coefficients. First, the limits of this integral are simply the domain boundaries shifted by $\ysh(x)$ but we use periodicity to simply wrap the integral around in $y$, indicated with $(\pm\Ly/2-\ysh)\mathrm{mod}\Ly$. Second, irrespective of the labels used for indices and variables, the donor field appears in its basic form (without dependencies on the shift), so we are simply performing a weighted integral of it. The weight however is a shifted basis function, which originally had $\mathrm{supp}_y(\psi_{j',k'}(x,y)) = [y_{j'-1/2},y_{j'+1/2}]$ but now has $\mathrm{supp}_y(\psi_{j',k'}(x,y+\ysh)) = [y_{j'-1/2}+\ysh(x),y_{j'+1/2}+\ysh(x)]$. The integral will be zero outside of this shifted support, so we can rewrite the $y$ limits as\footnote{The integral limits on the right side of equation~\ref{eq:interpFormula} also indicate how the blue lines in figure~\ref{fig:periodicityExample}(b) are defined: they are simply $y_{j-1/2}-\ysh(x)$ and $y_{j+1/2}-\ysh(x)$.}
\begin{eqnal} \label{eq:interpFormula}
f_{\mathrm{tar},j,k} &= \frac{4}{\Dx\Dy}\int_{x_{i-1/2}}^{x_{i+1/2}}\int_{(y_{j-1/2}-\ysh)\mathrm{mod}\Ly}^{(y_{j+1/2}-\ysh)\mathrm{mod}\Ly}\dx\,\mathrm{d}y'~\psi_{j,k}(x,y'+\ysh)\sum_{j',k'}f_{\mathrm{do},j',k'}\psi_{j',k'}(x,y')
\end{eqnal}
and thus our task consists of computing the inner product of $\psi_{j,k}(x,y+\ysh)$ and $\fdo$ over the shifted region. In general this contributing region does not consist of a single cell, multiple whole cells, or even rectangular sub-regions of a cell. It can consist of integrals over non-rectangular sub-regions of multiple cells. Therefore we must be able to compute the integral in the right side of equation~\ref{eq:interpFormula} adding up contributions from non-rectangular sub-cell regions coming from multiple cells. We can simplify this task by imposing two restrictions on the $y$-shift $\ysh$:
\begin{enumerate}
\item $\ysh(x)$ is monotonically increasing or decreasing.
\item $\ysh(x)\neq0$ and is not close to zero anywhere in the domain.
\end{enumerate}
The first of these constraints the set of sub-cell integrals the algorithm has to be capable of performing. The second restriction imposes limits on the shear (i.e. $d\ysh/dx$) or the $x$-domain, because they cannot be so large that somewhere in the domain $\ysh(x)$ goes to zero. It is also imposed to lessen the potential for floating point comparison errors. However, there are some scenarios in which we have successfully used the algorithm presented here using a $\ysh(x)$ that satisfies the first of these restrictions but not the second, and we provide an example in section~\ref{sec:results3x}.

The algorithm by which we compute the integrals in equation~\ref{eq:interpFormula} involves a series of steps described in more detail below and in~\ref{sec:addDetails}. That said, we could briefly summarize it with the following four steps:
\begin{enumerate}
    \item Construct a discrete representation of the shift $\ysh(x)$.
    \item For a given target cell identify all the donor cells.
    \item Use the intersection of the shifted $y$-boundaries of the target cell and the donor cell to recognize the type of sub-cell integral needed.
    \item Construct the sub-cell integral by locating key intersection points and projecting functions that describe integral limits onto a 1D basis.
    \item Perform and sum the sub-cell integrals from all donor cells.
\end{enumerate}
We dive into each of these next.

\subsection{Discrete approximation to the shift $\ysh(x)$} \label{sec:discreteS}

At various steps in our algorithm we will refer to the $y$-shift, $\ysh(x)$, implying that it is an analytic function or, if $\ysh(x)$ originates from a numerical solution (e.g. a meshing program or equilibrium solver), that a procedure for evaluating it at an arbitrary $x$ exists (e.g. via interpolation). Yet there are two places in the algorithm below where we will in fact use a discrete approximation to $\ysh(x)$. Let us then introduce the 1D polynomial space $\mathcal{V}^p_i=\{x^m~|~\mathrm{deg}(x^m)\leq p\}$ with cardinality $\abs{\mathcal{V}^p_i}=\numB^{1d}$ in the $i$-th cell such that we can represent the $y$-shift as the polynomial expansion
\begin{equation}
    \yshh(x) = \sum_{i=1}^{\Nx}\sum_{k=1}^{\numB^{1d}}\yshhik\varphi_{i,k}(x),
\end{equation}
where $\varphi_{i,k}(x)\in\mathcal{V}^p_i$. The DG coefficients $\yshhik$ are obtained by projecting $\ysh(x)$ onto the 1D polynomial basis $\varphi_i(x)$ in a manner that results in a continuous function across cells. The way to accomplish this is to, in every cell, evaluate $\ysh(x)$ at Gauss-Lobatto nodes and perform a nodal-to-modal transformation.

The discrete shift $\yshh(x)$ is primarily used in calculating sub-cell integrals and in finding the donor cells, although the latter could just as well use the analytic $\ysh(x)$. Elsewhere we employ the analytic $\ysh(x)$; how the algorithm performs were we to use $\yshh(x)$ everywhere could be explored in the future.

\subsection{Finding donor cells} \label{sec:getDonors}

\begin{figure}[h]
    \centering
    \includegraphics[width=0.6\textwidth]{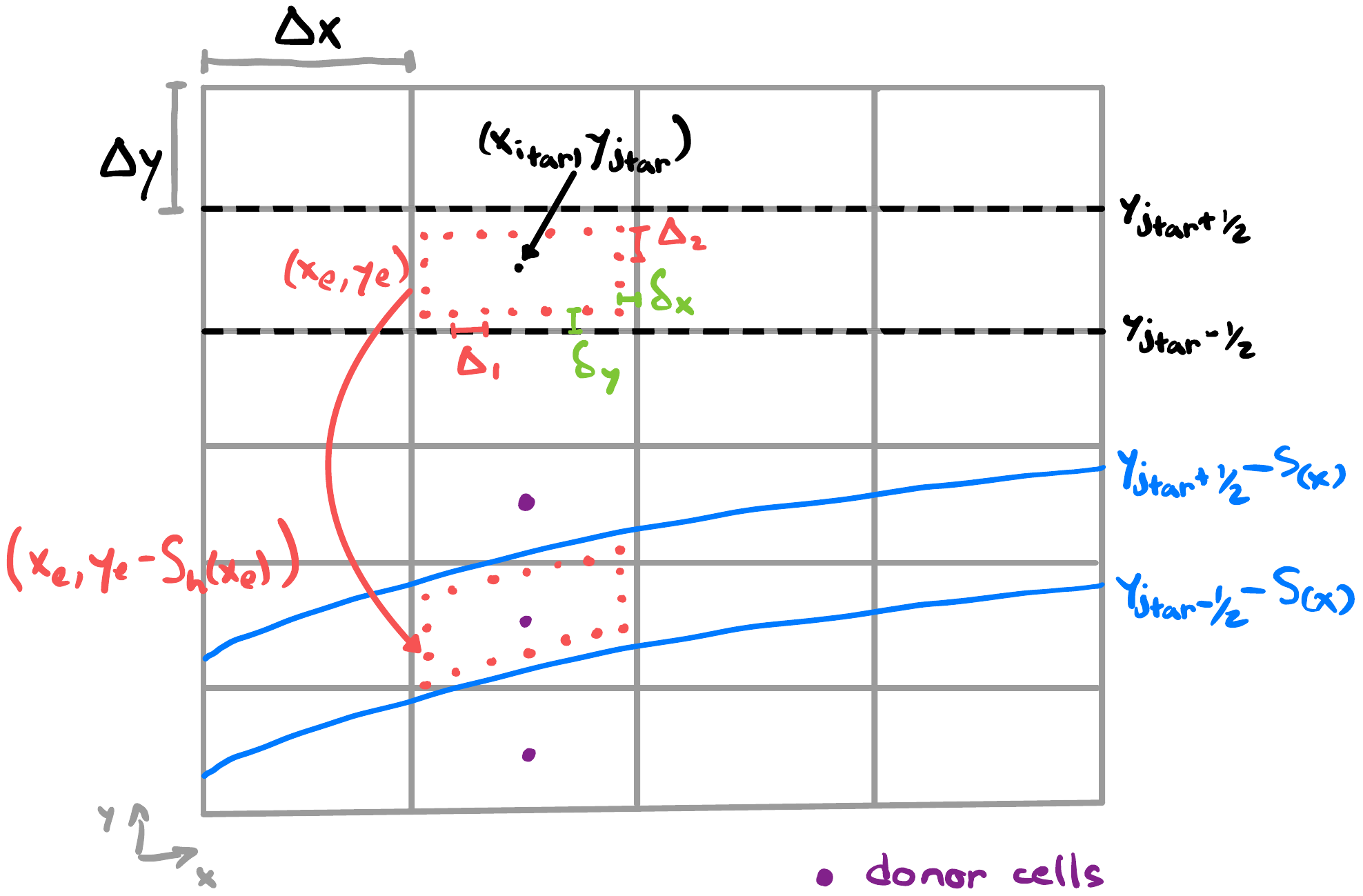}
   \caption{Sketch of the procedure for finding donor cells (centered on purple dots) for the target cell centered at $(x_{\itar},y_{\jtar})$. Fewer test points $(x_e,y_e)$ than actually used in the code are drawn for simplicity.}
   \label{fig:findDonor}
\end{figure}

For each cell in the 2D target grid we need to find the donor cells that will contribute to it. We do so with the following procedure, sketched out in figure~\ref{fig:findDonor}:
\begin{enumerate}
    \item Loop through the target cells.
    \item Given the target cell centered at $(x_{\itar},y_{\jtar})$, for example, select a number test points just inside of the cell boundaries, a distance $(\delta_x,\delta_y)=(10^{-9}\Dx,10^{-9}\Dy)$ away from those boundaries. Using inner points instead of boundary points reduces the possibility of floating point comparison errors in subsequent steps. These test points are separated by $(\Delta_1,\Delta_2)=(\Dx/10,\Dy/10)$, i.e. we consider ten points along each boundary.
    \item Loop through the test points.
    \item For test point $(x_e,y_e)$, for example, apply the
    shift to arrive at $(x_e,y_e-\yshh(x_e))$. Assume $y$-periodicity\footnote{Applying periodicity isn't entirely trivial because when a shifted test point is on the lower(upper) domain boundary we must be careful to select the proper cell depending on whether the other test points lie above or below it. That is, assuming positive $\ysh(x)$, if $(x_e,y_e)$ abuts the $y=y_{\jtar-1/2}$ line and $(x_e,y_e-\yshh(x_e))$ lands on the lower domain boundary we must select $\jdo=1$. But if $(x_e,y_e)$ abuts the $y=y_{\jtar+1/2}$ and $(x_e,y_e-\yshh(x_e))$ lands on the upper domain boundary, we must select $\jdo=\Ny$. We identify whether a shifted point lands on a boundary by checking if $\abs{y_e-\yshh(x_e)-(\pm\Ly/2)}<10^{-12}$.}.
    \item Find the cell that owns this shifted test point through a multidimensional binary search\footnote{At the heart of this binary search is a comparison like $x_{i-1/2} \leq x' \leq x_{i+1/2}$, however due to floating point comparison errors it seemed better for the cases tested so far to instead use comparisons like $x_{i-1/2}-\epsilon \leq x' \leq x_{i+1/2}+\epsilon$, where $\epsilon=10^{-14}$. It would perhaps be better to set this $\epsilon$ as a function of the cell length, e.g. $10^{-14}\Dx$, but the chosen number is already very small compared to the scales considered in flux-tube simulations.} and record its indices $(\ido,\jdo)$.
\end{enumerate}
After looping through all the test points in all the target cells, we will have compiled a list of donor cells for each target cell, i.e. for each $(\itar,\jtar)$ a list of $\Ndo$ $(\ido,\jdo)$ pairs. The number of donor cells for each target cell, $\Ndo$, depends on the number of cells, the domain and the $\ysh(x)$.

\subsection{Identifying sub-cell integral types} \label{sec:identifySubInt}

\begin{figure}[h]
    \centering
    \includegraphics[width=0.6\textwidth]{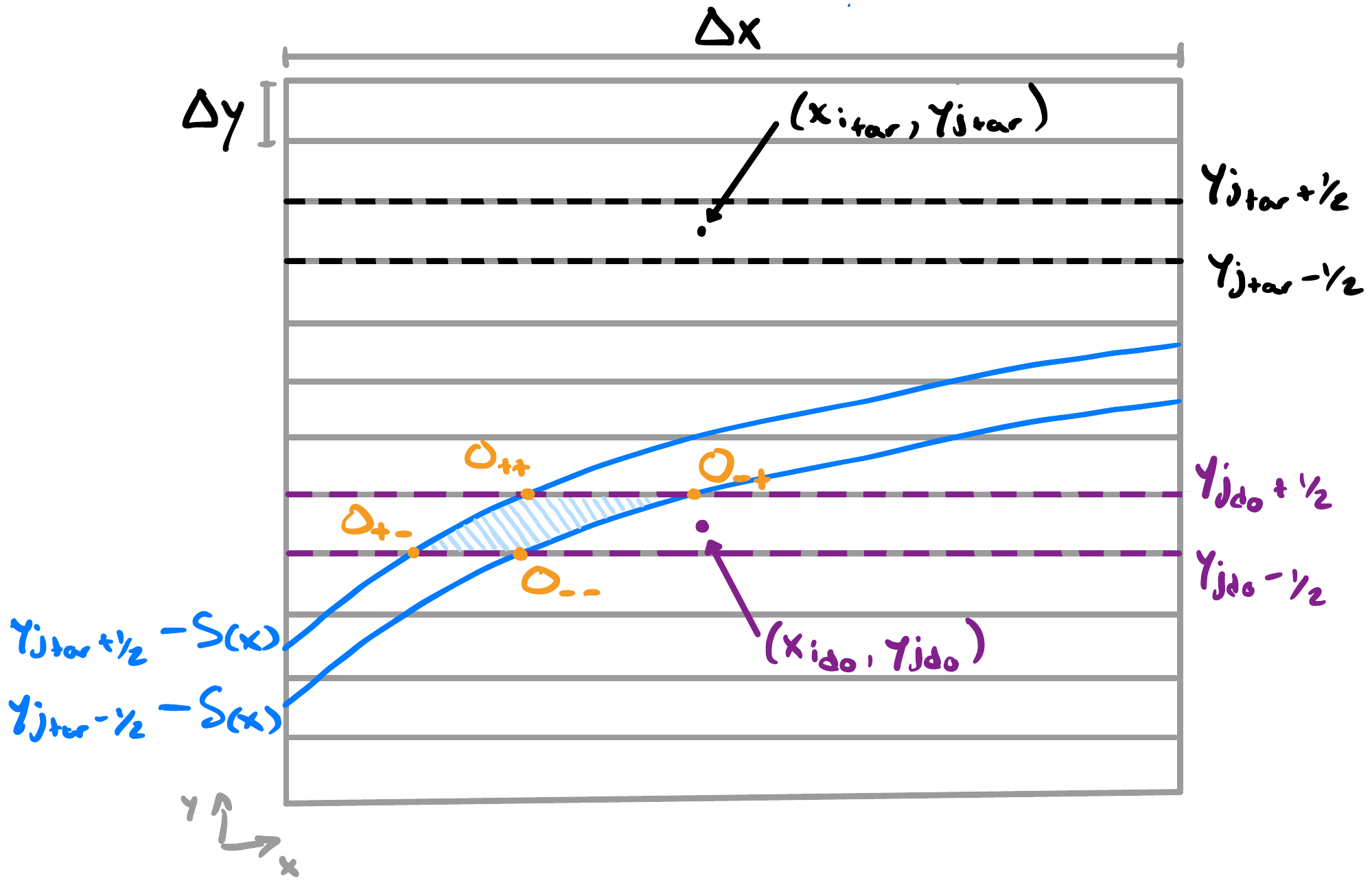}
   \caption{Sketch of scenario $sNi$, with the intersection points $\{O_{--},O_{-+},O_{+-},O_{++}\}$ we search for in order to identify the type of sub-cell integral required. A simplified case with a single cell in $x$ is used for demonstration.}
   \label{fig:findOpoints}
\end{figure}

Having found the donor cells for each target cell we then determine what kind of sub-cell integral is needed from each of those donor cells. The starting point for this step is sketched in figure~\ref{fig:findOpoints}. We need to find the $x$-coordinates $O=\{O_{--},O_{-+},O_{+-},O_{++}\}$ where $y_{\jtar-1/2}-\ysh(x)$ and $y_{\jtar+1/2}-\ysh(x)$ intersect the lines $y=y_{\jdo-1/2}$ and $y=y_{\jdo+1/2}$, since these four points are the corners of the sub-cell region that we must integrate over. We find them using a Ridders' root finding algorithm. For example, we find $O_{--}$ by looking for the roots of the function
\begin{equation} \label{eq:rootFind}
R(x) = y_{\jtar-1/2}-\ysh(x) - y_{\jdo-1/2}
\end{equation}
in the region $[x_{\itar-1/2},x_{\itar+1/2}]$ down to a tolerance of $10^{-13}$. Note that in equation~\ref{eq:rootFind} we use the analytic $\ysh(x)$ provided by the user and not its polynomial approximation, $\yshh(x)$. If $|R(x_{\itar-1/2})|<10^{-13}$ or $|R(x_{\itar+1/2})|<10^{-13}$ it is assumed that the root is at $x_{\itar-1/2}$ or $x_{\itar+1/2}$, respectively. If the root of $R(x)$ is not found at first, it could be because $y_{\jtar-1/2}-\ysh(x)$ lies in a periodic copy of this domain. For that reason we also look for the roots of the function
\begin{equation}
R(x) = y_{\jtar-1/2}-\ysh(x) - \left(y_{\jdo-1/2}-n\Ly\right),
\end{equation}
where $n\in\mathbb{Z}$ (which could be positive or negative, depending on the sign of $\ysh(x)$).

\begin{figure}[h]
    \centering
    \includegraphics[width=0.6\textwidth]{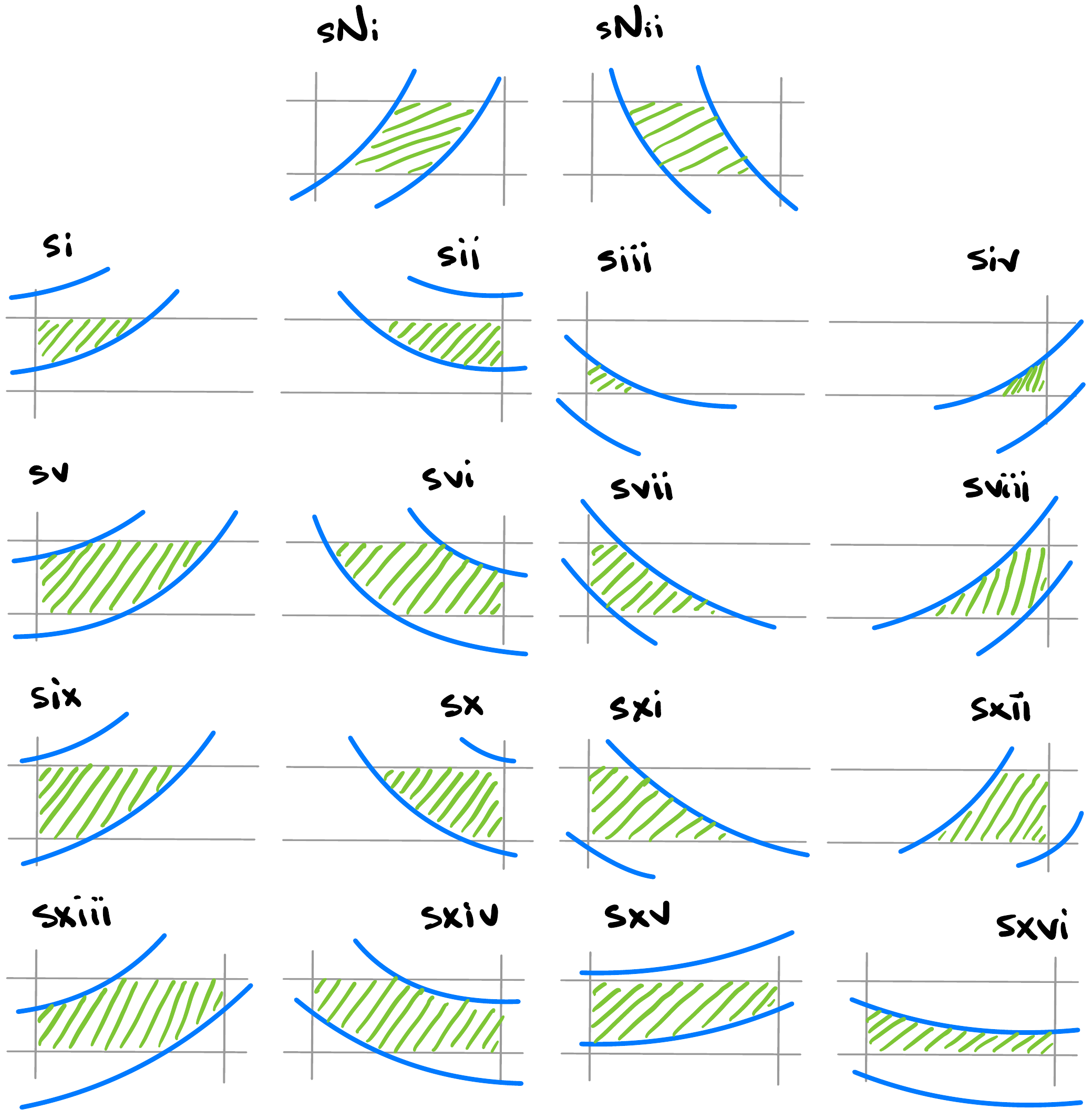}
   \caption{Example of sub-cell regions (green) over which we need to perform integrals. They are created by the space bound by the $y_{\jtar\pm1/2}-\ysh(x)$ lines (blue curves) overlapping with the donor cell (grey boundaries). In scenarios $sNi$-$sNii$ the sub-cell region is not in contact with $x$-boundaries of the cell, while in $si$-$sxii$ the sub-cell region abuts the left or the right boundary of the cell. Scenarios $sxiii$-$sxvi$ have sub-cell regions that meet both the left and right boundaries of the cell. Half of the first 16 scenarios correspond to monotonically increasing $\ysh(x)$, while the other half are for monotonically increasing $\ysh(x)$. Scenarios $sxv-sxvi$ are those in which either of the $y_{\jtar\pm1/2}-\ysh(x)$ lines intersect both $x$-boundaries.}
   \label{fig:subCellScenarios}
\end{figure}

Figure~\ref{fig:findOpoints} uses what we call scenario $sNi$ to illustrate the intersection points. This is the case in which all four intersection points are found. However changes to the grid or to $\ysh(x)$ can make it so that the $y_{\jtar\pm1/2}-\ysh(x)$ curves intersect $x=\mathrm{const.}$ lines instead of $y=\mathrm{const.}$ lines, or one of the curves could have no intersection with the donor cell boundaries. So far we have identified 18 possible sub-cell integral scenarios, depicted in figure~\ref{fig:subCellScenarios}. In this figure the $y_{\jtar\pm1/2}-\ysh(x)$ are shown in blue, and the intersection between the region bounded by $y_{\jtar\pm1/2}-\ysh(x)$ and the donor cell is shaded green. Of the 18 sub-cell integrals we have considered, 8 are for monotonically decreasing $\ysh(x)$, 8 for monotonically increasing $\ysh(x)$, and 2 of them for either. They are also qualified by the differences in how the $y_{\jtar\pm1/2}-\ysh(x)$ curves intersect (or not) the boundaries of the donor cell, summarized as:
\begin{itemize}
\item $sNi$-$sNii$: both $y_{\jtar\pm1/2}-\ysh(x)$ intersect $y_{\jdo\pm1/2}$.
\item $si$-$siv$: one of the $y_{\jtar\pm1/2}-\ysh(x)$ intersects one of $y_{\jdo\pm1/2}$.
\item $sv$-$sviii$: one of $y_{\jtar\pm1/2}-\ysh(x)$ intersects both $y_{\jdo\pm1/2}$ lines, while the other only intersects one.
\item $six$-$sxii$: one of $y_{\jtar\pm1/2}-\ysh(x)$ intersects both $y_{\jdo\pm1/2}$ lines, while the other doesn't intersect either.
\item $sxiii$-$sxiv$: $y_{\jtar-1/2}-\ysh(x)$ intersects $y_{\jdo-1/2}$ and $y_{\jtar+1/2}-\ysh(x)$ intersects $y_{\jdo+1/2}$, or viceversa.
\item $sxv$-$sxvi$: one of $y_{\jtar\pm1/2}-\ysh(x)$ intersects both $x$-boundaries of the donor cell.
\end{itemize}

In all scenarios aside from $sNi-sNii$ at least one of $O=\{O_{--},O_{-+},O_{+-},O_{++}\}$ lies outside of the donor cell, and therefore would not be found. By identifying which of them is exterior to the donor cell (and other considerations), we can classify the sub-cell integral in any given donor cell. We therefore categorize sub-cell integrals using the criteria outlined in~\ref{sec:scenarioClassify}.

\subsection{Performing sub-cell integrals} \label{sec:subInt}

Once we identify the type of sub-cell integral required we recourse to a series of function inversions, projections on basis functions and integrations in order to compute the contribution of a sub-cell region to equation~\ref{eq:interpFormula}. The various sub-cell integrals we need to perform are made up of simply 3 kinds of integrals: integrals with variable $x$-limits, integrals with variable $y$-limits, or integrals over the whole cell minus two integrals of the other two types\footnote{It may be possible to formulate this in terms of fewer or a even a single sub-cell integral, but we have not yet achieved that.}. Below we give two examples using the simplest sub-cell integrals. Additional details on other more complex scenarios are given in~\ref{sec:addDetails}.

\subsubsection{Sub-cell integrals with variable $y$-limits} \label{sec:subcellVarY}

\begin{figure}[h]
  \centering
  \begin{subfigure}[b]{0.53\textwidth}
    \centering
    \includegraphics[width=\textwidth]{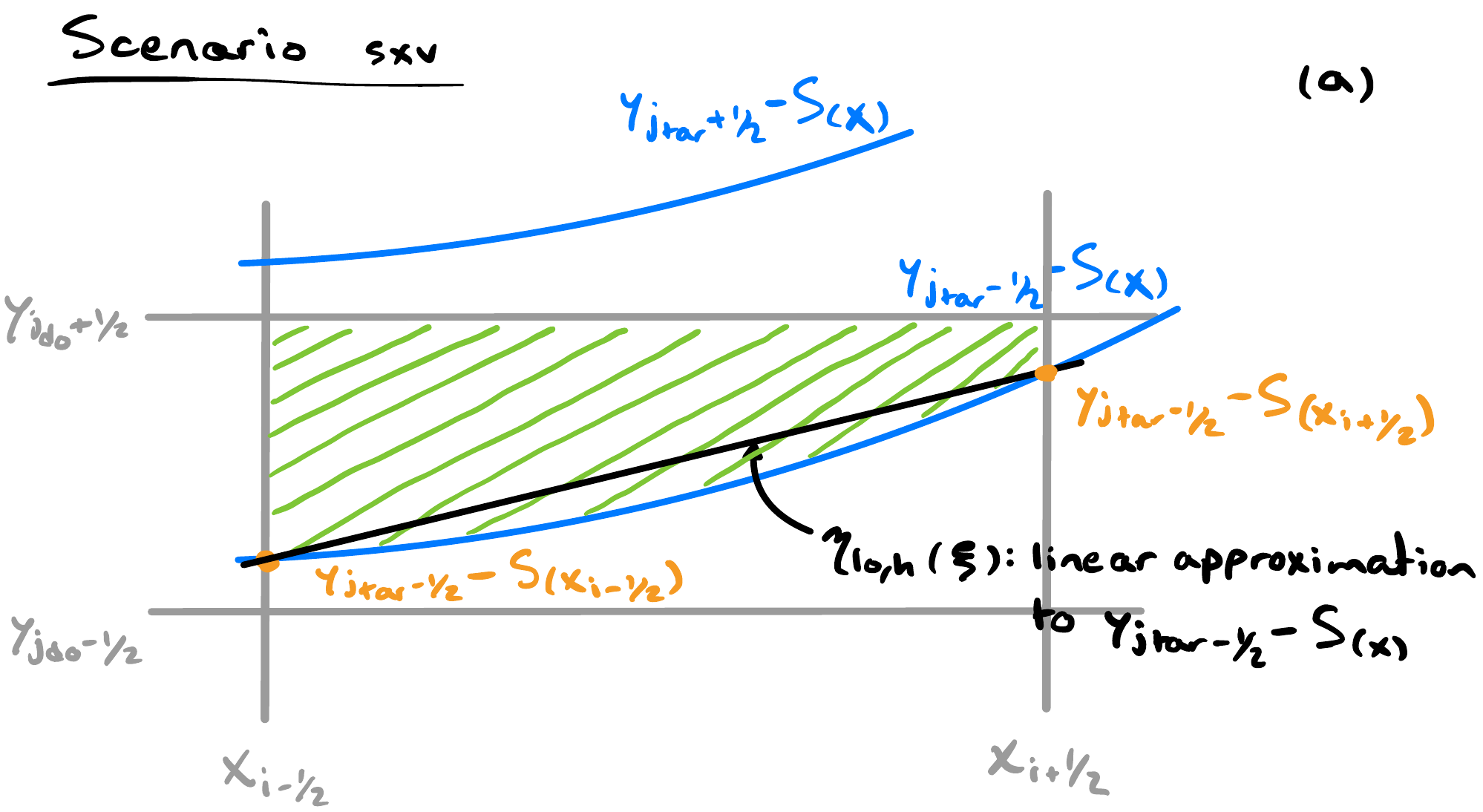}
  \end{subfigure}
    \begin{subfigure}[b]{0.45\textwidth}
    \centering
    \includegraphics[width=\textwidth]{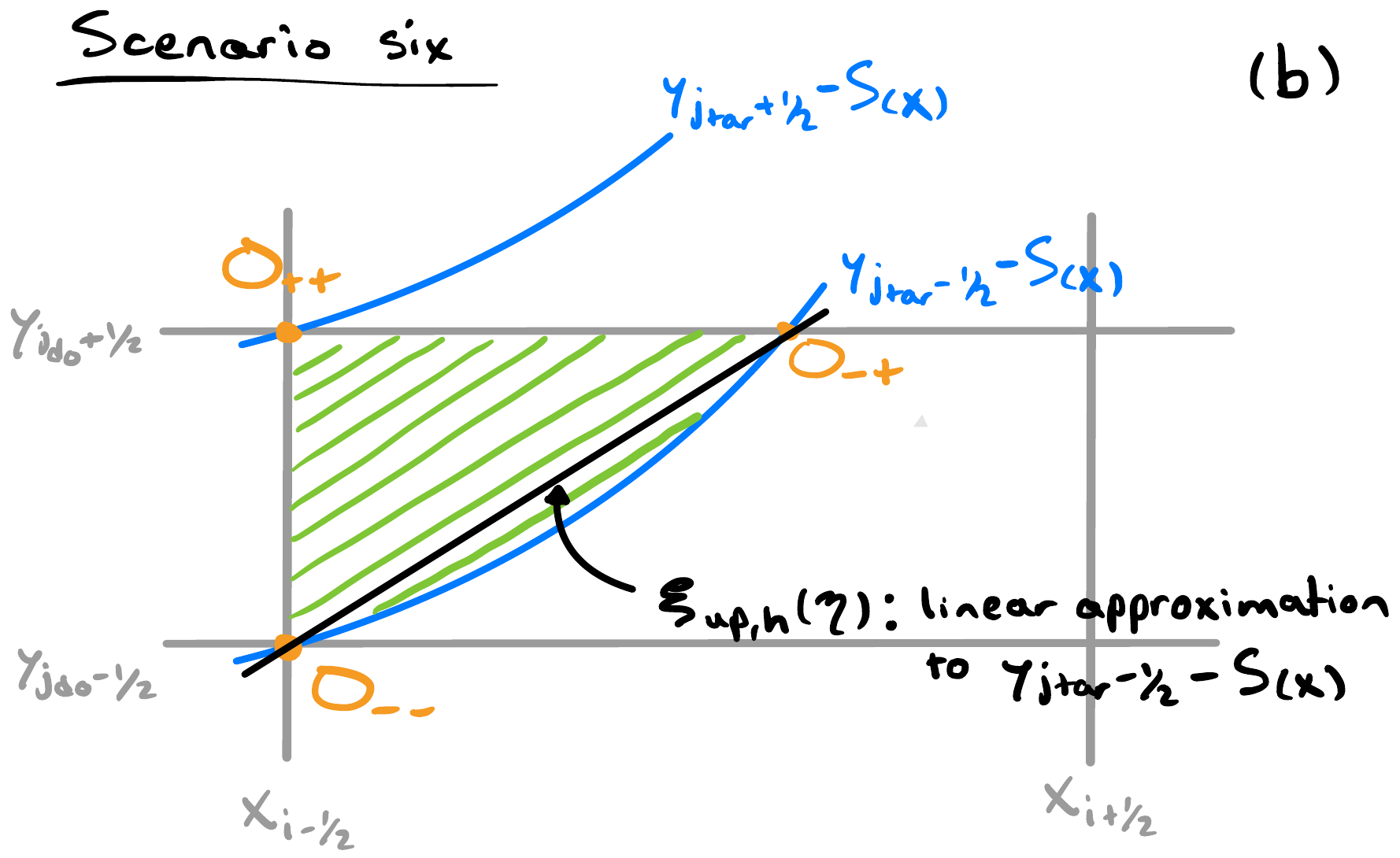}
  \end{subfigure}
   \caption{(a) Scenario $sxv$: the green striped area is the region in the donor cell we need to integrate over. The lower $\eta$ limit of the integral is given by the lower blue curve, but its discrete linear approximation is given by the black line. (b) Scenario $six$: the upper $\xi$ limit of the integral is given by the lower blue curve, but its discrete linear approximation is given by the black line.}
   \label{fig:scenarioSxvSix}
\end{figure}

We begin with an example of how to compute one of the simplest of the non-rectangular sub-cell integrals, that in scenario $sxv$ (see figure~\ref{fig:scenarioSxvSix}(a)). In this case we find that the integral spans the whole cell along $x$, it has a spatially varying lower $y$-limit and a fixed upper $y$-limit. We'll denote the contribution from this sub-cell scenario as $f_{\mathrm{tar},\jtar,k}^{sxv}$:
\begin{eqnal} \label{eq:sxvIntPhys}
f_{\mathrm{tar},\jtar,k}^{sxv} &= \frac{4}{\Dx\Dy}\int_{x_{i-1/2}}^{x_{i+1/2}}\int_{y_{\mathrm{lo}}(x)}^{y_{\jdo+1/2}}\psi_{\jtar,k}(x,y+\yshh(x))\sum_{k'}f_{\mathrm{do},\jdo,k'}\psi_{\jdo,k'}(x,y)\,\dy\,\dx.
\end{eqnal}
We wish to perform this integral analytically leveraging computer algebra systems (CAS). For that to be viable we use a suitable approximation to the lower limit $y_{\mathrm{lo}}(x)=(y_{\jtar}-\ysh(x))\mathrm{mod}\Ly$ which, after performing the $y$-integral, results in an $x$-analytically-integrable function. The same reasoning leads us to employ the polynomial approximation ($\yshh(x)$) to shift the basis function we are projecting on. Furthermore, we would like CAS to generate a kernel that can be applied in every cell; therefore we cast equation~\ref{eq:sxvIntPhys} in terms of logical coordinates (see equation~\ref{eq:logicalCoords})
\begin{eqnal} \label{eq:sxvIntLog}
f_{\mathrm{tar},\jtar,k}^{sxv} &= \int_{-1}^{1}\int_{\etaloh(\xi)}^{1}\psi_{\jtar,k}(\xi,\eta(y+\yshh(\xi)))\sum_{k'}f_{\mathrm{do},\jdo,k'}\psi_{\jdo,k'}(\xi,\eta)\,\deta\,\dxi, \\
&= \int_{-1}^{1}\int_{\etaloh(\xi)}^{1}\psi_{k}(\xi,\frac{y+\yshh(\xi)-y_{\jtar}}{\Dy/2})\sum_{k'}f_{\mathrm{do},\jdo,k'}\psi_{\jdo,k'}(\xi,\eta)\,\deta\,\dxi, \\
&= \int_{-1}^{1}\int_{\etaloh(\xi)}^{1}\psi_{\jdo,k}(\xi,\eta+\frac{\yshh(\xi)+y_{\jdo}-y_{\jtar}}{\Dy/2})\sum_{k'}f_{\mathrm{do},\jdo,k'}\psi_{\jdo,k'}(\xi,\eta)\,\deta\,\dxi.
\end{eqnal}
Notice that the last step changes the $j$ index of the basis we are projecting on from $\jtar$ to $\jdo$, because after adding and subtracting $y_{\jdo}$ to its argument we can define the logical coordinate $\eta$ in terms of the cell center of the donor cell. Equation~\ref{eq:sxvIntLog} is in a form that will look the same for any donor cell contributing via a scenario $sxv$ integral, and can thus be implemented in a single kernel.

It remains to define the numerical approximation to the lower limit $y_{\mathrm{lo}}(x)=(y_{\jtar}-\ysh(x))\mathrm{mod}\Ly$, whose approximation in logical coordinates we denoted $\etaloh(\xi)$. We compute this this quantity by projecting the function that describes that boundary onto a 1D polynomial basis along $x$, $\varphi_k(\xi)\in\mathcal{V}^p_{i}$ (see section~\ref{sec:discreteS}):
\begin{eqnal}
\eta_{\mathrm{lo},h,k}(\xi) = \int_{-1}^{+1}\varphi_k(\xi)\frac{2}{\Dy}\left(y_{\mathrm{lo}}(x_i+\frac{\Dx}{2}\xi)-y_{\jdo}\right)\,\dxi.
\end{eqnal}
Were this integral to be performed via (e.g. Gaussian) quadrature discontinuities would arise in the discrete representation of these integral limits from one cell to the next. Thus we perform this projection evaluating the function at nodal coordinates and using a nodal-to-modal transformation in order to obtain the DG expansion coefficients of $\eta_{\mathrm{lo},h}(\xi)$. Since we always have a node on the boundary, this gives a continuous representation of the integral limit from one cell to the next. An example of representing $\etalo(\xi)$ with a piecewise linear polynomial is depicted in figure~\ref{fig:scenarioSxvSix}(a).

\subsubsection{Sub-cell integrals with variable $x$-limits} \label{sec:subcellVarX}

The sub-cell integral procedure has some extra steps when the integral requires variable $x$-limits, as is the case with scenarios $sNi$-$sNii$ and $six$-$sxiv$. Scenarios $si$-$siv$ can also be done with variable $x$-limits, although it can be more robust to use variable $y$-limits. In order to illustrate how a variable $x$-limit sub-cell integral is computed we take scenario $six$ as a case study (figure~\ref{fig:scenarioSxvSix}(b)). In such a case the contribution to the right side of equation~\ref{eq:interpFormula} comes from
\begin{eqnal}
f_{\mathrm{tar},\jtar,k}^{six} &= \frac{4}{\Dx\Dy}\int_{y_{\jdo}-1/2}^{y_{\jdo}+1/2}\int_{x_{i-1/2}}^{x_{\mathrm{up}}(y)}\psi_{\jtar,k}(x,y+\yshh(x))\sum_{k'}f_{\mathrm{do},\jdo,k'}\psi_{\jdo,k'}(x,y)\,\dx\,\dy,
\end{eqnal}
or in logical coordinates:
\begin{eqnal} \label{eq:sixIntLog}
f_{\mathrm{tar},\jtar,k}^{six} &= \int_{-1}^{1}\int_{-1}^{\xiuph(\eta)}\psi_{\jdo,k}(\xi,\eta+\frac{\yshh(\xi)+y_{\jdo}-y_{\jtar}}{\Dy/2})\sum_{k'}f_{\mathrm{do},\jdo,k'}\psi_{\jdo,k'}(\xi,\eta)\,\dxi\,\deta.
\end{eqnal}

The sub-cell integral in equation~\ref{eq:sixIntLog} is not too dissimilar from the one used for scenario $sxv$ in equation~\ref{eq:sxvIntLog}, except that this time we need a function of $y$ that defines the upper $x$-limit. Such boundary is given by the $y_{\jtar-1/2}-\ysh(x)$ curve within the segment $x\in\left[O_{--},O_{-+}\right]$. But we wish to describe these curves as functions of the computational coordinate $\eta$ (see equation~\ref{eq:sixIntLog}). We obtain such functions by inverting the $y_{\jtar-1/2}-\ysh(x)$ function via root finding and translating it to logical space. That is, $\xiup(\eta)$ consists of the roots of
\begin{equation}
\mathcal{N}_{\xiup}(\xi) = \eta - \frac{2}{\Dy}\left[\left(y_{\jtar-1/2}-\ysh(x_i+\frac{\Dx}{2}\xi)\right)\mathrm{mod}\Ly - y_{\jdo}\right],
\end{equation}
or $\xiup(\eta)=\mathrm{roots}\left(\mathcal{N}_{\xiup}(\xi)\right)$, and we look for these roots in the $[O_{--}-\epsilon,O_{-+}+\epsilon]$ segment down to a tolerance of $10^{-10}$, where a small $\epsilon\sim10^{-11}$ may sometimes be needed to avoid floating point comparison errors. We then project $\xiup(\eta)$ onto the 1D basis along $\eta$, $\varphi_k(\eta)\in\mathcal{V}^p_{j}$ where $\mathcal{V}^p_j=\{\eta^n~|~\mathrm{deg}(\eta^n)\leq p\}$ and $\abs{\mathcal{V}^p_j}=\numB^{1d}$, using evaluation on nodes and a nodal-to-modal transformation to obtain its discrete approximation:
\begin{equation}
\xiuph(\eta)=\sum_{k=1}^{\numB^{1d}}\xi_{\mathrm{up},h,k}\varphi_k(\eta),
\end{equation}
which for a piecewise linear basis is represented by the black line in figure~\ref{fig:scenarioSxvSix}(b).

\subsection{Summing sub-cell integrals and applying the BC} \label{sec:timeDepApp}

Once integrals such as those in equations~\ref{eq:sxvIntLog} and \ref{eq:sixIntLog} (as well as any other sub-cell integral needed) are computed, their contributions are added up in order to compose the DG expansion coefficients of the target field, $f_{\mathrm{tar},\jtar,k}$. Overall though, the algorithm described in sections~\ref{sec:discreteS}-\ref{sec:subInt} involves many steps, complex pattern identification, root finders, and various projections onto basis functions. It would be expensive to carry out this task every time we need to compute the field in the $z$-ghost cells of a field in every single time step.

Examining sub-cell integral equations~\ref{eq:sxvIntLog},~\ref{eq:sixIntLog},~\ref{eq:subIntA},~\ref{eq:subIntV} and~\ref{eq:subIntB} we see that these operations are linear in the donor field DG coefficients. This means that each sub-cell integral can ultimately be expressed as a multiplication of a small matrix times the vector of DG coefficients of the donor cell, or as a linear stencil acting on the donor field. Furthermore, since we assume that $\ysh(x)$ is not changing in time we can pre-compute said matrices, and simply reuse them any time BCs are applied. If we write the matrix arising from the sub-cell integral in the $q$-th donor field as $M_q$ and express the DG coefficients of the target field in the $(i,\jtar)$-th cell as $\v{f}_{\mathbf{tar}i,\jtar}$, we can write the interpolation operation as
\begin{eqnal}
\v{f}_{\mathbf{tar}i,\jtar} = \sum_{q=1}^{\Ndo} M_q~ \v{f}_{\mathbf{do}i,\v{j}_{q}}
\end{eqnal}
where $\v{j}_q$ is the vector of $\jdo$ indices and $\v{f}_{\mathbf{do}i,\v{j}_q}$ is the vector of donor field DG coefficients in the $(i,\v{j}_q)$-th cells. The matrices are in general dense but small, and since we are using Galerkin projection the size of the matrix is equal to the number of basis functions, i.e. $\numB\times\numB$. Hence, the cost of the algorithm is that of $\Ndo$ matrix-vector multiplications with matrices containing $\numB\times\numB$ elements. Even in higher dimensions the size of these matrices remains $\numB\times\numB$; the fact that the shift only occurs in one direction may mean that in these cases the matrix is actually sparse, specially for higher dimensions, but we have not optimized this yet. 

\subsection{Twist shift in higher dimensions} \label{sec:algoHigherD}

Previous sections described the mechanics of shifting a DG field via two separate two-dimensional fields, $\fdo(x,y)$ and $\ftar(x,y)$. In reality we are interested in applying these boundary conditions in three dimensional fluid or five dimensional gyrokinetic simulations. Such simulations typically include a ghost cell on each $z$-side of our domain, abutting a layer of boundary cells inside the domain which we call skin cells. Our procedure in 3D and 5D is then to take the field in the skin cells, apply the twist-shift to it, and place the result in the ghost cells at the opposite boundary. Specifically, in a grid with cells of length $\Dz$ along $z$ we enforce the following condition at the lower-$z$ boundary:
\begin{eqnal} \label{eq:bc3xlo}
f_{z_\mathrm{ghost}^-}(x,y,z) &= f\left(x,y,-\frac{\Lz}{2}-\Dz\leq z\leq-\frac{\Lz}{2}\right) 
\\
&= f\left(x,y+\ysh(x),\frac{\Lz}{2}-\Dz\leq z\leq\frac{\Lz}{2}\right) = f_{z_\mathrm{skin}^+}(x,y+\ysh(x),z).
\end{eqnal}
Analogously, at the upper boundary we impose
\begin{eqnal} \label{eq:bc3xup}
f_{z_\mathrm{ghost}^+}(x,y,z) &= f\left(x,y,\frac{\Lz}{2}\leq z\leq\frac{\Lz}{2}+\Dz\right) \\
&= f\left(x,y-\ysh(x),-\frac{\Lz}{2}\leq z\leq-\frac{\Lz}{2}+\Dz\right) = f_{z_\mathrm{skin}^-}(x,y-\ysh(x),z).
\end{eqnal}

Equations~\ref{eq:bc3xlo}-\ref{eq:bc3xup} are enforced by an algorithm nearly identical to that described in sections~\ref{sec:discreteS}-\ref{sec:subInt}. The only difference is that the Galerkin projection is done in a higher dimensional space using a basis functions from a higher dimensional polynomial space. For example, for three-dimensions we introduce the polynomial space $\mathcal{V}^p_{i,j,k}$ containing the basis functions $\psi_{i,j,k,\ell}\in\mathcal{V}^p_{i,j,k}$ in the $(i,j,k)$-th cell on which we expand our dynamical fields. The Galerkin projection upon which the interpolation is founded is then
\begin{eqnal} \label{eq:bc3xloProjG}
\int_{-\Lx/2}^{\Lx/2}&\int_{-\Ly/2}^{\Ly/2}\int_{-\Lz/2}^{\Lz/2}\psi_{i,j,k,\ell}(x,y,z)\left(f_{z_\mathrm{ghost}^+}(x,y,z)-f_{z_\mathrm{skin}^-}(x,y-\ysh(x),z)\right)\,\dx\,\dy\,\dz = 0,
\end{eqnal}
The ensuing sub-cell integrals are constructed in the same manner as previously described. Once the interpolation is performed, the coefficients corresponding to basis functions with mixed monomials involving $y$ (e.g. that multiplying $y\,z$) will change due to the variable change $y\to y+\ysh$, but the integrals over higher dimensions involving monomials other than $y$ are unaffected. Similar arguments apply to 5D gyrokinetic simulations, and the latter implies that velocity moments of the distribution function ought to be preserved exactly, a property that we will confirm in section~\ref{sec:results}.

\section{Benchmarking results} \label{sec:results}

The algorithm described in section~\ref{sec:algorithm} has been implemented in the \gkeyll~computational plasma physics framework~\cite{gkeyllWeb}. In order to confirm the validity of the algorithm and the correctness of the implementation we performed tests of increasing complexity, starting with interpolations of 2D fields, followed by experiments with static 3D fields and time-dependent 3D and 5D problems employing twist-shift BCs. All tests employed serendipity basis sets since they are not as strongly afflicted by the curse of dimensionality as tensor product bases~\cite{Arnold2011}, a property of interest for 5D and 6D models. The results of these tests are given below and can be reproduced with the input files made available online (see section~\ref{sec:getGkeyll}).

\subsection{Interpolation of 2D fields}

\subsubsection{Constant shift and diffusion}

\begin{figure}
  \begin{subfigure}[b]{0.49\textwidth}
    \centering
    \includegraphics[width=\textwidth]{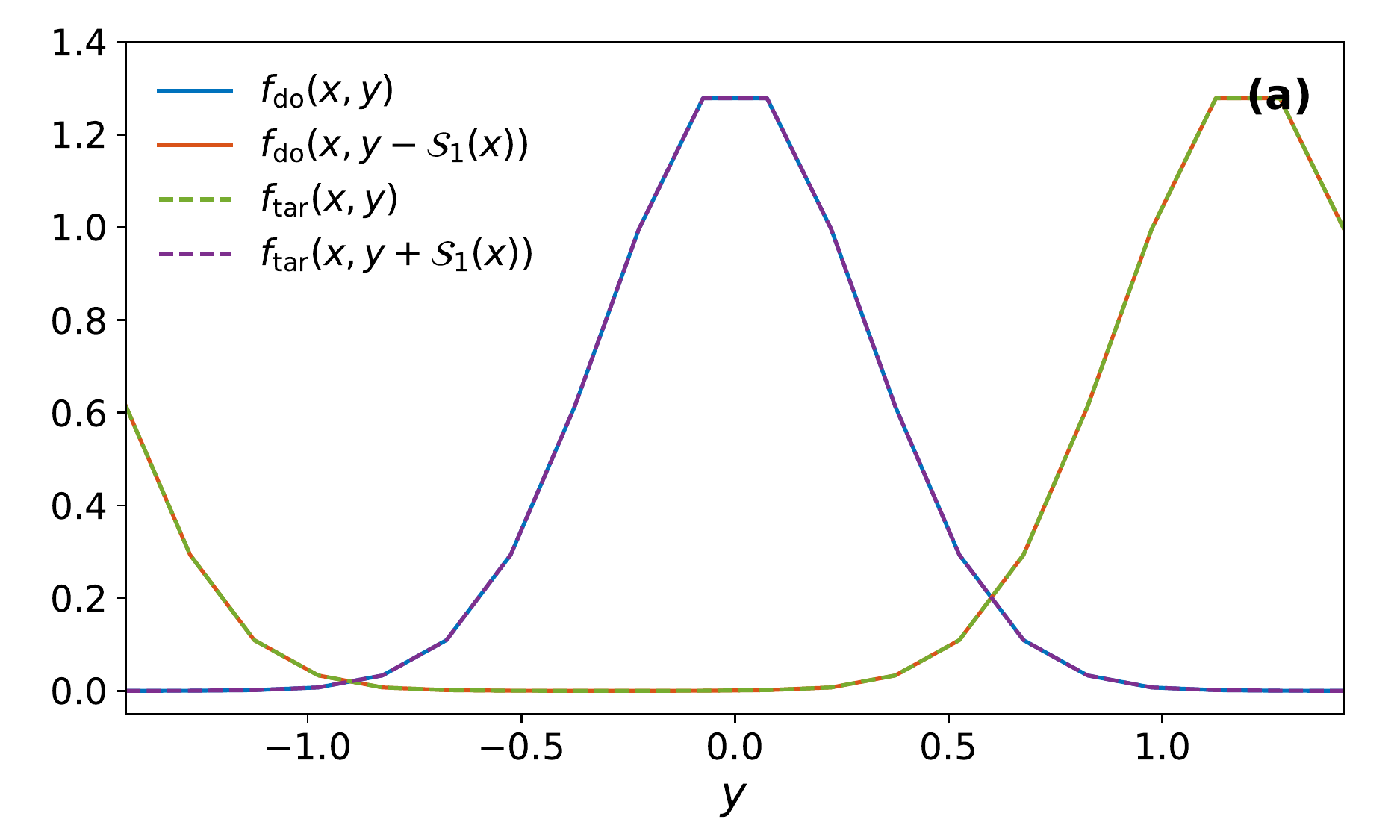}
  \end{subfigure}
  \begin{subfigure}[b]{0.49\textwidth}
    \centering
    \includegraphics[width=\textwidth]{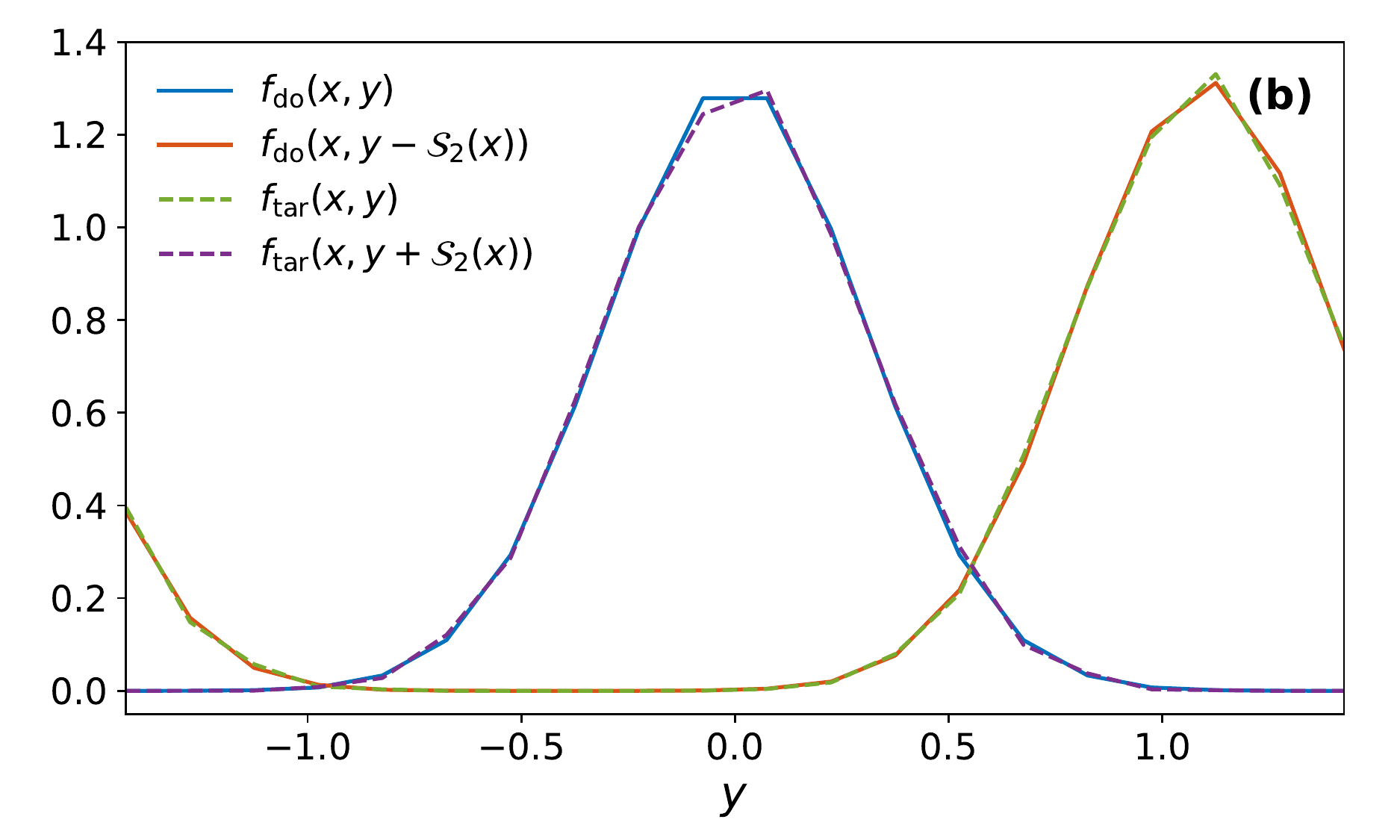}
  \end{subfigure}
  \caption{A donor field with a Gaussian variation along $y$ and constant in $x$ is shifted in a grid with $1\times10$ cells and $p=1$. Fields are shown at $x=0$. The donor field is shown in solid blue, the quadrature-shifted donor field in solid orange, the target field in dashed green, and the target field shifted back (should match $\fdo$) in dashed purple. (a) $\ysh=4\Dy=1.2$, (b) $\ysh=1.1$.}
  \label{fig:yGaussian}
\end{figure}

Consider a 2D $y$-periodic domain $(x,y)\in\left[-\Lx/2,\Lx/2\right]\times\left[-\Ly/2,\Ly/2\right]$ with $\Lx=4$ and $\Ly=3$, discretized with $1\times\Ny$ cells and a polynomial basis of order $p$. Take the donor field to be Gaussian distributed along $y$ and constant in $x$:
\begin{equation}
    \fdo(x,y) = \frac{1}{\sqrt{2\pi\sigma_y^2}}\exp\left[-\frac{\left(y-\mu_y\right)^2}{2\sigma_y^2}\right],
\end{equation}
with $\mu_y=0$ and $\sigma_y=0.3$. Our task is then to compute the target field, $\ftar$, and we begin with the very simple case of a shift that is a multiple of the cell length and constant in x: $\ysh_1(x)=4\Dy$. Using $\Ny=10$ and $p=1$ we obtain the donor and target fields shown in figure~\ref{fig:yGaussian}(a) with solid blue and dashed green lines, respectively. We can make sure we get the correct result by taking advantage of our analytic knowledge of $\fdo$ to compute the shifted donor field via quadrature. That is, in cell $j$ the $k$-th coefficient of the shifted donor cell is 
\begin{equation}
    f_{\mathrm{do},j,k}(x,y-\ysh(x)) = \int_{-\Lx/2}^{\Lx/2}\int_{y_{j-1/2}}^{y_{j+1/2}}\dx\,\dy\,\psi_{j,k}(x,y)\frac{1}{\sqrt{2\pi\sigma_y^2}}\exp\left[-\frac{\left(y-\ysh(x)-\mu_y\right)^2}{2\sigma_y^2}\right],
\end{equation}
and this integral is computed with Gaussian quadrature. The quadrature-shifted donor field is indicated in figure~\ref{fig:yGaussian}(a) with a solid orange line, and is seen to overlap with the target field. In this case computing the target field is equivalent to translating the DG coefficients of the donor field by 4 cells, and we see that the algorithm indeed has the intended effect. We can go a step further and apply the shift to the target field in the opposite direction ($-\ysh_1(x)$) and in order to test whether we get the donor field back, which we do (dashed purple line in figure~\ref{fig:yGaussian}(a)).

Translating the target field back in the case of $\ysh_1(x)=4\Dy$ would give the impression that the algorithm presented in this work is equivalent to the identity operator when applied a second time with the negated shift, i.e. $(f(x,y-\ysh))(x,y+\ysh) = f(x,y)$. But that is in general not the case. We can demonstrate the lack of such property by using a shift that is not a multiple of the cell length, e.g. $\ysh_2(x)=11\Dy/3=1.1$. Such case is illustrated in figure~\ref{fig:yGaussian}(b) with solid blue and dashed purple lines; after applying the shift a second time with the opposite sign we do not get the same field as the donor field. The asymmetry in the purple dashed line of figure~\ref{fig:yGaussian}(b) is caused by the algorithm and not the implementation, which we checked by carrying out this operation analytically (not shown here). The stencil resulting from the forward and backward interpolation is in general not symmetric and introduces diffusion, as we will see below.

\begin{figure}[h]
  \centering
  \includegraphics[width=0.7\textwidth]{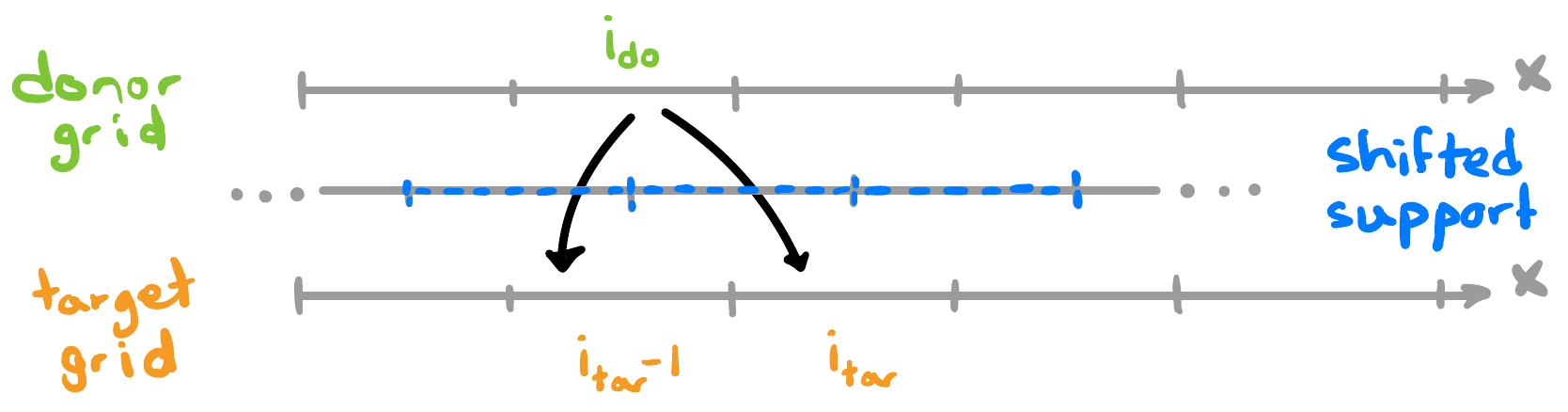}
  \caption{One-dimensional donor and target grids representing the interpolation of a donor field, non-zero only in cell $\ido$, onto the target grid. The shifted compact support of the basis functions in cells $\itar-1$ and $\itar$ is shown with dashed blue lines.}
  \label{fig:1DinterpDown}
\end{figure}

In order to further illustrate its diffusive property, we briefly consider a 1D donor field which is only non-zero in cell $\ido$ shifted by $\ysh=\Dx/2$, see figure~\ref{fig:1DinterpDown}. The shifted support of the bases in cells $\itar-1$ and $\itar$ overlap with cell $\ido$ and therefore only these two cells in the target field will be non-zero. Following equation~\ref{eq:interpFormula} but in 1D we would say that the DG coefficients for the target field $\ftar = \fdo(x-\ysh)$ in cell $\itar$ are
\begin{eqnal}
f_{\itar,k} &= \int_0^1\dxi\,\varphi_k\left(\frac{x+\ysh-x_{\itar}}{\Dx/2}\right)\sum_\ell f_{\ido,\ell}\varphi_\ell(\xi), \\
&= \int_0^1\dxi\,\varphi_k\left(\xi+\frac{\ysh-\left(x_{\itar}-x_{\ido}\right)}{\Dx/2}\right)\sum_\ell f_{\ido,\ell}\varphi_\ell(\xi).
\end{eqnal}
Since $\ysh=\Dx/2$ we know that $x_{\itar}-x_{\ido}=\Dx$ and thus
\begin{eqnal}
f_{\itar,k} &= \int_0^1\dxi\,\varphi_k\left(\xi-1\right)\sum_\ell f_{\ido,\ell}\varphi_\ell(\xi).
\end{eqnal}
Similarly, since $x_{\itar}-x_{\itar-1}=0$, we have that
\begin{eqnal}
f_{\itar-1,k} &= \int_{-1}^0\dxi\,\varphi_k\left(\xi+1\right)\sum_\ell f_{\ido,\ell}\varphi_\ell(\xi).
\end{eqnal}
For our $p=1$ orthonormal basis $\varphi_k\in\{1/\sqrt{2},\,\sqrt{3/2}\,\xi\}$ we can write these in the following form
\begin{eqnal} \label{eq:fitar}
\v{f_{i_{\mathrm{\bf tar}}}} = \frac{1}{4}\begin{bmatrix}2 & \sqrt{3}\\-\sqrt{3} & -1\end{bmatrix}\v{f_{i_{\mathrm{\bf do}}}},
\qquad
\v{f_{i_{\mathrm{\bf tar}}-1}} = \frac{1}{4}\begin{bmatrix}2 & -\sqrt{3}\\\sqrt{3} & -1\end{bmatrix}\v{f_{i_{\mathrm{\bf do}}}},
\end{eqnal}
where $\v{f_{i_{\mathrm{\bf tar}}}}$ and $\v{f_{i_{\mathrm{\bf do}}}}$ are the vector of DG coefficients of the target and donor field, respectively.

\begin{figure}[h]
  \centering
  \includegraphics[width=0.7\textwidth]{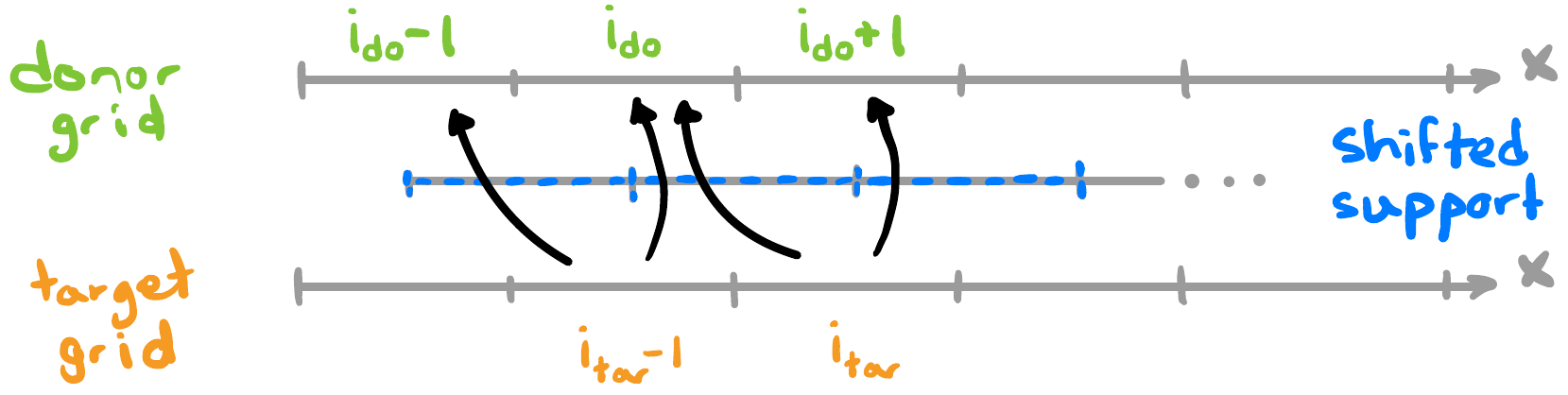}
  \caption{Representation of the reverse shift applied to the target field (non-zero only in cells $\itar-1$ and $\itar$) to compute the donor field again. In this case the top cells are the true target, and dashed blue lines show the shifted support of their bases.}
  \label{fig:1DinterpUp}
\end{figure}

\begin{figure}
  \centering
  \includegraphics[width=0.5\textwidth]{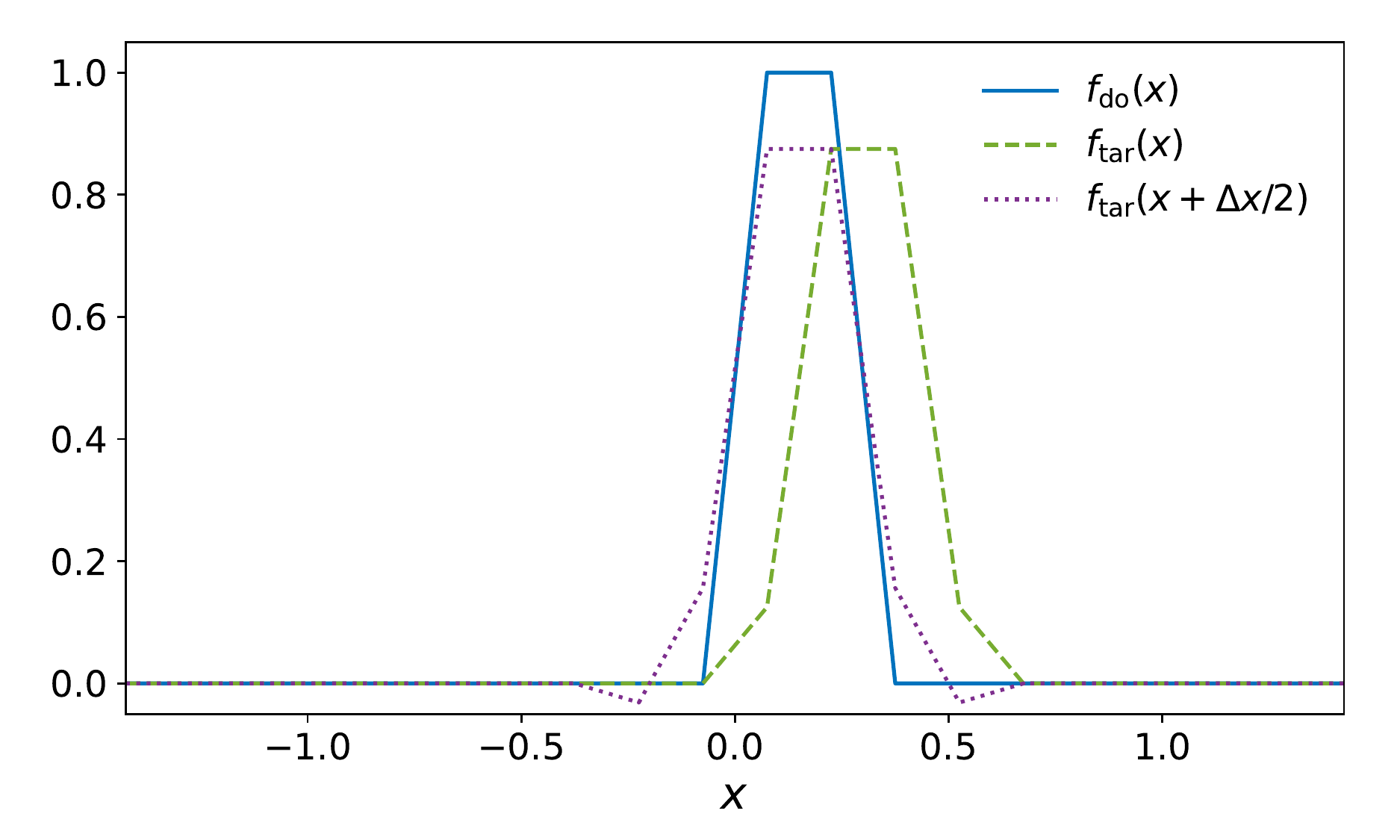}
  \caption{Shifting a 1D step function $\fdo=1$ for $0<x<\Dx=0.3$ (solid blue) in a domain with $\Lx=3$, 10 cells and piecewise linear basis functions. The target field obtained with a shift $\ysh=\Dx/2=0.15$ (dashed green) is later shifted again by applying the negated shift to it, in order to obtain the dotted purple line. These plots are produced by dividing each cell into $(p+1)$ subcells and plotting (sub)cell-center values of each function.}
  \label{fig:yStep}
\end{figure}

Having obtained the non-zero expansion coefficients of the target field we can turn the problem around and shift the target field back. That is, we wish to now compute $\fdo' = \ftar(x-\ysh)$ with $\ysh=-\Dx/2$. In cell $\ido$ the coefficients are given by
\begin{eqnal}
f_{\ido,k}' &= \int_{-1}^0\dxi\,\varphi\left(\frac{x+\ysh-x_{\ido}}{\Dx/2}\right)\sum_\ell f_{\itar,\ell}\varphi_\ell(\xi) + \int_0^1\dxi\,\varphi\left(\frac{x+\ysh-x_{\ido}}{\Dx/2}\right)\sum_\ell f_{\itar-1,\ell}\varphi_\ell(\xi), \\
&= \int_{-1}^0\dxi\,\varphi\left(\eta+\frac{\ysh-\left(x_{\ido}-x_{\itar}\right)}{\Dx/2}\right)\sum_\ell f_{\itar,\ell}\varphi_\ell(\xi) \\
&\quad+ \int_0^1\dxi\,\varphi\left(\eta+\frac{\ysh-\left(x_{\ido}-x_{\itar-1}\right)}{\Dx/2}\right)\sum_\ell f_{\itar-1,\ell}\varphi_\ell(\xi).
\end{eqnal}
Doing these integrals for $p=1$ and $\ysh=-\Dx/2$ we find
\begin{equation}
    \v{f_{i_{\mathrm{\bf do}}}'} = \frac{1}{4}\begin{bmatrix}2 & -\sqrt{3}\\\sqrt{3} & -1\end{bmatrix}\v{f_{i_{\mathrm{\bf tar}}}}
    + \frac{1}{4}\begin{bmatrix}2 & \sqrt{3}\\-\sqrt{3} & -1\end{bmatrix}\v{f_{i_{\mathrm{\bf tar}}-1}}.
\end{equation}
We can substitute what we had obtained for the DG coefficients of the target field in equations~\ref{eq:fitar} in order to obtain
\begin{equation} \label{eq:fido}
    \v{f_{i_{\mathrm{\bf do}}}'} = \begin{bmatrix}7/8 & 0\\0 & 1/2\end{bmatrix}\v{f_{i_{\mathrm{\bf do}}}}.
\end{equation}
This demonstrates that performing the shift back is indeed not equivalent to the inverse operator. Furthermore, where previously only the cell $\ido$ had non-zero coefficients, now the neighboring cells $\ido-1$ and $\ido+1$ also have non-zero coefficients. We can calculate these with the same procedure as above, yielding
\begin{eqnal} \label{eq:fidopmO}
    \v{f_{i_{\mathrm{\bf do}}\pm1}'} = \frac{1}{16}\begin{bmatrix}1 & \pm\sqrt{3}\\\mp\sqrt{3} & -2\end{bmatrix}\v{f_{i_{\mathrm{\bf do}}}}.
\end{eqnal}
The coefficients in equations~\ref{eq:fido}-\ref{eq:fidopmO} is indeed what the implemented code yields. The result of this exercise in a domain $x\in[-1.5,1.5]$ with 10 cells and piecewise linear polynomial basis ($p=1$) is shown in figure~\ref{fig:yStep}. Notice how the field shifted twice (dotted purple line) does not equal our original donor field (solid blue line), even though the volume integral of the function is preserved to machine precision ($\mathcal{O}(10^{-15})$).

Both the one and the two dimensional tests provided above raise two additional concerns. First, notice how in figure~\ref{fig:yStep} the shifted target field contains regions with negativity, i.e. regions where $\ftar(x+\Dx/2)<0$. This can be detrimental or even lead to instability in simulations where the scalar field must stay positive, e.g. particle density or distribution function. Second, the fact that the negated shift does not invert the operator can lead to unphysical diffusion or drifts. Notice that, upon applying the shift followed by the negated shift, the density increases at locations where previously it had been zero, suggesting that the algorithm presented here introduces a certain amount of diffusion.

\begin{figure}
  \begin{subfigure}[b]{0.49\textwidth}
    \centering
    \includegraphics[width=\textwidth]{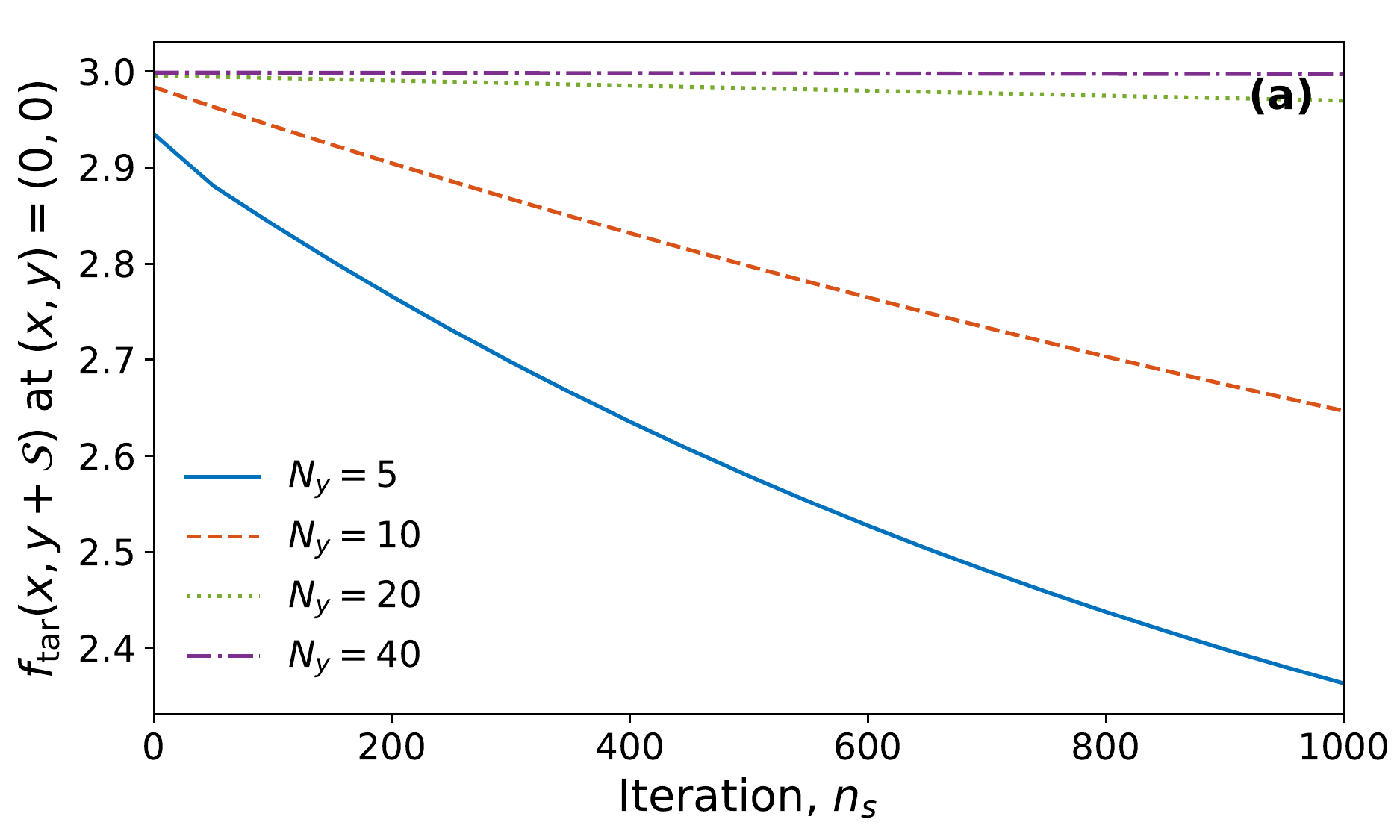}
  \end{subfigure}
  \begin{subfigure}[b]{0.49\textwidth}
    \centering
    \includegraphics[width=\textwidth]{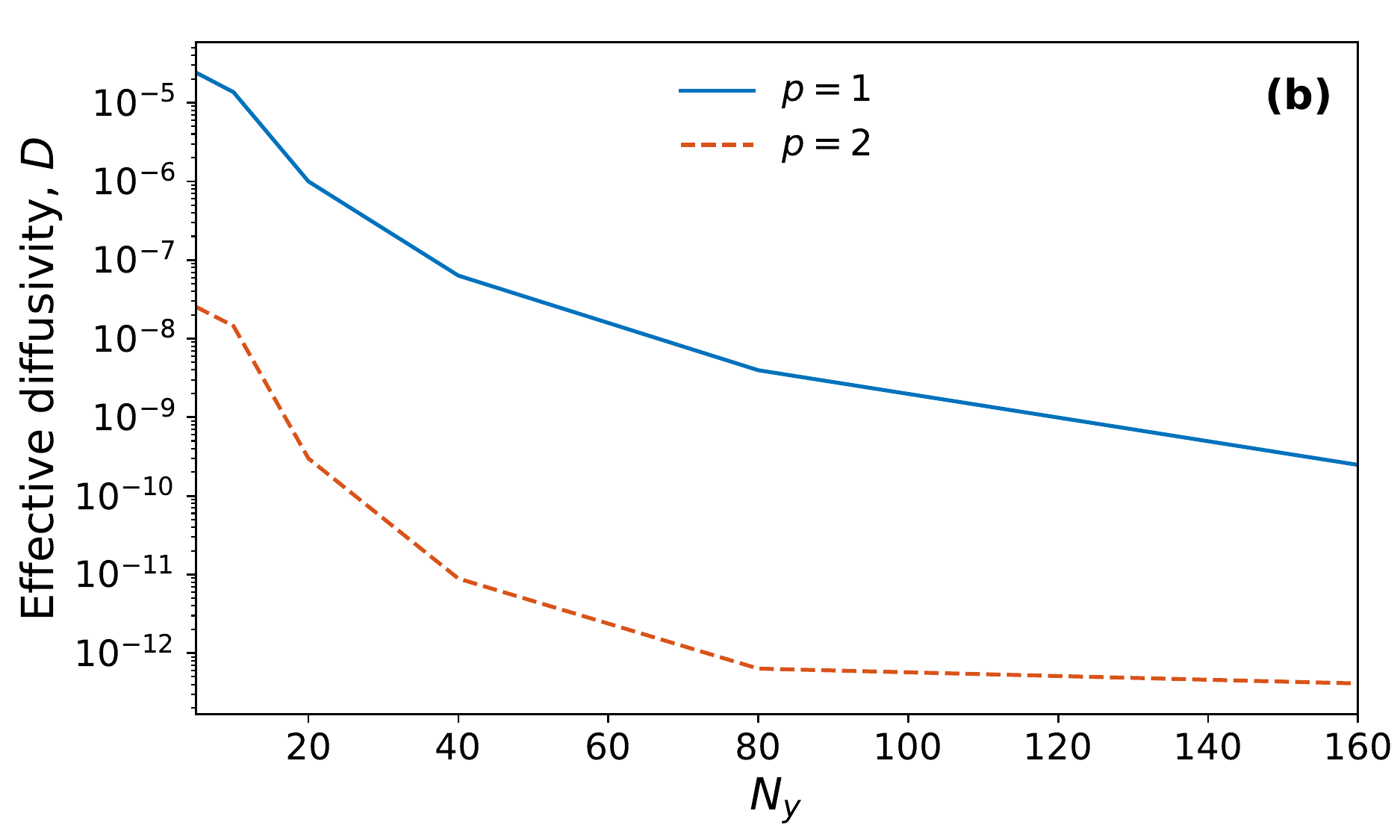}
  \end{subfigure}
  \caption{Given a sine donor field $\fdo(x,y)$ we applied the forward shift to obtain $\ftar=\fdo(x,y-\ysh)$ followed by the backward shift $\ftar(x,y+\ysh)$ repeatedly. (a) Value of $\ftar(x,y+\ysh)$ at the origin as a function of iteration number $n_s$ and resolution. (b) Effective diffusivity of the twist-shift operation in this test, computed by fitting $\exp(-2k_y^2 D n_s)$ to the data in (a).}
  \label{fig:ySineDiffusion}
\end{figure}

One could get a sense of how diffusive the algorithm is by applying the shift followed by the negated shift numerous times and measuring the effective diffusion coefficient of this operator. In a time dependent simulation one would not immediately apply the forward and backward shifts to a single 2D plane, but if we envision a perturbation rapidly advected along the field line it could make it from one $z$-boundary to the other relatively unchanged, and we would like to know how much the twist-shift BCs alone would diffuse such perturbation. So we enlist a donor field of the form $\fdo(x,y)=2+\cos(k_y y)$ ($k_y=2\pi/\Ly$), defined on a domain with $\Lx=2$, $\Ly=3$ and discretized using $1\times\Ny$ cells and a polynomial basis of order $p$. Assuming the shift $\ysh=0.9+2\Dy/3$ we compute the target field $\ftar=\fdo(x,y-\ysh)$ followed by an application of the negated shift, i.e. $\ftar(x,y+\ysh)$, and we do this $n_s$ times. As we iterate through the forward and backward shift pairs, we see the amplitude of the sine function decrease. For example, the value of $\ftar(x,y+\ysh)$ at the origin as we iteratively shift and shift back is given in figure~\ref{fig:ySineDiffusion}(a), showing that the amplitude of the function decreases rapidly for coarse meshes but very slowly for well resolved simulations. We can quantify this effective diffusion coefficient $D$ by fitting the exponential $\exp\left(-2k_y^2Dn_s\right)$ (the factor of 2 is to account for the fact that two shifts take place) and plot it against the resolution, as carried out in figure~\ref{fig:ySineDiffusion}(b) for $\Ny={5,10,20,40,80,160}$. The diffusivity of the operator drops by several orders of magnitude with only a few mesh refinements. We have also performed this test with a piecewise quadratic basis function ($p=2$), which figure~\ref{fig:ySineDiffusion}(b) indicates has much lower levels of diffusion than $p=1$.

\subsubsection{Sheared shifts and accuracy}

\begin{figure}
  \centering
  \includegraphics[width=0.8\textwidth]{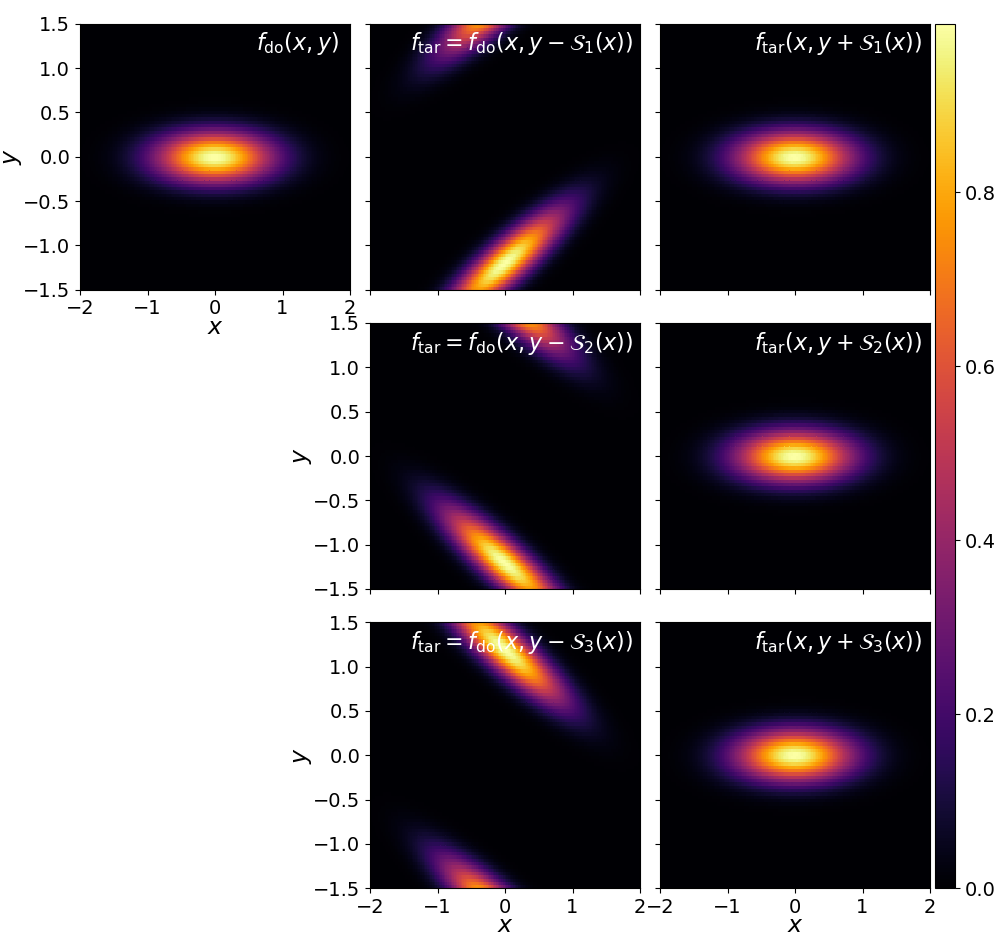}
  \caption{Shift the 2D anisotropic Gaussian defined in equation~\ref{eq:2dGaussian} (top row, left) by $\ysh_1=0.6x+1.8$ (top row center), $\ysh_2(x)=-0.6x+1.8$ (middle row center) and $\ysh_3(x)=-0.6x-1.8$ (bottom row center) along $y$ to obtain the target fields $\ftar$. Then apply the backward shift to the target field (right column), to confirm that we approximately recover the donor field.}
  \label{fig:yGaussianSheared}
\end{figure}

So far we have only performed tests with a single cell in $x$ and a constant shift in $y$. We can also demonstrate that the algorithm performs as expected when $\Nx>1$ and when the shift is sheared, i.e. $\ysh=\ysh(x)$. We now employ an anisotropic 2D Gaussian donor field
\begin{equation} \label{eq:2dGaussian}
    \fdo(x,y) = \exp\left[-\frac{\left(x-\mu_x\right)^2}{2\sigma_x^2}-\frac{\left(y-\mu_y\right)^2}{2\sigma_y^2}\right],
\end{equation}
with $\mu_x=\mu_y=0$, $\sigma_x=0.45$, $\sigma_y=0.3$ again on a $(\Lx,\Ly)=(4,3)$ domain but this time using $80\times40$ cells and $p=1$ basis functions. For each of the three shifts  $\ysh_1(x)=0.6x+1.8$, $\ysh_2(x)=-0.6x+1.8$ and $\ysh_3(x)=-0.6x-1.8$ we compute the target field $\ftar=\fdo(x,y-\ysh)$ and we apply the opposite negated shift to the target field (i.e. $\ftar(x,y+\ysh)$) to check that it approximately yields the donor field. The results are given in figure~\ref{fig:yGaussianSheared}; examining the center column we note that target field appears qualitatively correct for the cases of a positive shift with positive shear, a positive shift with a negative shear, and a negative shift with a negative shear. Furthermore, upon applying the opposite shift to the target field we approximately recover the donor field (right column in figure~\ref{fig:yGaussianSheared}).

\begin{figure}
  \begin{subfigure}[b]{0.49\textwidth}
    \centering
    \includegraphics[width=\textwidth]{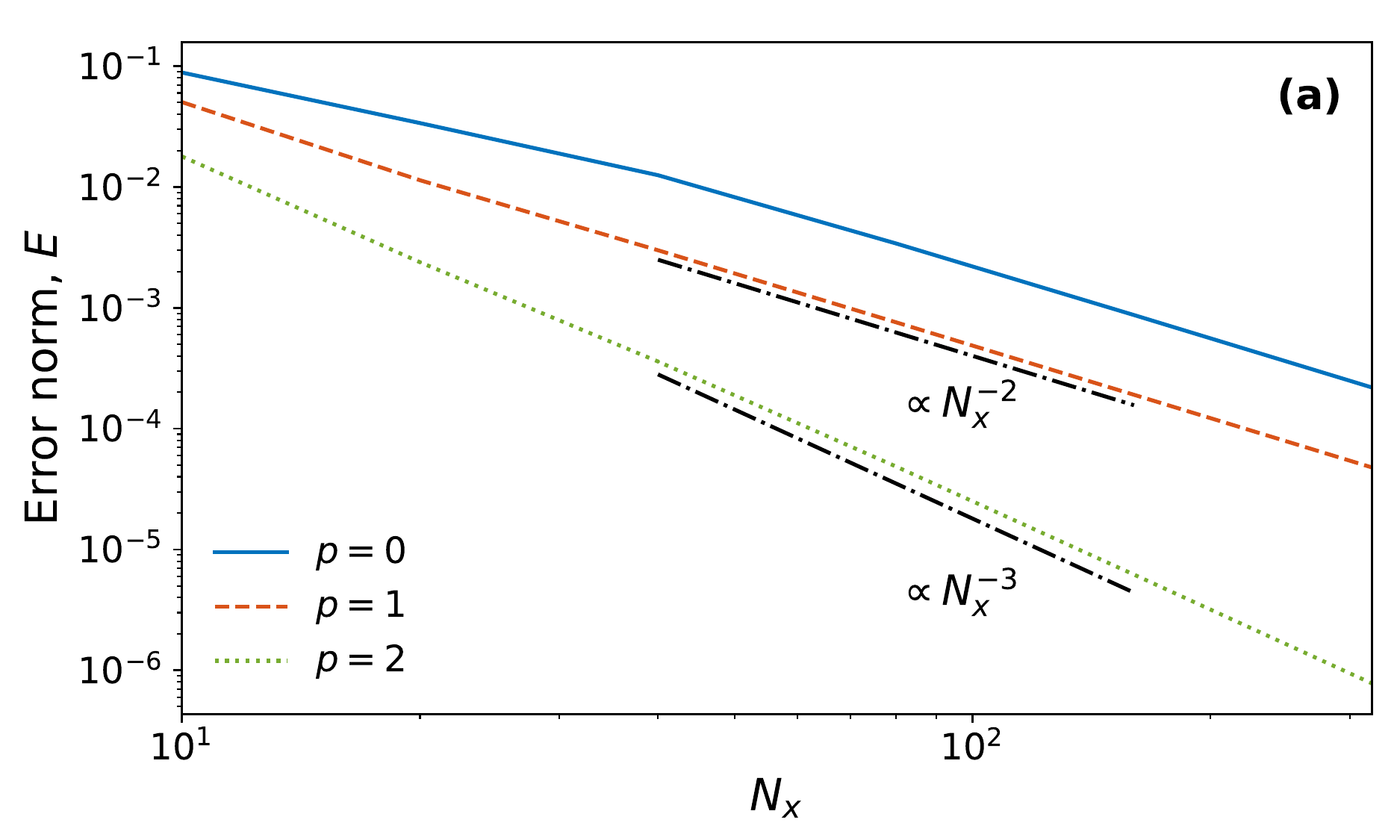}
  \end{subfigure}
  \begin{subfigure}[b]{0.49\textwidth}
    \centering
    \includegraphics[width=\textwidth]{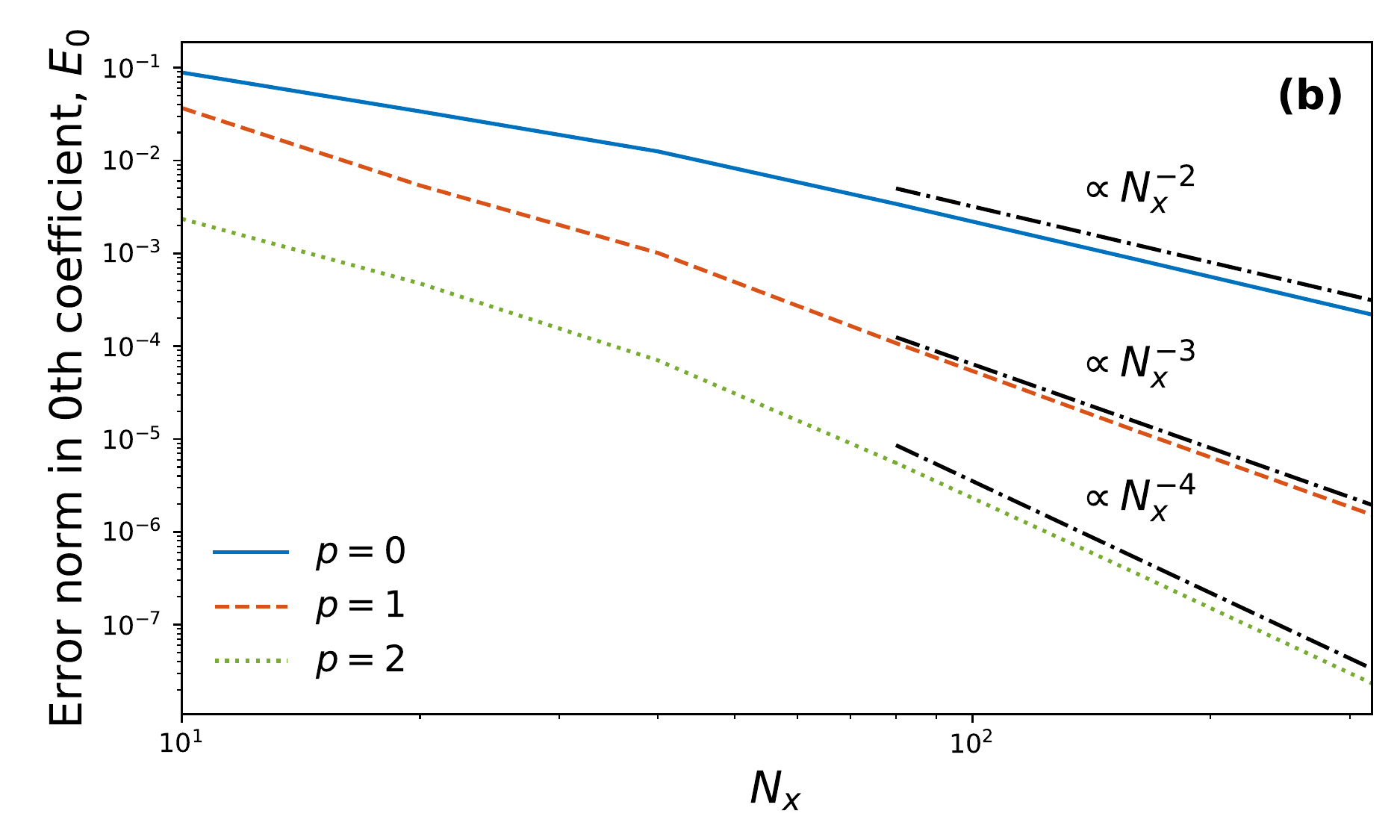}
  \end{subfigure}
  \caption{(a) Error (norm) in the field obtained after shifting the donor field in equation~\ref{eq:2dGaussian} and subsequently applying the negated shift. The error norm is defined in equation~\ref{eq:errorNorm} and is plotted as a function of number of cells along $x$, $\Nx$. (b) Error norm of the 0th DG coefficient only, which is proportional to the cell average.}
  \label{fig:yGaussianShearedError}
\end{figure}

It would be good to do more than qualitatively assess the correctness of the algorithm and provide some quantitative characterization of the error. Computing errors however is not trivial because we do not have an analytic discrete target field. As was done for figure~\ref{fig:yGaussian} one could project the shifted donor onto the basis using quadrature or evaluation at Gauss-Lobatto nodes followed by nodal-to-modal conversion; however that either incurs errors or produces a function that is not the weak (Galerkin) equivalent of the shifted donor field. For lack of a better option we opt for defining the error as the algorithm's inability to obtain the donor field again after shifting the target field back. That is, if $g(x,y) = \ftar(x,y+\ysh)$ we compute the error via
\begin{equation} \label{eq:errorNorm}
    E = \sum_{i,j,k}\sqrt{\frac{\Dx\Dy}{4}\left[\frac{1}{2}\left(f_{\mathrm{do},i,j,k}-g_{i,j,k}\right)\right]^2},
\end{equation}
where $i$ labels the cell along $x$, $j$ the cell along $y$, $k$ the basis function (coefficient), and $N$ is the total number of cells. In equation~\ref{eq:errorNorm} there's an addition factor of 1/2 to account for the fact that two shifts are performed. We examined the convergence of this error on grids with $N=\Nx\times\Ny=10c\times5c$ cells ($c=\{1,2,4,8,16,32\}$) using the $\ysh_1(x)=0.6x+1.8$ shift, and obtained the results in figure~\ref{fig:yGaussianShearedError}(a). This suggests that the algorithm's ability to invert by negating the shift only improves quadratically with the number of cells for $p=1$, while it exhibits cubic convergence for piecewise quadratic basis ($p=2$). That said the convergence in the cell-average, which is just the zeroth DG coefficient times a constant, is of order $p+2$ (figure~\ref{fig:yGaussianShearedError}(b)). It is in principle possible to obtain higher order convergence in the DG representation by taking the values in neighboring cells either before or after the twist-shift is applied in order to also obtain $(p+2)$-order accuracy in the DG representation.

There are two additional tests that we carried out with these 2D twists and shifts of a Gaussian donor. The first is that we also tested the algorithm with nonlinear shifts, e.g. $\ysh(x)=0.09(x-2.5)^2+1$. In that case the outcome is qualitatively similar to that depicted in figure~\ref{fig:yGaussianSheared}, and measuring the error in retrieving the donor field after a forward and a backward shift yields nearly the same picture as in figure~\ref{fig:yGaussianShearedError}. The second experiment we carried out was to use a higher order polynomial to represent $\ysh(x)$ and the boundaries of the sub-cell integrals. That is, we used a $p=1$ donor field to obtain a $p=1$ target field, but using a $p=2$ $\yshh(x)$. This allows a more accurate representation of sub-cell boundaries than, for example, what is depicted with a black line in figure~\ref{fig:scenarioSxvSix} for a $p=1$ $\yshh(x)$. Unfortunately for this test we saw no improvement in accuracy; it's possible that the shift profile was not non-linear enough, but also as we refine the mesh a piecewise linear approximation to sub-cell boundaries becomes increasingly accurate, such that there's less incentive for using a higher order $\yshh(x)$. The option to use a higher-order $\yshh(x)$ does add support for $p=0$ (FV), since we can represent the $y$-shift with a $p=1$ basis but the field with cell-average values only. We confirmed that the implementation works with $p=0$ and quantified its error convergence as well (solid blue lines in figure~\ref{fig:yGaussianShearedError}).

\subsection{Static and time-dependent 3D tests} \label{sec:results3x}


As explained in section~\ref{sec:algoHigherD}, three dimensional time-dependent simulations use one ghost cell on each $z$-side of the domain ($\abs{z}>\Lz/2$). So our application of the BCs consists of populating the ghost cell with the field on the opposite skin cell and twist-shifting it. More precisely, if $\Dz$ is the cell length along $z$, the lower ghost cell ($z\in\left[-\Lz/2-\Dz,-\Lz/2\right]$) will receive the twist-shifted field in the upper skin cell ($z\in\left[\Lz/2-\Dz,\Lz/2\right]$) while the upper ghost cell ($z\in\left[\Lz/2,\Lz/2+\Dz\right]$) will receive the twist-shifted field in the lower skin cell ($z\in\left[-\Lz/2,-\Lz/2+\Dz\right]$). We test this operation by creating a 3D field with the following profile
\begin{equation} \label{eq:2dGaussian}
    f(x,y,z) = \exp\left[-\frac{\left(x-\mu_x\right)^2}{2\sigma_x(z)^2}-\frac{\left(y-\mu_y\right)^2}{2\sigma_y(z)^2}\right]
\end{equation}
within the domain, i.e. $z\in\left[-\Lz/2,\Lz/2\right]$. We allowed for a Gaussian width that varies with $z$ according to $\sigma_x(z)=0.3(\Lz+z)/\Lz$ and $\sigma_y(z)=0.3(\Lz-z)/\Lz$. This time the Gaussian is not centered at the origin; $\mu_x=0.5$ and $\mu_y=0$.

\begin{figure}[h]
  \centering
  \includegraphics[width=\textwidth]{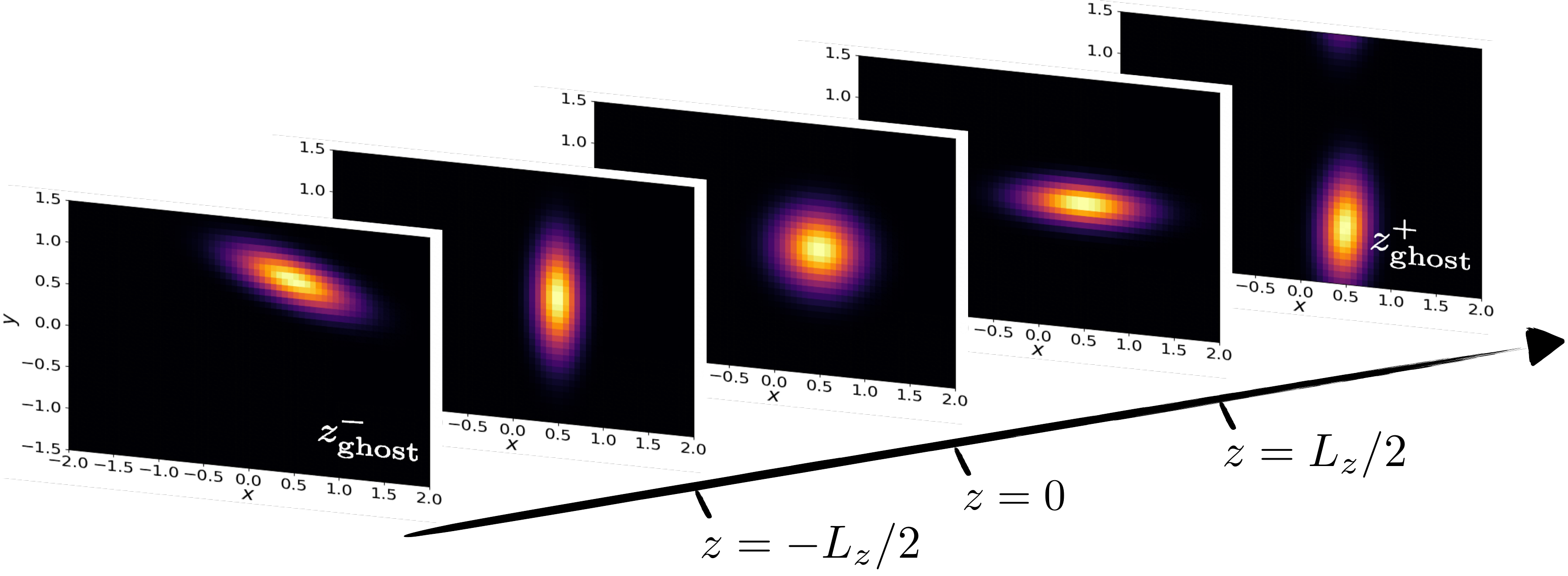}
  \caption{Gaussian with $\sigma_x=\sigma_x(z)$ and $\sigma_y=\sigma_y(z)$ at 5 $z$-planes, including one plane in each $z$-ghost cell.}
  \label{fig:staticGaussian3x}
\end{figure}

Recall that in accordance with equations~\ref{eq:bc3xlo}-\ref{eq:bc3xup} the twist-shift happens in different directions at either $z$-end of the box. We can qualitatively confirm this by plotting the field in the skin and the ghost cells. Figure~\ref{fig:staticGaussian3x} presents 5 slices of the field defined on a grid with $\Lx=4$, $\Ly=3$, $\Lz=6$, $32\times20\times8$ cells and a $p=1$ basis. We used the shift $\ysh(x)=-0.3x+0.97$. From left to right we show the lower $z$-ghost plane, the lower $z$-skin plane, the center plane ($z=0$), the upper $z$-skin plane and the upper $z$-ghost plane. The structure twists in opposite directions in going from the top skin plane to the bottom ghost plane than in going from the bottom skin plane to the top ghost plane.

In addition to applying this operation to a static 3D field once, we can test the twist-shift BCs in a time-dependent simulation of a passively advected scalar field . That is, we can solve
\begin{eqnal} \label{eq:passiveScalar}
    \pd{f}{t} + \div{(f\v{u})} = 0,
\end{eqnal}
with $\v{u}=(0,0,u_z)$ applying the twist-shift BCs at the $z$-ends of the box, and regular periodicity along $x$ and $y$. We do this in a unit cube domain ($\Lx=\Ly=\Lz=1$) with $16^3$ cells and $p=1$ polynomial basis functions. The discretization of~\ref{eq:passiveScalar} follows the DG scheme in \gkeyll~documented in, for example,~\cite{Hakim2020}. Basically one can multiply equation~\ref{eq:passiveScalar} by a basis function $\psi_{\ell}$ in the $(i,j,k)$-th cell to obtain the weak form
\begin{equation}
    \int_{K_{i,j,k}}\,\psi_\ell\pd{f}{t}\,\mathrm{d}\v{x} + \oint_{\partial K_{i,j,k}}\psi_\ell^-\uv{n}\cdot\v{\hat{F}}\,\mathrm{d}S - \int_{K_{i,j,k}}\grad{\psi_\ell}\cdot f\v{u}\,\mathrm{d}\v{x}= 0,
\end{equation}
where $\hat{n}\cdot\v{\hat{F}}=\hat{n}\cdot\v{\hat{F}}(f^-\v{u}^-,f^+\v{u}^+)$ is a numerical flux depending on the values of $f$ and $\v{u}$ on either side of the cell surface perpendicular to $\uv{n}$, which is up-winded based on the value of $\v{u}$ at Gaussian quadrature points. The $-/+$ superscript signals evaluation at the lower/upper side of the surface, respectively. The integrals in this last equation are computed exactly using kernels generated with computer algebra systems. The results presented here use a strong-stability-preserving (SSP) Runge-Kutta third-order time marching scheme. More details can be found in other \gkeyll~works~\cite{Hakim2020,Mandell2020}.

We solve equation~\ref{eq:passiveScalar} beginning with the following rectangular initial condition
\begin{eqnal} \label{eq:passiveAdvIC}
    f(x,y,z,t=0) = \begin{cases}
    1 & \left|x_\nu-L_{x_\nu}/2\right|<L_{x_\nu}/4\quad\forall\nu\\
    10^{-10}
    \end{cases}
\end{eqnal}
where $x_\nu\in\{x,y,z\}$ and $L_{x_\nu}\in\{\Lx,\Ly,\Lz\}$. We set $u_z=1$ and use the linear shift $\ysh(x)=x-0.5$; note that in the unit cube domain this $y$-shift goes through zero in the center of the $x$-domain. This is one of the few scenarios in which the implementation works despite violating the first of the restrictions on $\ysh(x)$ stated in section~\ref{sec:algorithm}. The implementation may allow $\ysh(x)=0$ at some $x$, as long as this takes place at a cell boundary and not within a cell. As the rectangular IC is advected in the $\uv{z}$ direction it sees no impact by the BC at $z=\Lz/2$ due to upwinding. It is only twisted and sheared by the BC in equation~\ref{eq:bc3xlo}, which for $\ysh=x-0.5$ causes the left half of the rectangular $f$ to shift upwards and the right half to shift downwards.

\begin{figure}
  \centering
  \includegraphics[width=\textwidth]{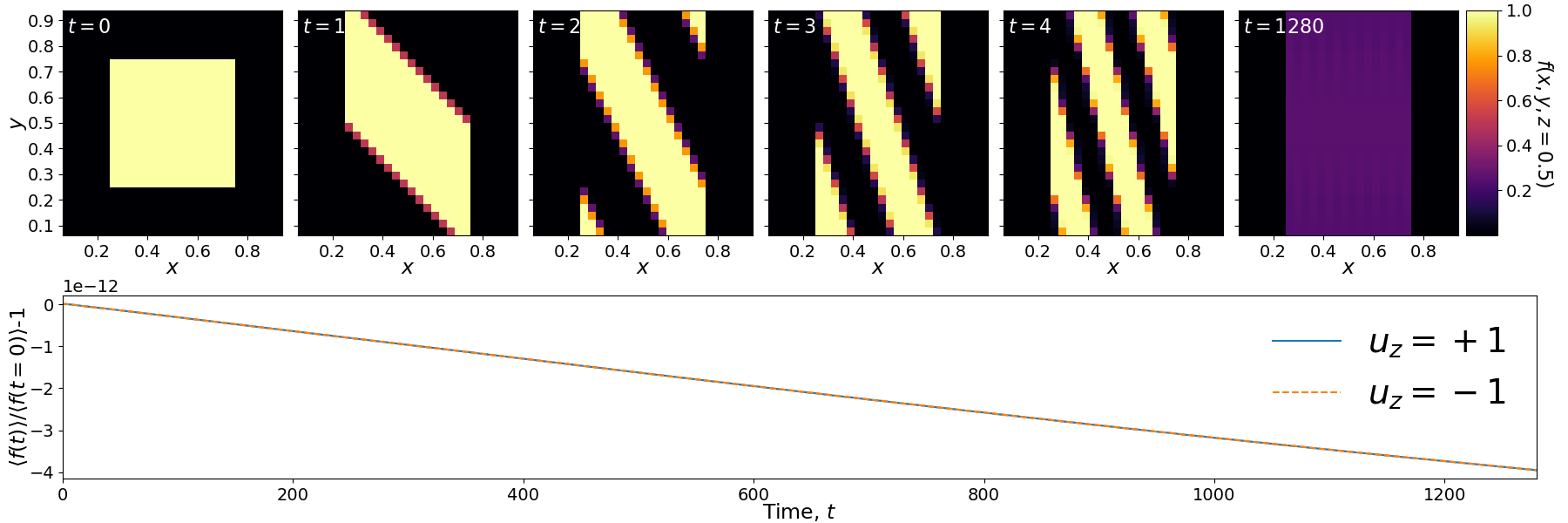}
  \caption{Top: snapshots of the $z=0.5$ plane of the passively advected 3D scalar at times $t=0,\,1,\,2,\,3,\,4,\,1280$. Bottom: Volume integral of the advected scalar over time.}
  \label{fig:passiveAdv3x}
\end{figure}

In figure~\ref{fig:passiveAdv3x} we display six snapshots of $f(x,y,z=0.5,t)$ at $t=0,1,2,3,4,1280$, as well as the relative error in the volume integral of $f$ ($\langle f\rangle$) showing that this quantity is conserved to machine precision. A movie of this simulation is also provided in the supplemental materials. We see that as time goes by and the field is repeatedly advected through the lower $z$-boundary, it is increasingly sheared. Eventually the structures have such a small wavelength in $x$ (large $k_x$) that they cannot be resolved ($t\simeq 18$). Aliasing causes these high $k_x$ modes to re-enter the simulation at lower $k_x$. The process continues until eventually the diffusion in the interpolation algorithm produces a solution uniform in $y$ ($t=1280$ in figure~\ref{fig:passiveAdv3x}). Note that the algorithm's diffusion is only in the $y$-direction; there is no diffusion in the direction perpendicular to the shift.

In spectral codes~\cite{Jenko2000,Dorland2000} this large shearing eventually causes some structures to exit the finite $k_x$-grid, and the mode is simply lost. Physically that mode would likely be diffused once it reaches the viscous range anyway, so one is justified in dropping it. But for real-space codes the ever shearing structures cause aliasing, also referred to as recurrence in kinetic simulation or carbuncles in shock and accretion disk modeling~\cite{Xu2019}. Proposed solutions to this problem include adding artificial dissipation or using numerical fluxes that have sufficient intrinsic diffusion to destroy structures with a $k_x$ higher than what the grid can support. However the diffusion must be introduced in the $x$-direction; for the present test advection was solely in the $\hat{z}$ direction so upwinding introduced no $x$-diffusion. If we instead use $\v{u}=(0.2,0,1)$ we find that the $x$-diffusion introduced by upwinding along $x$ quickly ($t\sim10$) dissipates the structure once its $k_x$ is above the maximum $k_x$ of the grid, $\sim (p+1)\pi/\Dx$. The result is then a structure that moves in both $\uv{x}$ and $\uv{z}$, and has been diffused slightly along $x$ and strongly along $y$. A movie of this scenario is provided in the supplemental materials. In turbulence simulation there is advection in all three (or 5) dimensions, so we expect some amount of diffusion in all of these. However it may still be necessary to either add additional artificial diffusion or to improve the interpolation algorithm to limit aliasing. Exploring these strategies will be the subject of future work.

\subsection{Conservation in 5D and linear ITG benchmark}

As discussed in sections~\ref{sec:intro}-\ref{sec:coords} twist-shift BCs are common amongst gyrokinetic solvers for magnetized plasma turbulence modeling. These codes evolve the 5D guiding center distribution function $f_s(\v{R},\vpar,\mu)$ of species $s$ (e.g. electrons, ions) having mass $m_s$, with $\v{R}=(x,y,z)$ referring to the guiding center position. In keeping with physical conservation laws we would like our interpolation and BC algorithm to conserve the first three integrated velocity moments of the distribution function, which are (dropping the species label $s$)
\begin{eqnal}
    \avg{M_0} &= \left\langle (2\pi/m)\int Bf\,\dvpar\dmu\right\rangle, \\
    \avg{M_1} &= \left\langle (2\pi/m)\int \vpar\,Bf\,\dvpar\dmu\right\rangle, \\
    \avg{M_2} &= \left\langle (2\pi/m)\int \left(\vpar^2+2\mu B/m\right)Bf\,\dvpar\dmu\right\rangle.
\end{eqnal}
The velocity moments $M_0$, $M_1$ and $M_2$ are not necessarily conserved because, being functions of $y$, they are also shifted by $\ysh(x)$. But as shown in the previous section with 3D fields, the shift should be area (or volume) preserving, so we are lead to expect that the integrated velocity moments should remain constant to machine precision.

\begin{figure}
  \centering
  \includegraphics[width=0.8\textwidth]{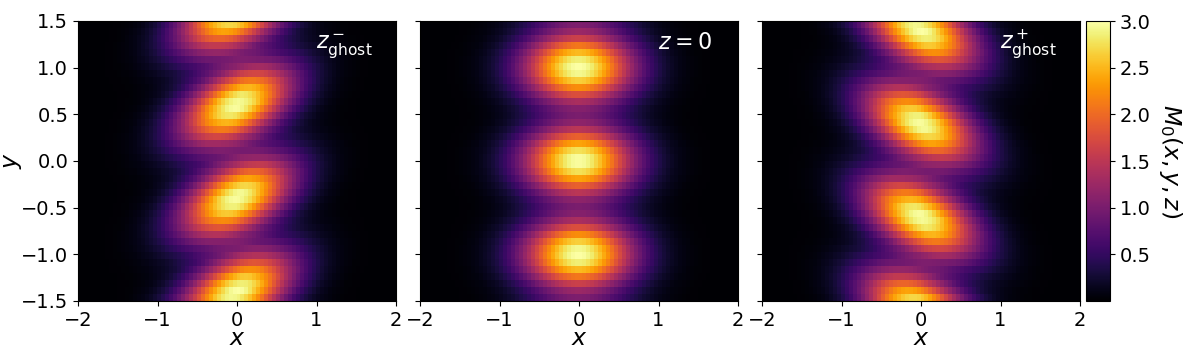}
  \caption{Number density $M_0(x,y,z)$ of the 5D distribution function in equation~\ref{eq:static3x2vDistF} in the lower-$z$ ghost plane (left), the $z=0$ plane (center), and the upper-$z$ ghost plane (right), after applying the twist-shift BC (equations~\ref{eq:bc3xlo}-\ref{eq:bc3xup}) to the distribution function with the $y$-shift $\ysh(x)=-0.3x+1.4$.}
  \label{fig:static3x2v}
\end{figure}

We test this property by initializing a 5D field with a Maxwellian dependence in $\vpar$-$\mu$ space:
\begin{equation} \label{eq:static3x2vDistF}
    f(x,y,z,\vpar,\mu) = \frac{n(x,y)}{\left(2\pi v_t^2\right)^{3/2}}\exp\left[-\frac{\left(\vpar-\upar\right)^2+2\mu B/m}{2v_t^2}\right]
\end{equation}
with number density $n(x,y) = \left[2+\cos\left(2\pi y\right)\right]\exp\left[-\left(x-\mu_x\right)^2/(2\sigma_x^2)\right]$ where $\sigma_x=0.5$, $\mu_x=0$, $B=v_t=m=1$ and $\upar=1.2$. The position space domain consists of $\v{R}\in\left[-2,2\right]\times\left[-1.5,1.5\right]\times\left[-3,3\right]$ while velocity space is $\left(\vpar,\mu\right)\in\left[-5v_t,5v_t\right]\times\left[0,25mv_t^2/(2B)\right]$. We discretize the distribution function using $\Nx\times\Nx/2\times4\times\Nvpar\times\Nmu$ cells with a $p=1$ Serendipity basis, and apply the twist-shift BCs by populating the $z$ ghost cells according to equations~\ref{eq:bc3xlo}-\ref{eq:bc3xup} with $\ysh(x)=-0.3x+1.4$. An example of this operation with $(\Nx,\Nvpar,\Nmu)=(40,16,12)$ is shown in figure~\ref{fig:static3x2v}, which shows the number density $M_0$ at $z=0$ along side the $M_0$ in the lower and upper $z$ ghost planes, $(2\pi/m)\int Bf_{z_{\mathrm{ghost}}^\mp}\,\dvpar\dmu$. The shift is applied to $f(\v{R},\vpar,\mu)$, but we see that the velocity moments are shifted as expected as well; at the upper boundary they are shifted by $\ysh(x)$ and at the lower boundary they are shifted by $-\ysh(x)$.

\begin{figure}
  \centering
  \includegraphics[width=0.6\textwidth]{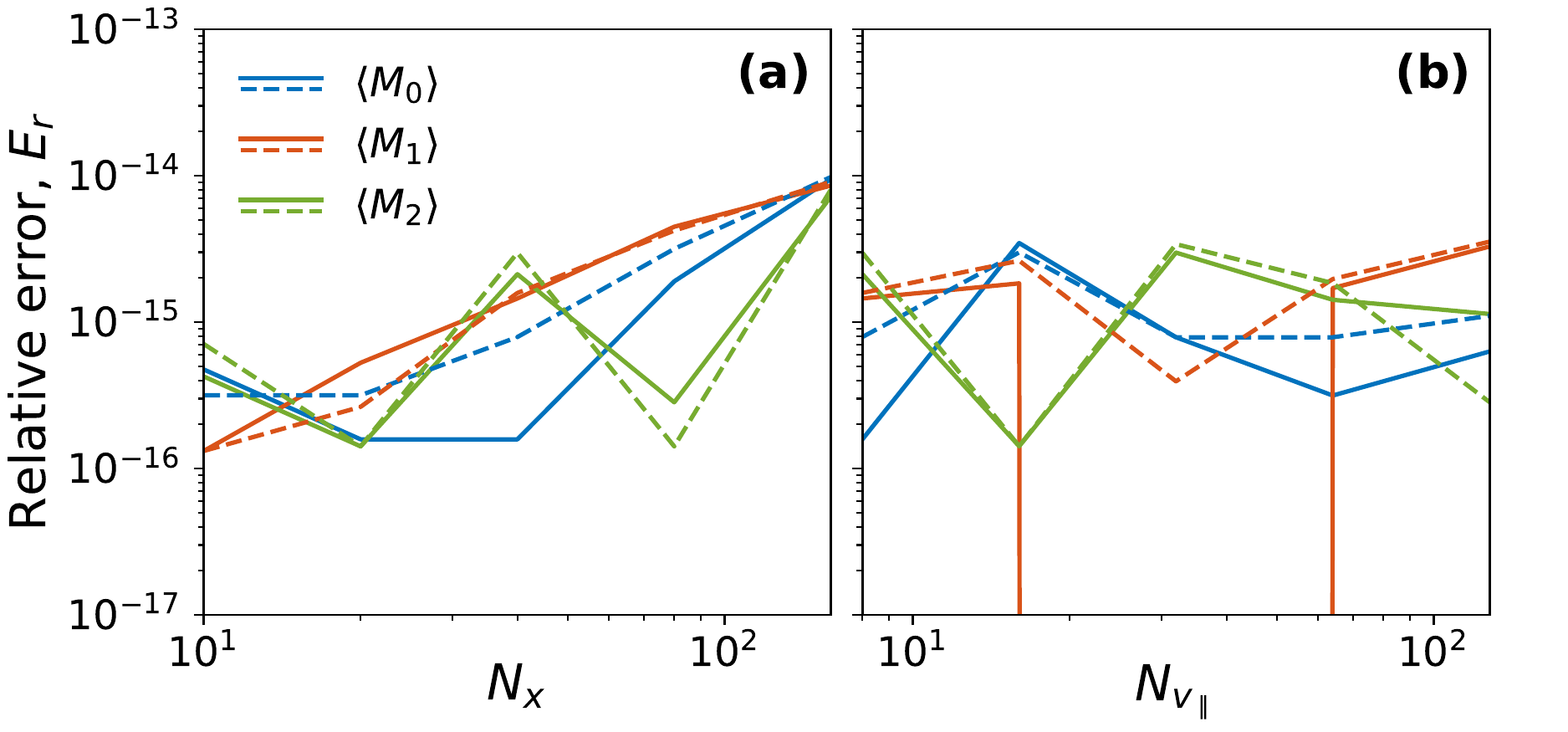}
  \caption{Relative error (equation~\ref{eq:relativeIntMerr}) in the volume integrated velocity moments $M_0$ (blue), $M_1$ (orange) and $M_2$ (green) in the lower-$z$ (solid lines) and upper-$z$ (dashed lines) cells after applying the twist-shift BC to the distribution function in equation~\ref{eq:static3x2vDistF} as we vary the $x-y$ resolution (a) or the $\vpar-\mu$ resolution (b).}
  \label{fig:static3x2vError}
\end{figure}

We confirmed that the volume integrated velocity moments are conserved to machine precision by the twist-shift interpolation algorithm in both static and time-dependent tests. For example, we carried out the same operation as that used to produce figure~\ref{fig:static3x2v} followed by an integral of the velocity moments in the ghost cells, and computed the relative error using the velocity moments of integrated over the corresponding skin cells. That is, we compute the relative $M_0$ error, and similarly for $M_{1,2}$, in the lower-$z$ ghost plane as
\begin{equation} \label{eq:relativeIntMerr}
    E_r = \abs{\avg{M_0}_{z_{\mathrm{ghost}}^-}-\avg{M_0}_{z_{\mathrm{skin}}^+}}/\avg{M_0}_{z_{\mathrm{skin}}^+},
\end{equation}
where $\avg{\cdot}_{z_{\mathrm{ghost}}^-}$ is the volume integral of in the $-\Lz/2-\Dz\leq z \leq -\Lz/2$ range, and $\avg{\cdot}_{z_{\mathrm{skin}}^+}$ is the volume integral of in $\Lz/2-\Dz\leq z \leq\Lz/2$. The relative errors in the integrated moments are shown in figure~\ref{fig:static3x2vError} for $N_x\in\{10,20,40,80,160\}$ keeping $(\Nvpar,\Nmu)=(8,6)$ fixed, and for $(\Nvpar,\Nmu)=(8c,6c)$, $c\in\{1,2,4,8,16\}$ keeping $N_x=40$ fixed (figure~\ref{fig:static3x2vError}(b)). We see that in all cases the relative error in the integrated moments is $\mathcal{O}(10^{-14})$ or smaller.

\begin{figure}
  \centering
  \begin{subfigure}[b]{0.49\textwidth}
    \centering
    \includegraphics[width=\textwidth]{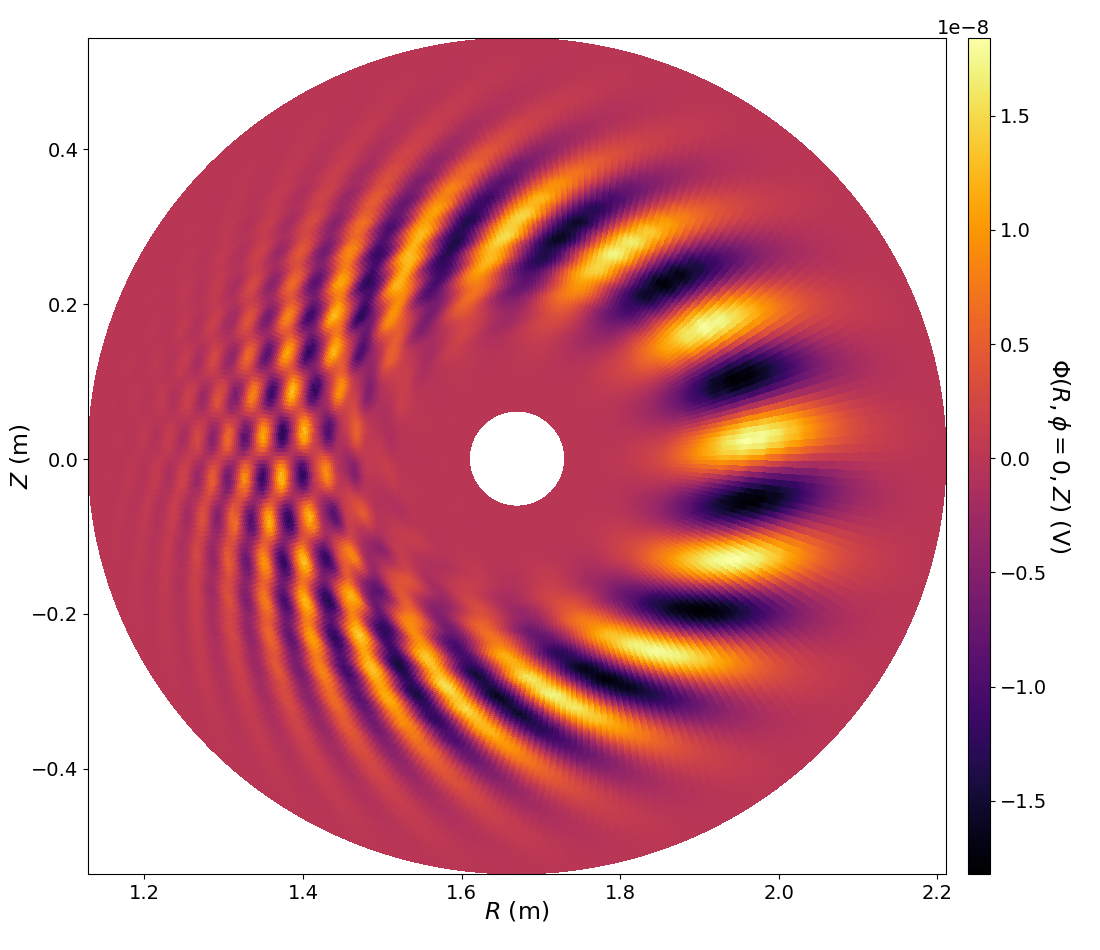}
  \end{subfigure}
  \begin{subfigure}[b]{0.49\textwidth}
    \centering
    \raisebox{4em}{\includegraphics[width=\textwidth]{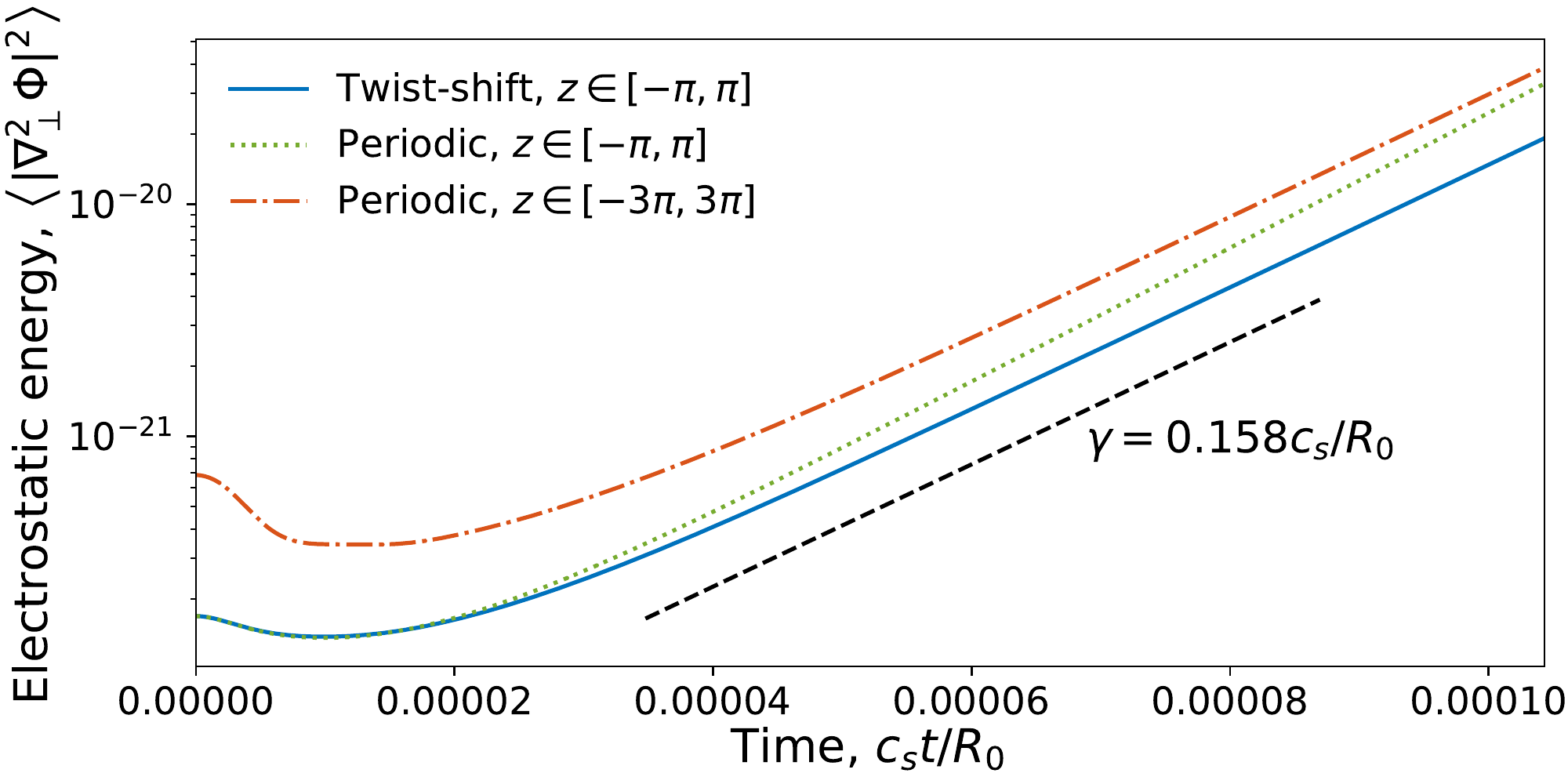}}
  \end{subfigure}
  \caption{Left: electrostatic potential ($\Phi(R,\phi=0,Z)$) at $c_st/R_0=2$ in global, electrostatic delta-$f$ gyrokinetic simulation with twist-shift BCs. Right: time trace of the electrostatic energy for simulations with twist-shift BCs (solid blue), and periodic BCs with $\Lz=2\pi$ (dotted green) and $\Lz=6\pi$ (orange dash-dot). The dashed black line is a reference exponential growing at the rate $\gamma=0.158c_s/R_0$.}
  \label{fig:itg}
\end{figure}

In addition to static twist-and-shifts we could also perform a time-dependent passive advection test in 5D as in the previous section. The results would be essentially identical to those presented in section~\ref{sec:results3x}, and for that reason we opt for a more complex experiment instead. We simulate the exponential growth of an ion-temperature-gradient (ITG) driven instability in a tokamak with circular flux-surfaces~\cite{Beer1996}. ITG modes tap the free energy stored in the temperature gradient to grow perturbations that are elongated in the radial direction at the outboard midplane ($k_x\approx0$) and acquire finite $k_x$ as one traverses the poloidal angle ($\theta$) because they are elongated along a helical, sheared magnetic field (see figure~\ref{fig:itg}(left)). We examine this initial growth phase with a version of the gyrokinetic solver in the \gkeyll~framework~\cite{Mandell2021} that solves the electrostatic, linear delta-$f$ gyrokinetic equations in the long-wavelength limit. This model consists of the following equation for the perturbed gyrocenter distribution function, $\delta f_s(\v{R},\vpar,\mu)$,
\begin{equation} \label{eq:deltafLinearGK}
    \pd{\delta f_s}{t} + \{H_{0s},\delta f_s\} + \{H_{1s},f_{0s}\} = 0,
\end{equation}
where the gyrokinetic Poisson bracket is defined by
\begin{equation}
    \{F,G\} = \frac{\v{B}^*}{m_s B_\parallel^*} \cdot \left(\nabla F \frac{\partial G}{\partial v_\parallel} - \frac{\partial F}{\partial v_\parallel}\nabla G\right) - \frac{ \uv{b}}{q_s B_\parallel^*}\times \nabla F \cdot \nabla G, \label{gkpb}
\end{equation}
with $\v{B}^*=\v{B} + (m_sv_\parallel/q_s)\nabla\times\uv{b}$, $\uv{b}=\v{B}/B$, and $B_\parallel^*=\uv{b}\cdot\v{B}^*\approx B$. The zeroth and first order Hamiltonians are, respectively,
\begin{gather}
    H_{0s} = \frac{1}{2}m_s v_\parallel^2 + \mu B, \\
    H_{1s} = q_s\Phi.
\end{gather}
In these equations $\mu$ is the adiabatic moment, $\vpar$ the particle velocity along the magnetic field, and $q_s$ and $m_s$ are the charge and mass of species $s$. 
The electrostatic potential $\Phi$ is obtained from the long-wavelength gyrokinetic Poisson equation
\begin{equation} \label{eq:poisson}
    -\nabla \cdot (\epsilon_\perp \nabla_\perp \Phi) = \sum_s q_s \int \delta f_s\, \mathrm{d}^3 v
\end{equation}
with $\epsilon_\perp=\sum_sm_sn_{0s}/B^2$.
Additional details may be found in~\cite{Mandell2021}. 

Equations~\ref{eq:deltafLinearGK}-\ref{eq:poisson} are solved in a radially-wide flux-tube as is done in standard benchmarks for global gyrokinetic codes using Cyclone parameters~\cite{Gorler2016}. This test is carried out assuming the electrons are adiabatic, meaning that we only evolve the perturbed ion distribution function $\delta f_i$ and assume $n_e=n_{0e}\left(1+e\phi/T_{e0}\right)$ with a quasineutral background ($n_{0i}=n_{0e}$). The background ion distribution function $f_{0i}$ is taken to be a Maxwellian with density and temperature profiles given by
\begin{gather}
    A(r) = A_\mathrm{ref} \exp\left[-\kappa_A w_A \frac{a}{R_0}\tanh\left(\frac{r-r_0}{w_A a}\right) \right]
\end{gather}
for $A = (n,T)$. Following \cite{Gorler2016}, we take $\kappa_n = 2.23$, $\kappa_T = 6.96$, and $w_{n}=w_{T}=0.3$. The remaining physical parameters are also taken to be the same as in \cite{Gorler2016}.
The domain spans 80\% of the minor radius ($\Lx=0.8a$), has a $z$-extent of $\Lz=2\pi$, is narrow in $y$ ($\Ly=2\pi r_0/(n_0q_0)$), and uses $\ysh(x)=\Lz(r_0/q_0)q(x)$. Here we limit ourselves to studying the toroidal mode number $n_0=10$, since its growth rate is reported in previous benchmarks~\cite{Gorler2016} and we can only accurately model low mode numbers due to the long-wavelength assumption. Therefore the 5D ion phase space consists of $\Lx\times\Ly\times\Lz\times[-3v_{t0i},3v_{t0i}]\times[0,9T_{i0}/B_0]$ given in terms of the reference ion thermal speed $v_{t0i}=\sqrt{T_{i0}/m_i}$ and magnetic field $B_0$. The domain is discretized using $96\times16\times16\times16\times8$ cells and a piecewise linear basis ($p=1$).

These simulations are initialized with a number density containing a sinusoidal (in $y$) perturbation of $\mathcal{O}(10^{-10}\rho_s/a)$, where $\rho_s=c_s/\Omega_i=\sqrt{m_iT_{e0}}/{eB_0}$, $c_s$ is the ion acoustic speed and $\Omega_i$ is the ion cycloctron frequency. As time progresses one watches these perturbations grow and twist with the sheared magnetic field, as shown in figure~\ref{fig:itg}(left). As described in sections~\ref{sec:intro}-\ref{sec:coords}, the the ends of the domain do not necessarily correspond to the same physical location, calling for the use of twist-shift BCs. When these BCs are correctly implemented one can recover the precise growth rate of this mode, $\gamma=0.158c_s/R_0$, which has been computed by multiple other gyrokinetic codes~\cite{Gorler2016}. The solid blue line in figure~\ref{fig:itg}(right), giving the time trace of the electrostatic field energy, confirms that our algorithm and code to solve the linearized delta-$f$ gyrokinetic model with twist-shift BCs is capable of reproducing such evolution. Were we to have simply used regular periodicity along $z$ the exponential growth of this mode would have occurred an an erroneous rate (dotted green line in figure~\ref{fig:itg}(right)). However one can perform an additional check by running the same simulation using periodic BCs in a much longer box (e.g. $\Lz=6\pi$), in which case the mode is insensitive to the details of the boundary conditions and once again exhibits the correct growth rate (orange dash-dot line in figure~\ref{fig:itg}(right)).

\section{Lessons for other applications} \label{sec:beyond}

\begin{figure}
  \centering
  \begin{subfigure}[b]{0.54\textwidth}
    \centering
    \includegraphics[width=\textwidth]{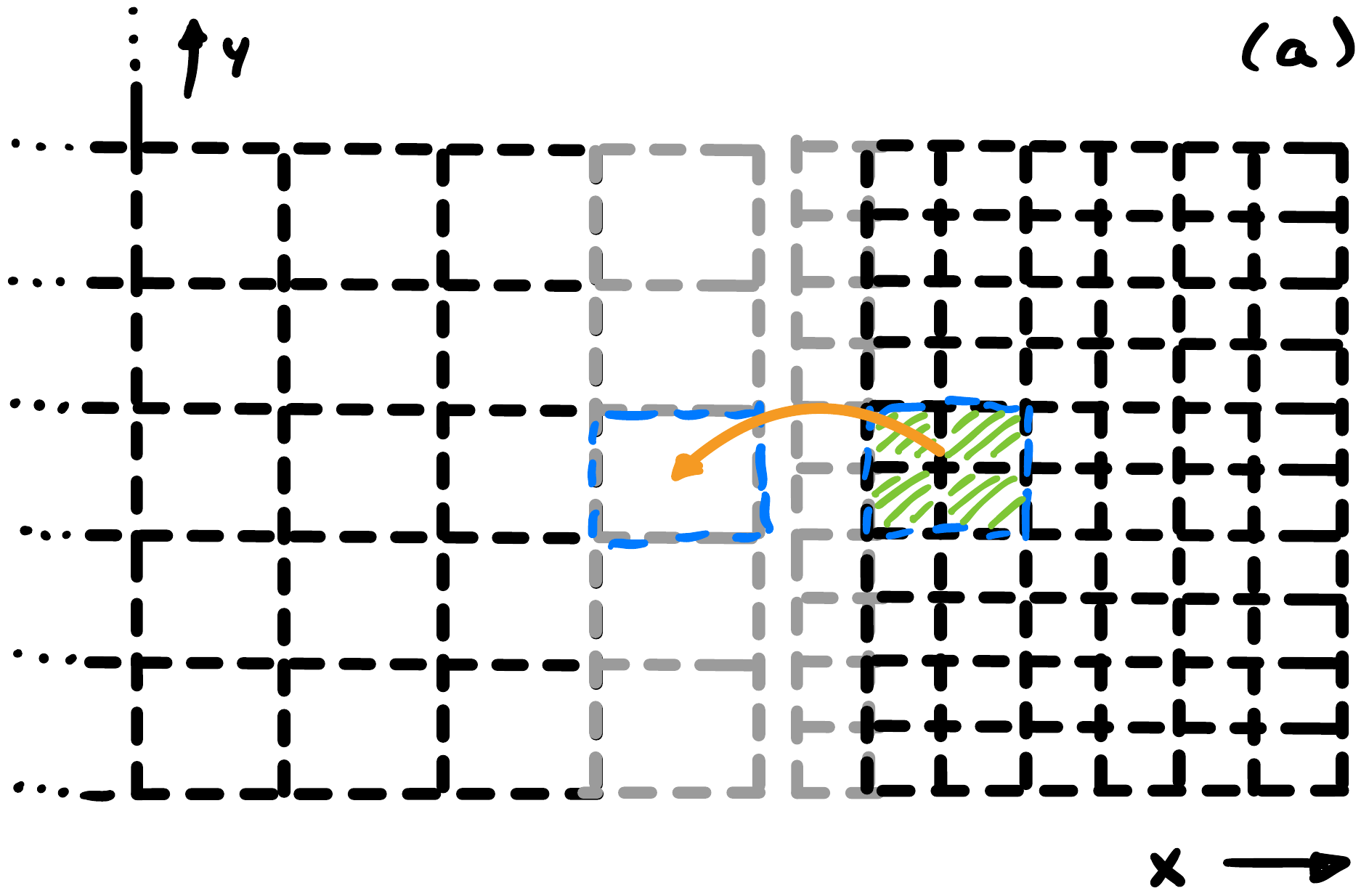}
  \end{subfigure}
  \begin{subfigure}[b]{0.44\textwidth}
    \centering
    \includegraphics[width=\textwidth]{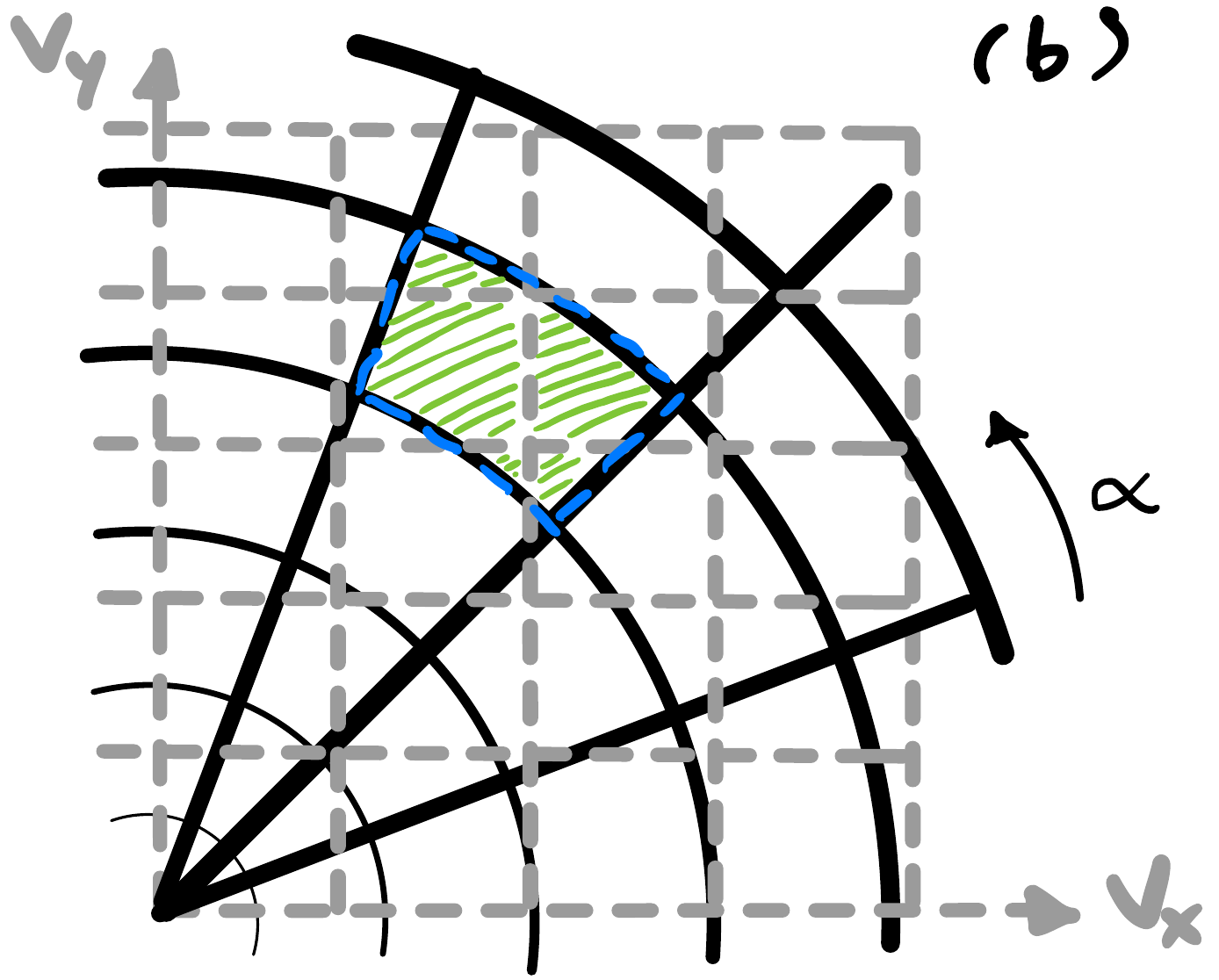}
  \end{subfigure}
  \caption{(a) Two adjacent 2D meshes (black), each with a layer of ghost cells (grey). The right mesh has twice as many cells along $x$ and $y$. (b) A Cartesian velocity-space mesh (grey) overlayed on a cyclindrical velocity-space mesh (black).}
  \label{fig:beyond1}
\end{figure}

The algorithm presented in this work is based on the very general concept of Galerkin projection. This starting point, along with the ideas presented in previous sections, may provide the basis for other operations arising in the solution of PDEs with DG methods. For example, interpolations between non-conforming adjacent grids like those arising when multiblock methods are used to refine parts of the simulation domain can be formulated in terms of a Galerkin projection in order to, for example, populate the ghost cells of the coarse mesh with integrals over the neighboring cells in the fine mesh as sketched in figure~\ref{fig:beyond1}(a). A similar procedure takes place during prolongation and coarsening of a field in a multigrid solver. Both of these ideas have been tested in \gkeyll.

It may also be possible that Galerkin projections lay the foundation for other coordinate transformations of interest. One of those is, for example, the translation between Cartesian $(v_x,v_y)$ and cylindrical $(v,\alpha)$ coordinates in velocity-space of kinetic simulations. Ignoring what happens near the boundaries for now, if we wanted to compute the field in the cell that is circumscribed by the dashed blue line in figure~\ref{fig:beyond1}(b), we would have to compute an integral of the field defined on the Cartesian grid over the striped green region. More precisely, this coordinate transformation could be formulated as
\begin{equation}
    \int_{v_{i-1/2}}^{v_{i+1/2}}\int_{\alpha_{j-1/2}}^{\alpha_{j+1/2}}\mathrm{d}v\,\mathrm{d}\alpha\,v\,\psi_{\itar,\jtar,k}(v,\alpha)\ftar(v,\alpha) = \int_{v_{i-1/2}}^{v_{i+1/2}}\int_{\alpha_{j-1/2}}^{\alpha_{j+1/2}}\mathrm{d}v\,\mathrm{d}\alpha\,v\,\psi_{\itar,\jtar,k}(v,\alpha)\fdo(v_x,v_y).
\end{equation}
The left side of this equation would simplify due to the orthonormality and compact support of the basis functions. But on the right side it may be possible to use a coordinate transformation (e.g. $v^2=v_x^2+v_y^2$, $\tan\alpha=v_y/v_x$) in order to perform the integral in $v_x$-$v_y$ space, leveraging the ability to approximate complex sub-cell integrals described in this manuscript.

\begin{figure}
  \centering
  \includegraphics[width=0.8\textwidth]{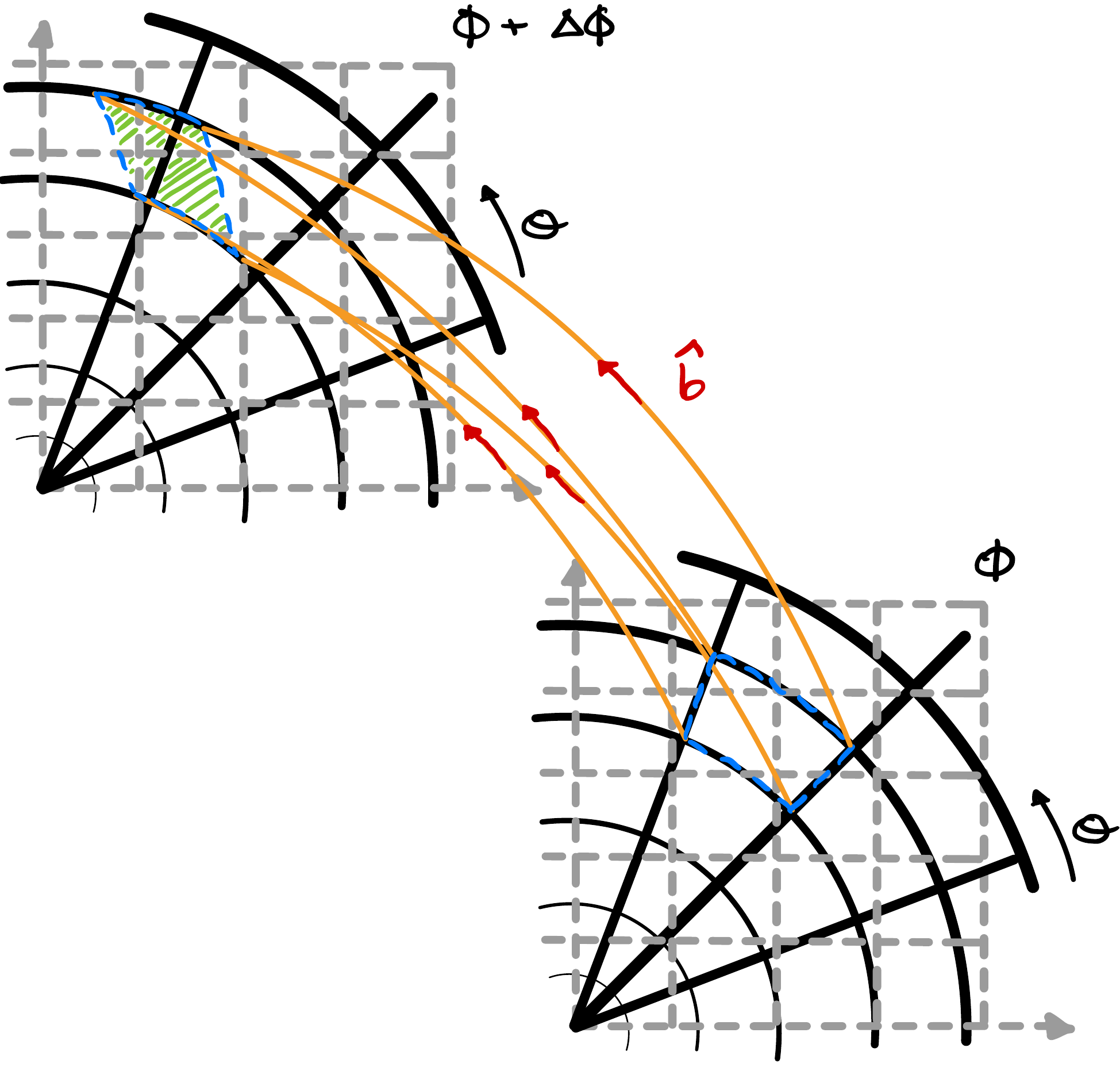}
  \caption{A portion of two consecutive poloidal planes at toroidal angles $\phi$ and $\phi+\Delta\phi$. Orange lines trace the magnetic field through the corners of a cell in the $\phi$-plane, and form a different quadrilateral when crossing the $(\phi+\Delta\phi)$-plane, filled with green lines.}
  \label{fig:beyond2}
\end{figure}

Lastly, the fact that these interpolations were performed while exactly respecting the conservation properties of the physical models hints at the possibility of developing conservative FCI approaches~\cite{Stegmeir2018,Hariri2013} for DG~\cite{Held2016,Dingfelder2020} or even FV~\cite{Dorf2021} simulations of laboratory plasmas. Take for example a grid that is aligned with toroidal coordinates $(r,\theta,\phi)$, a portion of which is depicted in figure~\ref{fig:beyond2} for two consecutive toroidal planes, $\phi$ and $\phi+\Delta\phi$. As one traces magnetic field lines from one cell on the $\phi$-plane to the $(\phi+\Delta\phi)$-plane, the intersection of the magnetic field lines with the latter plane do not trace a quadrilateral that aligns with the $(r,\theta)$ grid (dashed blue contour in the $(\phi+\Delta\phi)$-plane of figure~\ref{fig:beyond2}). Consider a continuity equation with advection in the parallel direction as an example. Its weak form in cell $(i,j,k)$ of the grid stems from:
\begin{equation} \label{eq:continuity}
    \int\mathrm{d}^3\v{x}\,\psi\pd{n}{t}+\int\dx\,\dy\, n\upar\Big|^{z_{k+1/2}}_{z_{k-1/2}} - \int\mathrm{d}^3\v{x}\,n\upar\bhat\cdot\grad{\psi}=0
\end{equation}
where $\bhat=\v{B}/B$, $\upar=\bhat\cdot\v{u}$ and $(x,y)=(r,\theta)$ and $z$ is locally field aligned. The finite difference version of this FCI approach was utilized in the \gdb~code~\cite{Zhu2018}, for example. Equation~\ref{eq:continuity} suggests that an integral over the upper and lower $z$-boundaries of a cell would be required, one of which, as illustrated in figure~\ref{fig:beyond2}, would not be aligned with the grid. In this case we may be able to recourse to the strategies presented here for performing integrals over multiple non-rectangular sub-cell regions and construct a conservative algorithm.

\section{Conclusion} \label{sec:conclusion} 

In this work we presented an algorithm for performing interpolations between Cartesian and curvilinear grids when using a discontinuous Galerkin discretization. The algorithm originates from a Galerkin projection of the solution on each grid, identification and construction of sub-cell integrals, and polynomial approximations to sub-cell integral boundaries. These ideas were formulated in the context of sheared boundary conditions (BCs), which arise in the simulation of plasma turbulence in fusion devices and accretion disks.

Our results show that the algorithm produces results that match our qualitative expectations when applied to 2D, 3D and 5D fields (the latter is relevant to the gyrokinetic simulation of magnetized plasmas). Two-dimensional tests shifting a donor function forward to obtain the target field, and shifting the target field back (i.e. compute $\ftar=\fdo(x,y-\ysh(x))$, followed by $\ftar(x,y+\ysh(x))$), indicate that a certain amount of diffusion 
is introduced by the operator. For higher-order discretizations, however, the effective diffusivity decreases rapidly with resolution. We were also able to quantify the accuracy of the operation by calculating the difference between $\fdo$ and $\ftar(x,y+\ysh(x))$, indicating that the algorithm is second-order accurate for piecewise constant basis functions ($p=0$) and $(p+1)$-order accurate in the DG representation and $(p+2)$-order accurate in the cell averages for $p\geq1$. It is still possible to improve the accuracy and obtain $(p+2)$-order accuracy in the DG representation by using cell average values from enough neighboring cells in order to interpolate a higher order solution.

Our tests in 2D and 3D demonstrate that despite the finite diffusion, we are able to conserve the total volume integral of the shifted function. In 5D this translates to conserving the number of particles, momentum and energy, which is desirable in codes aiming to use coarse resolutions or simulate extremely long time periods. Our 3D passive advection test also made it evident that as structures get more and more sheared by the twist-and-shift BCs, they alias to lower mode-numbers and introduce unphysical oscillations. The diffusion inherent to upwinded numerical fluxes perpendicular to the shift is able to mitigate this effect, but in the future we would like to pursue an alias-free algorithm that does not rely on the direction of the flows in a manner analogous to spectral formulations~\cite{Beer1995}.

Lastly, by combining these twist-shift BCs with \gkeyll's gyrokinetic solver we are able to accurately reproduce the linear growth phase of an electrostatic ITG mode in the Cyclone benchmark that is commonly used by other gyrokinetic codes~\cite{Gorler2016}. This proof of principle signals the possibility of carrying out core, gyrokinetic simulations with \gkeyll~in the near future, especially once gyroaveraging is implemented. These new capabilities will allow benchmarking \gkeyll~against other gyrokinetic codes and potentially pave the way for simulations spanning both the core and the edge of fusion devices.

\appendix

\section{Getting \gkeyll~and reproducing results} \label{sec:getGkeyll}

Readers may reproduce our results and also use \gkeyll~for their applications. The code and input files used here are available online. Full installation instructions for \gkeyll~are provided on the \gkeyll~website~\cite{gkeyllWeb}. The code can be installed on Unix-like operating systems (including Mac OS and Windows using the Windows Subsystem for Linux) either by installing the pre-built binaries using the conda package manager (\url{https://www.anaconda.com}) or building the code via sources. The input files used here are under version control and can be obtained from the repository at \url{https://github.com/ammarhakim/gkyl-paper-inp/tree/master/2021_JCP_TwistShift}.

\section{Additional details on classifying and computing sub-cell integrals} \label{sec:addDetails}

The in-depth details of how sub-cell integrals are classified and how more complex integrals are computed are not necessary for a conceptual understanding of the algorithm presented in this work. For that reason we omitted further explanations on these topics from the main text. Nevertheless, we provide them in this appendix for completeness.

\subsection{Sub-cell integral scenario classification criteria} \label{sec:scenarioClassify}

After looking for the intersection points $O=\{O_{--},O_{-+},O_{+-},O_{++}\}$ corresponding to the intersection of the curves $y_{\jtar}\mp\ysh(x)$ and the lines $y_{\jdo}\mp\Dy/2$ we identify which sub-cell integral scenario is required by checking which of the $O$ points was found, their location relative to each other and whether $\ysh(x)$ is increasing or decreasing. For compactness we symbolize a monotonically increasing $\ysh(x)$ with $\ysh\uparrow$, and a monotonically decreasing $\ysh(x)$ with $\ysh\downarrow$. Then we identify the sub-cell integral scenario with the following criteria:
\begin{itemize}
    \item All $O$ points found $\Rightarrow$ scenario $sNi$ or $sNii$.
    \begin{itemize}
        \item $O_{--}>O_{+-}$ means $\ysh\downarrow~\Rightarrow$ scenario $sNi$ (figure~\ref{fig:findOpoints}).
        \item $O_{--}<O_{+-}$ means $\ysh\uparrow~\Rightarrow$ scenario $sNii$.
    \end{itemize}
    \item 3 $O$ points missing $\Rightarrow$ scenarios $si-siv$.
    \begin{itemize}
        \item $O_{--}$ or $O_{-+}$ are found:
        \begin{itemize}
            \item $-\ysh(x_{i+1/2}) \geq -\ysh(O_{-+})$ means $\ysh\downarrow~\Rightarrow$ scenario $si$. 
            \item $-\ysh(x_{i+1/2}) < -\ysh(O_{-+})$ means $\ysh\uparrow~\Rightarrow$ scenario $sii$. 
        \end{itemize}
        \item Neither $O_{--}$ nor $O_{-+}$ are found:
        \begin{itemize}
            \item $-\ysh(x_{i+1/2}) \leq -\ysh(O_{+-})$ means $\ysh\uparrow~\Rightarrow$ scenario $siii$.
            \item $-\ysh(x_{i+1/2}) > -\ysh(O_{+-})$ means $\ysh\downarrow~\Rightarrow$ scenario $siv$.
        \end{itemize}
    \end{itemize}
    \item 1 $O$ point missing $\Rightarrow$ scenarios $sv-sviii$.
    \begin{itemize}
        \item $O_{+-}$ is missing:
        \begin{itemize}
            \item $O_{-+}>O_{++}~\Rightarrow$ scenario $sv$.
            \item $O_{-+}<=O_{++}~\Rightarrow$ scenario $svi$.
        \end{itemize}
        \item $O_{-+}$ is missing:
        \begin{itemize}
            \item $O_{--}<O_{+-}~\Rightarrow$ scenario $svii$.
            \item $O_{--}>=O_{+-}~\Rightarrow$ scenario $sviii$.
        \end{itemize}
    \end{itemize}
    \item 2 $O$ points missing $\Rightarrow$ scenarios $six-sxiv$.
    \begin{itemize}
        \item $\{O_{+-},O_{++}\}$ or $\{O_{--},O_{-+}\}$ are missing:
        \begin{itemize}
            \item $\{O_{+-},O_{++}\}$ are missing:
            \begin{itemize}
                \item $O_{--}<O_{-+}~\Rightarrow$ scenario $six$.
                \item $O_{--}>=O_{-+}~\Rightarrow$ scenario $sx$.
            \end{itemize}
            \item $\{O_{--},O_{-+}\}$ are missing:
            \begin{itemize}
                \item $O_{++}<O_{+-}~\Rightarrow$ scenario $sxi$.
                \item $O_{++}>=O_{+-}~\Rightarrow$ scenario $sxii$.
            \end{itemize}
        \end{itemize}
        \item $\{O_{-+},O_{+-}\}$ are missing:
        \begin{itemize}
            \item $-\ysh(x_{i-1/2})<-\ysh(x_{i+1/2})$ means $\ysh\downarrow~\Rightarrow$ scenario $sxiii$.
            \item $-\ysh(x_{i-1/2})\geq-\ysh(x_{i+1/2})$ means $\ysh\uparrow~\Rightarrow$ scenario $sxiv$.
        \end{itemize}
    \end{itemize}
    \item All $O$ points missing $\Rightarrow$ scenarios $sxv-sxvi$.
    \begin{itemize}
        \item $y_{\jdo-1/2}\leq y'_{x_c}$ and $y'_{x_c}\leq y_{\jdo+1/2}~\Rightarrow$ scenario $sxv$.
        \item $y_{\jdo-1/2}> y'_{x_c}$ or $y'_{x_c}> y_{\jdo+1/2}~\Rightarrow$ scenario $sxvi$,
    \end{itemize}
\end{itemize}
where $y'_{x_c}=\left(y_{\jtar-1/2}-\ysh(x_i)\right)\mathrm{mod}\Ly$.

\subsection{More complex sub-cell integrals}

In sections~\ref{sec:subcellVarY}-\ref{sec:subcellVarX} we described how two simple sub-cell integrals with variable $y$ or $x$ limits are performed. Those sections focused on scenarios $six$-$sxii$ and $sxv$-$sxvi$ which are some of the simplest because they involved a single sub-cell integral with one variable limit over the whole $x$ or $y$ extent of the cell, respectively. There are other more complex sub-cell integrals which we expand on below.

\subsubsection{Scenarios $sNi$-$sNii$ and $si$-$siv$} \label{sec:addSiSiv}

The integrals in scenarios $sNi$-$sNii$, $si$-$siv$, and $six$-$sxii$ can be constructed with contributions from integrals with variable $x$-limits. For example, we can write the contribution of a $sNi$ integral as
\begin{equation} \label{eq:subIntA}
f_{\mathrm{tar},\jtar,k}^{sNi} = \int_{\etalo}^{\etaup}\int_{\xiloh(\eta)}^{\xiuph(\eta)}\psi_{\jdo,k}(\xi,\eta+\frac{\yshh(\xi)+y_{\jdo}-y_{\jtar}}{\Dy/2})\sum_{k'}f_{\mathrm{do},\jdo,k'}\psi_{\jdo,k'}(\xi,\eta)\,\dxi\,\deta,
\end{equation}
where $\etalo,\,\etaup$ are fixed values and $\xiuph(\eta),\,\xiloh(\eta)$ are discrete approximations to the limits of the $\xi$ integral as described in section~\ref{sec:subcellVarX}. In the case of $sNi$-$sNii$ the $y$-integral spans the whole cell, so $\etaup=-\etalo=1$. On the other hand, scenarios $si$-$siv$ only span a fraction of the $y$-extent of the cell and have $\etalo,\,\etaup$ that are other than $\pm1$. In fact, scenarios $si$-$siv$ can also be formulated in terms of an integral that uses fixed $\xi$-limits but variable $\eta$-limits. It is actually advantageous to do it that way since the $y_{\jtar\pm1/2}-\ysh(x)$ does not have to be inverted. This is indeed what the implementation in \gkeyll~does, but for now we stick to variable $x$-limits for demonstration purposes.

\begin{figure}
    \centering
    \includegraphics[width=0.5\textwidth]{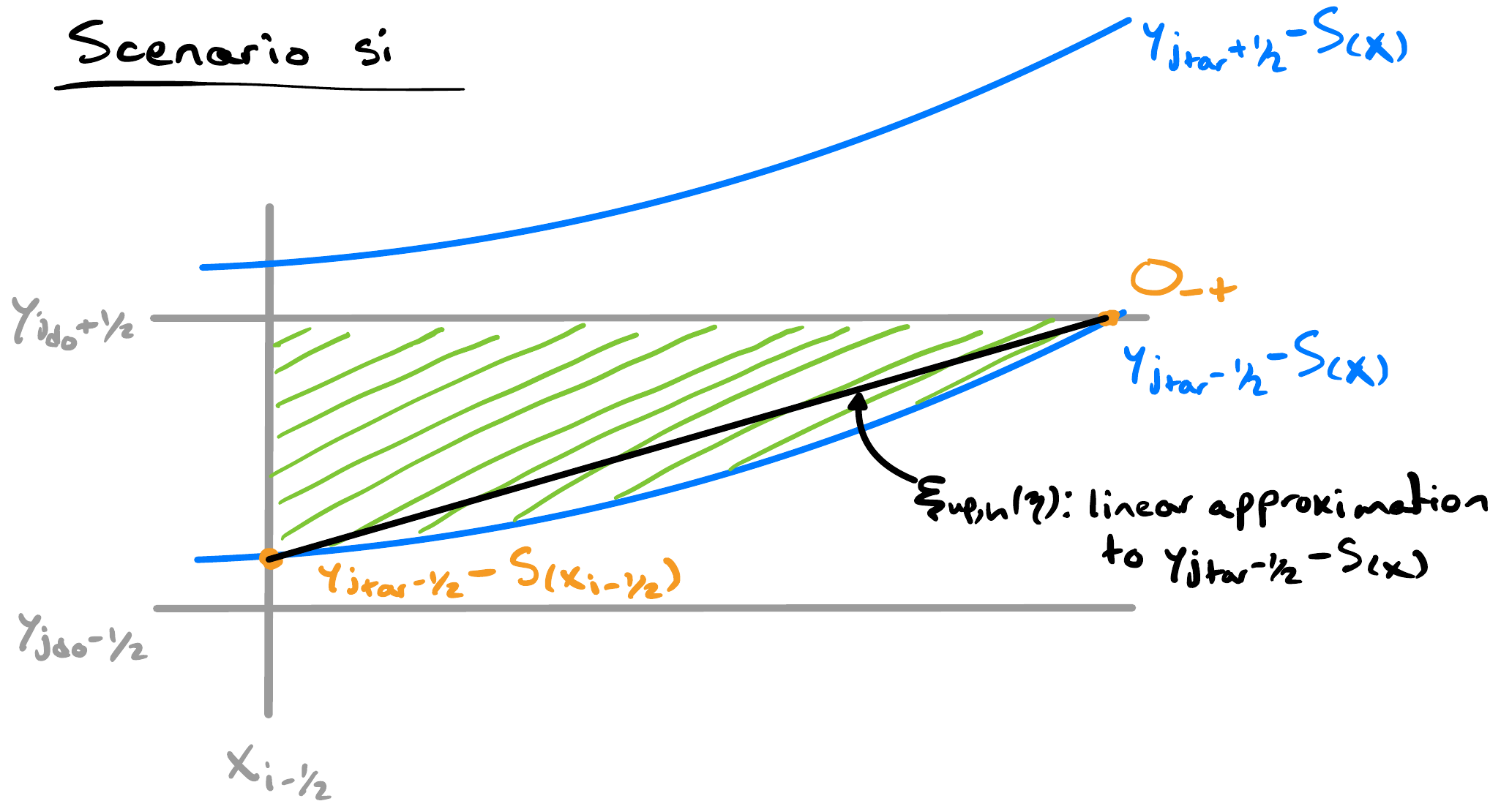}
   \caption{Sketch of scenario $si$. The green striped area is the sub-cell region we need to integrate over. The upper $\xi$ limit of the integral is given by the lower blue curve, but the discrete linear approximation to it is given by the black line.}
   \label{fig:scenarioSi}
\end{figure}

Since the $y$-integral of scenarios $si$-$siv$ does not necessarily extend over the whole cell the discrete approximation to the $\xi$ integral limits (e.g. $\xiuph(\eta)$) are only defined in a fraction of the $\eta\in[-1,1]$ logical space that the donor field is defined on. Take scenario $si$ as an example (figure~\ref{fig:scenarioSi}). The upper $\xi$ limit of the integral is $\xiup(\eta)$, and its projection onto a 1D basis function (in order to obtain $\xiuph(\eta)$ thus takes place over the segment $\eta\in\left[\etalo,1\right]$ where
\begin{equation}
    \etalo = \frac{2}{\Dy}\left[\left(y_{\jtar-1/2}-\ysh(x_{i-1/2})\right)\mathrm{mod}\Ly-y_{\jdo}\right]
\end{equation}
is the lower limit of the $y$-integral translated to the logical coordinates of the donor cell (lower left orange point in figure~\ref{fig:scenarioSi}). To be more precise, the approximation to this upper $\xi$-limit has the form
\begin{equation}
    \xiuph(\eta) = \sum_{k}\xiuphk\varphi_k(u)
\end{equation}
where $u$ is the logical coordinate in the segment $\eta\in\left[\etalo,1\right]$ and can be written as a function of $\eta$ using
\begin{eqnal} \label{eq:etau}
    \eta &= \frac{1}{2}\left(1+\etalo\right)+\frac{1}{2}\left(1-\etalo\right)u = \eta_c^{si} + \frac{\Delta\eta^{si}}{2}u.
\end{eqnal}
The coefficients $\xiuphk$ are then calculated using nodal evaluation followed by a nodal-to-modal transformation. That is, we select nodes in the $\eta\in\left[\etalo,1\right]$ segment, evaluate $\xiup(\eta)$ at those nodes, and then perform a nodal-to-modal transformation to obtain $\xiuphk$. This gives the expansion coefficients multiplying basis functions of $u$, which then have to be re-written in terms of $\eta$ using equation~\ref{eq:etau} before performing the $\eta$ integral in equation~\ref{eq:subIntA}. If one uses a $p=1$ expansion of $\ysh(x)$ the upper limit function $\xiup$ is then approximated by a linear polynomial; compare the lower blue curve and the black line in figure~\ref{fig:scenarioSi}.

\subsubsection{Scenarios $sv$-$sviii$} \label{sec:addSvSviii}

\begin{figure}[h]
    \centering
    \includegraphics[width=0.7\textwidth]{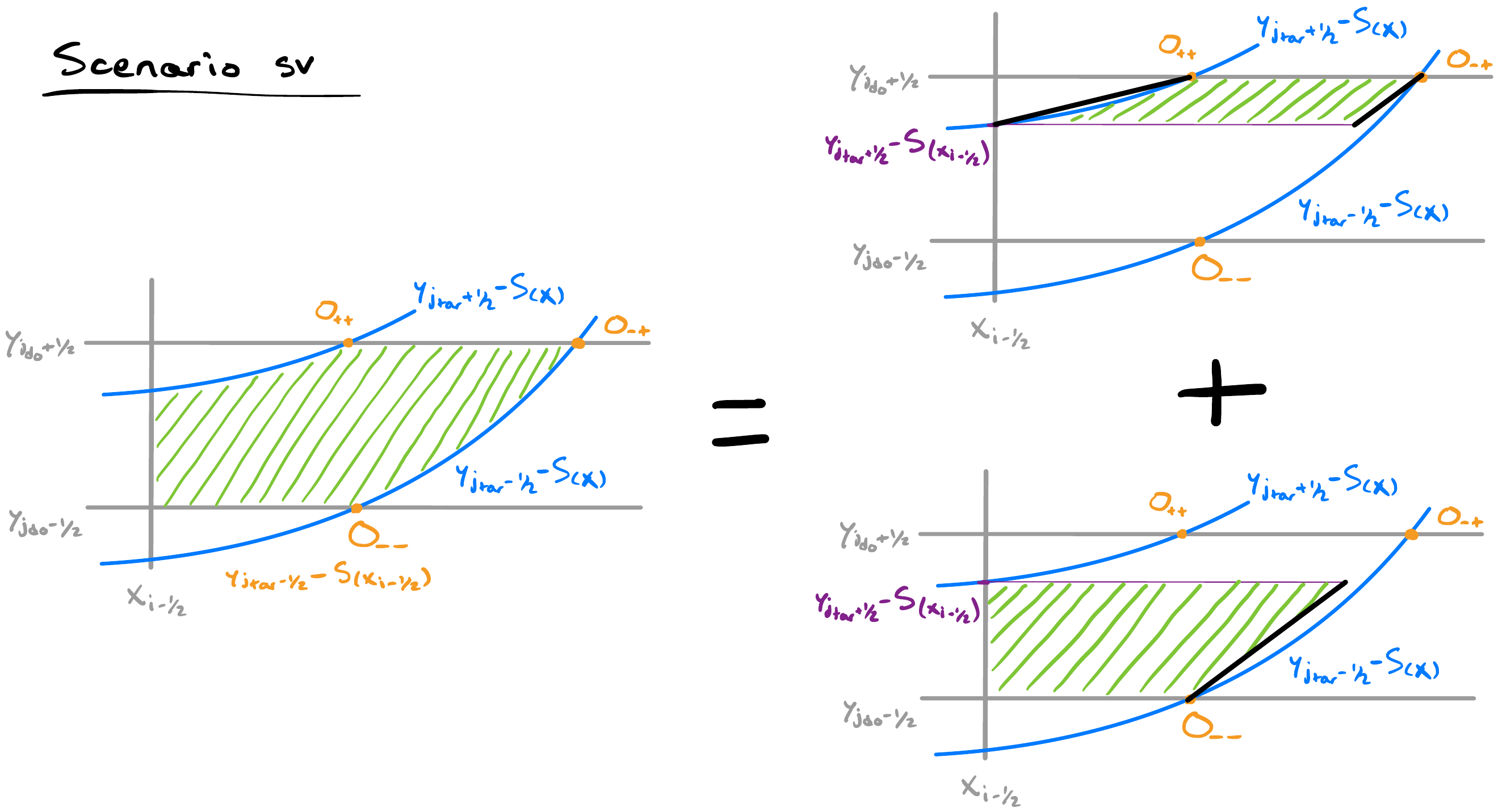}
   \caption{For the integral in scenario $sv$ we add up the contribution from two sub-cell regions. The first (top right) from an integral with fixed lower and upper $\eta$ limits and $\xi$-limits functions of $\eta$, and the second from (bottom right) an integral with fixed upper and lower $\eta$-limits, a fixed lower $\xi$-limit but a variable upper $\xi$-limit.}
   \label{fig:scenarioV}
\end{figure}

The integrals in scenarios $sv$-$sviii$ are composed of two contributions with the form of equation~\ref{eq:subIntA}. Specifically, take scenario $sv$ as an example. We construct this sub-cell integral with an integral similar to that in scenario $sNi$ but with a lower $\eta$-limit greater than $-1$, and an integral similar to that in scenario $siii$ but with a lower $\xi$-limit of $-1$ and a variable upper limit (see figure~\ref{fig:scenarioV}). Mathematically we write this as
\begin{eqnal} \label{eq:subIntV}
f_{\mathrm{tar},\jtar,k}^{sv} &= \int_{\eta_{\mathrm{lo}}^{sN}}^{1}\int_{\xiloh^{sN}(\eta)}^{\xiuph^{sN}(\eta)}\psi_{\jdo,k}(\xi,\eta+\frac{\yshh(\xi)+y_{\jdo}-y_{\jtar}}{\Dy/2})f_{\mathrm{do},\jdo}(\xi,\eta)\,\text{d}\xi\,\text{d}\eta \\
&\quad+ \int_{-1}^{\etaup^{iii}}\int_{-1}^{\xiuph^{iii}(\eta)}\psi_{\jdo,k}(\xi,\eta+\frac{\yshh(\xi)+y_{\jdo}-y_{\jtar}}{\Dy/2})f_{\mathrm{do},\jdo}(\xi,\eta)\,\text{d}\xi\,\text{d}\eta,
\end{eqnal}
where the limits $\etalo^{sN}=\etaup^{iii}$ are just a translation of $y_{\jtar+1/2}-\ysh(x_{i-1/2})$ to the logical space of the donor cell:
\begin{equation}
\etalo^{sN}=\etaup^{iii} = \frac{2}{\Dy}\left(y_{\jtar+1/2}-\ysh(x_{i-1/2}) - y_{\jdo}\right).
\end{equation}
On the other hand the functions defining the $\xi$-limits, $\xilo^{sN}(\eta)$, $\xiup^{sN}(\eta)$ and $\xilo^{iii}(\eta)$, are obtained by inverting and translating to logical space the $y_{\jtar\pm1/2}-\ysh(x)$ functions, as described in section~\ref{sec:subcellVarX}. Then we can project them onto a 1D polynomial basis (in a fraction of the $\eta$-space) to obtain $\xiloh^{sN}(\eta)$, $\xiuph^{sN}(\eta)$ and $\xiloh^{iii}(\eta)$ as described in section~\ref{sec:addSiSiv}. The linear ($p=1$) approximation to the curved integral boundaries are shown with straight black lines in figure~\ref{fig:scenarioV}. Notice in such figure that the discrete approximation to the upper $\xi$-limits do not terminate where the purple line ($y_{\jtar+1/2}-\ysh(x_{i-1/2})$) meets the $y_{\jtar-1/2}-\ysh(x)$ curve; refining this detail could be explored in the future.

\subsubsection{Scenarios $sxiii$-$sxiv$} \label{sec:addSxiiiSxiv}

Scenarios $sxiii-sxiv$ are computed by subtracting from the inner product over the whole cell two $si-siv$-like integrals, e.g.
\begin{equation} \label{eq:subIntB}
f_{\mathrm{tar},\jtar,k}^{sxiii} = \int_{-1}^{1}\int_{-1}^{1}\psi_{\jdo,k}(\xi,\eta+\frac{\yshh(\xi)+y_{\jdo}-y_{\jtar}}{\Dy/2})f_{\mathrm{do},\jdo}\,\dxi\,\deta - f_{\mathrm{tar},\jtar,k}^{si} - f_{\mathrm{tar},\jtar,k}^{siv}.
\end{equation}

\bibliography{main.bib}

\begin{thebibliography}{10}
\expandafter\ifx\csname url\endcsname\relax
  \def\url#1{\texttt{#1}}\fi
\expandafter\ifx\csname urlprefix\endcsname\relax\def\urlprefix{URL }\fi
\expandafter\ifx\csname href\endcsname\relax
  \def\href#1#2{#2} \def\path#1{#1}\fi

\bibitem{Steger1987}
J.~L. Steger, J.~A. Benek,
  \href{https://www.sciencedirect.com/science/article/pii/0045782587900454}{On
  the use of composite grid schemes in computational aerodynamics}, Computer
  Methods in Applied Mechanics and Engineering 64~(1) (1987) 301--320.
\newblock \href {https://doi.org/https://doi.org/10.1016/0045-7825(87)90045-4}
  {\path{doi:https://doi.org/10.1016/0045-7825(87)90045-4}}.
\newline\urlprefix\url{https://www.sciencedirect.com/science/article/pii/0045782587900454}

\bibitem{Sherer2005}
S.~E. Sherer, J.~N. Scott,
  \href{https://www.sciencedirect.com/science/article/pii/S0021999105002366}{High-order
  compact finite-difference methods on general overset grids}, Journal of
  Computational Physics 210~(2) (2005) 459--496.
\newblock \href {https://doi.org/https://doi.org/10.1016/j.jcp.2005.04.017}
  {\path{doi:https://doi.org/10.1016/j.jcp.2005.04.017}}.
\newline\urlprefix\url{https://www.sciencedirect.com/science/article/pii/S0021999105002366}

\bibitem{Barth1993}
T.~Barth, \href{https://arc.aiaa.org/doi/abs/10.2514/6.1993-668}{Recent
  developments in high order K-exact reconstruction on unstructured meshes},
  1993.
\newblock \href
  {http://arxiv.org/abs/https://arc.aiaa.org/doi/pdf/10.2514/6.1993-668}
  {\path{arXiv:https://arc.aiaa.org/doi/pdf/10.2514/6.1993-668}}, \href
  {https://doi.org/10.2514/6.1993-668} {\path{doi:10.2514/6.1993-668}}.
\newline\urlprefix\url{https://arc.aiaa.org/doi/abs/10.2514/6.1993-668}

\bibitem{Nejat2008}
A.~Nejat, C.~Ollivier-Gooch,
  \href{https://www.sciencedirect.com/science/article/pii/S0021999107004834}{A
  high-order accurate unstructured finite volume {Newton–Krylov} algorithm
  for inviscid compressible flows}, Journal of Computational Physics 227~(4)
  (2008) 2582--2609.
\newblock \href {https://doi.org/https://doi.org/10.1016/j.jcp.2007.11.011}
  {\path{doi:https://doi.org/10.1016/j.jcp.2007.11.011}}.
\newline\urlprefix\url{https://www.sciencedirect.com/science/article/pii/S0021999107004834}

\bibitem{Hall2013}
D.~M. Hall, R.~D. Nair,
  \href{https://journals.ametsoc.org/view/journals/mwre/141/1/mwr-d-12-00108.1.xml}{Discontinuous
  {Galerkin} transport on the spherical yin–yang overset mesh}, Monthly
  Weather Review 141~(1) (2013) 264 -- 282.
\newblock \href {https://doi.org/10.1175/MWR-D-12-00108.1}
  {\path{doi:10.1175/MWR-D-12-00108.1}}.
\newline\urlprefix\url{https://journals.ametsoc.org/view/journals/mwre/141/1/mwr-d-12-00108.1.xml}

\bibitem{Marshall2014}
M.~C. Galbraith, P.~D. Orkwis, J.~A. Benek,
  \href{https://arc.aiaa.org/doi/abs/10.2514/6.2014-0776}{A 3-D Discontinuous
  {Galerkin} Chimera Overset Method}, 2014.
\newblock \href
  {http://arxiv.org/abs/https://arc.aiaa.org/doi/pdf/10.2514/6.2014-0776}
  {\path{arXiv:https://arc.aiaa.org/doi/pdf/10.2514/6.2014-0776}}, \href
  {https://doi.org/10.2514/6.2014-0776} {\path{doi:10.2514/6.2014-0776}}.
\newline\urlprefix\url{https://arc.aiaa.org/doi/abs/10.2514/6.2014-0776}

\bibitem{Stegmeir2018}
A.~Stegmeir, D.~Coster, A.~Ross, O.~Maj, K.~Lackner, E.~Poli,
  \href{https://doi.org/10.1088/1361-6587/aaa373}{{GRILLIX}: a 3d turbulence
  code based on the flux-coordinate independent approach}, Plasma Physics and
  Controlled Fusion 60~(3) (2018) 035005.
\newblock \href {https://doi.org/10.1088/1361-6587/aaa373}
  {\path{doi:10.1088/1361-6587/aaa373}}.
\newline\urlprefix\url{https://doi.org/10.1088/1361-6587/aaa373}

\bibitem{Dorf2018}
M.~Dorf, M.~Dorr,
  \href{https://onlinelibrary.wiley.com/doi/abs/10.1002/ctpp.201700137}{Continuum
  kinetic modelling of cross-separatrix plasma transport in a tokamak edge
  including self-consistent electric fields}, Contributions to Plasma Physics
  58~(6-8) (2018) 434--444.
\newblock \href
  {http://arxiv.org/abs/https://onlinelibrary.wiley.com/doi/pdf/10.1002/ctpp.201700137}
  {\path{arXiv:https://onlinelibrary.wiley.com/doi/pdf/10.1002/ctpp.201700137}},
  \href {https://doi.org/https://doi.org/10.1002/ctpp.201700137}
  {\path{doi:https://doi.org/10.1002/ctpp.201700137}}.
\newline\urlprefix\url{https://onlinelibrary.wiley.com/doi/abs/10.1002/ctpp.201700137}

\bibitem{Hariri2013}
F.~Hariri, M.~Ottaviani,
  \href{https://www.sciencedirect.com/science/article/pii/S0010465513001999}{A
  flux-coordinate independent field-aligned approach to plasma turbulence
  simulations}, Computer Physics Communications 184~(11) (2013) 2419--2429.
\newblock \href {https://doi.org/https://doi.org/10.1016/j.cpc.2013.06.005}
  {\path{doi:https://doi.org/10.1016/j.cpc.2013.06.005}}.
\newline\urlprefix\url{https://www.sciencedirect.com/science/article/pii/S0010465513001999}

\bibitem{Hammett1993}
G.~W. Hammett, M.~A. Beer, W.~Dorland, S.~C. Cowley, S.~A. Smith,
  \href{https://doi.org/10.1088/0741-3335/35/8/006}{Developments in the
  gyrofluid approach to tokamak turbulence simulations}, Plasma Physics and
  Controlled Fusion 35~(8) (1993) 973--985.
\newblock \href {https://doi.org/10.1088/0741-3335/35/8/006}
  {\path{doi:10.1088/0741-3335/35/8/006}}.
\newline\urlprefix\url{https://doi.org/10.1088/0741-3335/35/8/006}

\bibitem{Beer1995}
M.~A. Beer, S.~C. Cowley, G.~W. Hammett,
  \href{https://doi.org/10.1063/1.871232}{Field‐aligned coordinates for
  nonlinear simulations of tokamak turbulence}, Physics of Plasmas 2~(7) (1995)
  2687--2700.
\newblock \href {http://arxiv.org/abs/https://doi.org/10.1063/1.871232}
  {\path{arXiv:https://doi.org/10.1063/1.871232}}, \href
  {https://doi.org/10.1063/1.871232} {\path{doi:10.1063/1.871232}}.
\newline\urlprefix\url{https://doi.org/10.1063/1.871232}

\bibitem{Rogers1998}
B.~N. Rogers, J.~F. Drake, A.~Zeiler,
  \href{https://link.aps.org/doi/10.1103/PhysRevLett.81.4396}{Phase space of
  tokamak edge turbulence, the $\mathit{L}\ensuremath{-}\mathit{H}$ transition,
  and the formation of the edge pedestal}, Phys. Rev. Lett. 81 (1998)
  4396--4399.
\newblock \href {https://doi.org/10.1103/PhysRevLett.81.4396}
  {\path{doi:10.1103/PhysRevLett.81.4396}}.
\newline\urlprefix\url{https://link.aps.org/doi/10.1103/PhysRevLett.81.4396}

\bibitem{Beer1996}
M.~A. Beer, G.~W. Hammett, \href{https://doi.org/10.1063/1.871538}{Toroidal
  gyrofluid equations for simulations of tokamak turbulence}, Physics of
  Plasmas 3~(11) (1996) 4046--4064.
\newblock \href {http://arxiv.org/abs/https://doi.org/10.1063/1.871538}
  {\path{arXiv:https://doi.org/10.1063/1.871538}}, \href
  {https://doi.org/10.1063/1.871538} {\path{doi:10.1063/1.871538}}.
\newline\urlprefix\url{https://doi.org/10.1063/1.871538}

\bibitem{Dorland2000}
W.~Dorland, F.~Jenko, M.~Kotschenreuther, B.~N. Rogers,
  \href{https://link.aps.org/doi/10.1103/PhysRevLett.85.5579}{Electron
  temperature gradient turbulence}, Phys. Rev. Lett. 85 (2000) 5579--5582.
\newblock \href {https://doi.org/10.1103/PhysRevLett.85.5579}
  {\path{doi:10.1103/PhysRevLett.85.5579}}.
\newline\urlprefix\url{https://link.aps.org/doi/10.1103/PhysRevLett.85.5579}

\bibitem{Candy2016}
J.~Candy, E.~Belli, R.~Bravenec,
  \href{https://www.sciencedirect.com/science/article/pii/S0021999116303400}{{A
  high-accuracy Eulerian gyrokinetic solver for collisional plasmas}}, Journal
  of Computational Physics 324 (2016) 73--93.
\newblock \href {https://doi.org/https://doi.org/10.1016/j.jcp.2016.07.039}
  {\path{doi:https://doi.org/10.1016/j.jcp.2016.07.039}}.
\newline\urlprefix\url{https://www.sciencedirect.com/science/article/pii/S0021999116303400}

\bibitem{Jenko2000}
F.~Jenko, W.~Dorland, M.~Kotschenreuther, B.~N. Rogers,
  \href{https://doi.org/10.1063/1.874014}{Electron temperature gradient driven
  turbulence}, Physics of Plasmas 7~(5) (2000) 1904--1910.
\newblock \href {http://arxiv.org/abs/https://doi.org/10.1063/1.874014}
  {\path{arXiv:https://doi.org/10.1063/1.874014}}, \href
  {https://doi.org/10.1063/1.874014} {\path{doi:10.1063/1.874014}}.
\newline\urlprefix\url{https://doi.org/10.1063/1.874014}

\bibitem{Watanabe2005}
T.-H. Watanabe, H.~Sugama,
  \href{https://doi.org/10.1088/0029-5515/46/1/003}{Velocity{\textendash}space
  structures of distribution function in toroidal ion temperature gradient
  turbulence}, Nuclear Fusion 46~(1) (2005) 24--32.
\newblock \href {https://doi.org/10.1088/0029-5515/46/1/003}
  {\path{doi:10.1088/0029-5515/46/1/003}}.
\newline\urlprefix\url{https://doi.org/10.1088/0029-5515/46/1/003}

\bibitem{Hawley1995}
J.~F. {Hawley}, C.~F. {Gammie}, S.~A. {Balbus},
  \href{https://ui.adsabs.harvard.edu/abs/1995ApJ...440..742H}{Local
  three-dimensional magnetohydrodynamic simulations of accretion disks}, The
  Astrophysical Journal 440 (1995) 742.
\newblock \href {https://doi.org/10.1086/175311} {\path{doi:10.1086/175311}}.
\newline\urlprefix\url{https://ui.adsabs.harvard.edu/abs/1995ApJ...440..742H}

\bibitem{Ball2021}
J.~Ball, S.~Brunner, \href{https://doi.org/10.1088/1361-6587/abf8f4}{A
  non-twisting flux tube for local gyrokinetic simulations}, Plasma Physics and
  Controlled Fusion 63~(6) (2021) 064008.
\newblock \href {https://doi.org/10.1088/1361-6587/abf8f4}
  {\path{doi:10.1088/1361-6587/abf8f4}}.
\newline\urlprefix\url{https://doi.org/10.1088/1361-6587/abf8f4}

\bibitem{Lapillonne2009}
X.~Lapillonne, S.~Brunner, T.~Dannert, S.~Jolliet, A.~Marinoni, L.~Villard,
  T.~G\"orler, F.~Jenko, F.~Merz,
  \href{https://doi.org/10.1063/1.3096710}{Clarifications to the limitations of
  the $s-\alpha$ equilibrium model for gyrokinetic computations of turbulence},
  Physics of Plasmas 16~(3) (2009) 032308.
\newblock \href {http://arxiv.org/abs/https://doi.org/10.1063/1.3096710}
  {\path{arXiv:https://doi.org/10.1063/1.3096710}}, \href
  {https://doi.org/10.1063/1.3096710} {\path{doi:10.1063/1.3096710}}.
\newline\urlprefix\url{https://doi.org/10.1063/1.3096710}

\bibitem{Mandell2021}
N.~Mandell, Magnetic fluctuations in gyrokinetic simulations of tokamak
  scrape-off layer turbulence (2021).
\newblock \href {http://arxiv.org/abs/2103.16062} {\path{arXiv:2103.16062}}.

\bibitem{gkeyllWeb}
{The Gkeyll team}, {The Gkeyll code}, \url{http://gkeyll.readthedocs.io}
  (2020).

\bibitem{Arnold2011}
D.~N. Arnold, G.~Awanou, {The Serendipity Family of Finite Elements},
  Foundations of Computational Mathematics 11~(3) (2011) 337--344.

\bibitem{Hakim2020}
A.~H. Hakim, N.~R. Mandell, T.~N. Bernard, M.~Francisquez, G.~W. Hammett, E.~L.
  Shi, \href{https://doi.org/10.1063/1.5141157}{Continuum electromagnetic
  gyrokinetic simulations of turbulence in the tokamak scrape-off layer and
  laboratory devices}, Physics of Plasmas 27~(4) (2020) 042304.
\newblock \href {http://arxiv.org/abs/https://doi.org/10.1063/1.5141157}
  {\path{arXiv:https://doi.org/10.1063/1.5141157}}, \href
  {https://doi.org/10.1063/1.5141157} {\path{doi:10.1063/1.5141157}}.
\newline\urlprefix\url{https://doi.org/10.1063/1.5141157}

\bibitem{Mandell2020}
N.~R. Mandell, A.~Hakim, G.~W. Hammett, M.~Francisquez, Electromagnetic
  full-$f$ gyrokinetics in the tokamak edge with discontinuous galerkin
  methods, Journal of Plasma Physics 86~(1) (2020) 905860109.
\newblock \href {https://doi.org/10.1017/S0022377820000070}
  {\path{doi:10.1017/S0022377820000070}}.

\bibitem{Xu2019}
W.~Xu, J.~M. Stone,
  \href{https://doi.org/10.1093/mnras/stz2002}{{Bondi–Hoyle–Lyttleton
  accretion in supergiant X-ray binaries: stability and disc formation}},
  Monthly Notices of the Royal Astronomical Society 488~(4) (2019) 5162--5184.
\newblock \href
  {http://arxiv.org/abs/https://academic.oup.com/mnras/article-pdf/488/4/5162/29173240/stz2002.pdf}
  {\path{arXiv:https://academic.oup.com/mnras/article-pdf/488/4/5162/29173240/stz2002.pdf}},
  \href {https://doi.org/10.1093/mnras/stz2002}
  {\path{doi:10.1093/mnras/stz2002}}.
\newline\urlprefix\url{https://doi.org/10.1093/mnras/stz2002}

\bibitem{Gorler2016}
T.~G\"orler, N.~Tronko, W.~A. Hornsby, A.~Bottino, R.~Kleiber, C.~Norscini,
  V.~Grandgirard, F.~Jenko, E.~Sonnendr\"ucker,
  \href{https://doi.org/10.1063/1.4954915}{Intercode comparison of gyrokinetic
  global electromagnetic modes}, Physics of Plasmas 23~(7) (2016) 072503.
\newblock \href {http://arxiv.org/abs/https://doi.org/10.1063/1.4954915}
  {\path{arXiv:https://doi.org/10.1063/1.4954915}}, \href
  {https://doi.org/10.1063/1.4954915} {\path{doi:10.1063/1.4954915}}.
\newline\urlprefix\url{https://doi.org/10.1063/1.4954915}

\bibitem{Held2016}
M.~Held, M.~Wiesenberger, A.~Stegmeir,
  \href{https://www.sciencedirect.com/science/article/pii/S0010465515003896}{{Three
  discontinuous Galerkin schemes for the anisotropic heat conduction equation
  on non-aligned grids}}, Computer Physics Communications 199 (2016) 29--39.
\newblock \href {https://doi.org/https://doi.org/10.1016/j.cpc.2015.10.009}
  {\path{doi:https://doi.org/10.1016/j.cpc.2015.10.009}}.
\newline\urlprefix\url{https://www.sciencedirect.com/science/article/pii/S0010465515003896}

\bibitem{Dingfelder2020}
B.~Dingfelder, F.~J. Hindenlang,
  \href{https://www.sciencedirect.com/science/article/pii/S0021999120300474}{{A
  locally field-aligned discontinuous Galerkin method for anisotropic wave
  equations}}, Journal of Computational Physics 408 (2020) 109273.
\newblock \href {https://doi.org/https://doi.org/10.1016/j.jcp.2020.109273}
  {\path{doi:https://doi.org/10.1016/j.jcp.2020.109273}}.
\newline\urlprefix\url{https://www.sciencedirect.com/science/article/pii/S0021999120300474}

\bibitem{Dorf2021}
M.~Dorf, M.~Dorr, \href{https://doi.org/10.1063/5.0039169}{Continuum
  gyrokinetic simulations of edge plasmas in single-null geometries}, Physics
  of Plasmas 28~(3) (2021) 032508.
\newblock \href {http://arxiv.org/abs/https://doi.org/10.1063/5.0039169}
  {\path{arXiv:https://doi.org/10.1063/5.0039169}}, \href
  {https://doi.org/10.1063/5.0039169} {\path{doi:10.1063/5.0039169}}.
\newline\urlprefix\url{https://doi.org/10.1063/5.0039169}

\bibitem{Zhu2018}
B.~Zhu, M.~Francisquez, B.~N. Rogers,
  \href{https://www.sciencedirect.com/science/article/pii/S001046551830208X}{{GDB:
  A global 3D two-fluid model of plasma turbulence and transport in the tokamak
  edge}}, Computer Physics Communications 232 (2018) 46--58.
\newblock \href {https://doi.org/https://doi.org/10.1016/j.cpc.2018.06.002}
  {\path{doi:https://doi.org/10.1016/j.cpc.2018.06.002}}.
\newline\urlprefix\url{https://www.sciencedirect.com/science/article/pii/S001046551830208X}

\end{thebibliography}

\end{document}